\documentclass[useAMS]{mn2e}

\usepackage{graphicx}

\newcommand{\MU}{[$\mu$]~}

\newcommand{\CCC}{$^{13}$C$^{16}$O~}
\newcommand{\CDC}{$^{12}$C/$^{13}$C~}
\newcommand{\Vt}{$V_{t}$~}
\newcommand{\lH}{$l/H$~}
\newcommand{\Tef}{T$_{\rm eff}$~}
\newcommand{\HHO}{H$_2$O~}

\newcommand{\mum}{$\mu$m~}

\title[
Carbon abundances and $^{12}$C/$^{13}$C from
globular cluster giants
]{
Carbon abundances and $^{12}$C/$^{13}$C from
globular cluster giants
}
\author[Yakiv V.Pavlenko et al.]{
Yakiv V.Pavlenko$^{1}$\thanks{E-mail:
yp@mao.kiev.ua}, Hugh
R.A.Jones$^2$\thanks{E-mail:hraj@astro.livjm.ac.uk} and
Andrew J. Longmore$^3$\thanks{E-mail:ajl@roe.ac.uk} \\
$^1$Main Astronomical Observatory of Ukrainian Academy of Sciences,
Golosiiv woods, 03680 Kyiv-127, Ukraine\\
$^2$Astrophysics Research Institute, Liverpool John Moores University,
Egerton Wharf, Birkenhead CH41 1LD, UK\\
$^3$UK Astronomy Technology Centre, Royal Observatory Edinburgh,
Blackford Hill, Edinburgh EH9 3HJ, Scotland}

\begin{document}

\date{Accepted 1988 December 15. Received 1988 December 14; in original form 1988 October 11}

\pagerange{\pageref{firstpage}--\pageref{lastpage}} \pubyear{2003}

\maketitle

\label{firstpage}

\begin{abstract}

The behaviour of the $\Delta\nu$=2 CO bands around 2.3 $\mu$m was examined by
comparing observed and synthetic spectra in
stars in globular clusters of different metallicity.
Changes in the $^{12}$C/$^{13}$C isotopic ratio and the carbon abundances
were investigated in stars from 3500--4900 K in the galactic globular
clusters M71, M5, M3 and M13, covering the metallicity range
from --0.7 to --1.6.
We found relatively low carbon abundances that
are not affected by the value of oxygen abundance. For
most giants
the $^{12}$C/$^{13}$C ratios determined are consistent with the
equilibrium value for the CN cycle. This suggests complete mixing on the
ascent of the red giant branch, in contrast to the substantially higher values
predicted across this range of parameters by the current
generation of models. We found some evidence for a larger dispersion of
\CDC in giants of M71 of metallicity [$\mu]$ = [M/H]
= -0.7 in comparison with the
giants of M3, M5 and M13, which are more metal deficient.
 Finally, we show  evidence for lower
\CDC in giants of globular clusters with lower metallicities, as
 predicted by theory.

\end{abstract}


\section{Introduction}

Abundance patterns are sensitive indicators of the history of
nuclear processing
and mixing processes inside stars.
During the main sequence
phase of its lifetime, a star of mass above around 0.8 M$_\odot$
will convert $^{12}$C into $^{13}$C (and also $^{14}$C) via the CN cycle.
If material which has undergone any degree of CN cycle burning is mixed
to the surface of a star, the carbon isotope ratio $^{12}$C/$^{13}$C will
decrease.

The equilibrium \CDC value for the CN cycle is $\sim$ 3.5. Standard
models of the mixing occurring at the first ascent of the Pop I red
giant branch find \CDC ratios in the range from 90
(solar value, Anders \& Grevesse 1989) to 20 (see Iben 1964,
Iben \& Renzini 1983, Sneden et al. 1986, Charbonnel 1994). The
first measurements of the \CDC in the atmospheres of field red giants
showed lower than predicted values.
Population I giants have carbon isotope
ratios in the vicinity of 5--40 (Lambert \& Ries 1981, see also
references in Briley et al. 1997). This discrepancy cannot be
explained by low initial isotope ratios (Sneden at al. 1986,
Gilroy \& Brown 1991).

In general, the situation
with the theory of \CDC evolution in Pop II giants is even more
problematic than for Pop I stars: results of modelling depend on metallicity.
Due to the lower opacity inside these stars, they are
more luminous.  In general, they evolve on a shorter
time scale than do Population I stars.
Thus, we expect that initial
\CDC (and carbon abundance, log N(C)) values may vary with
age and thus metallicity {\MU}.
For our range of masses, effective temperature (\Tef) and metallicities conventional theory
predicts rather high ratios of \CDC $>$ 20 (see Fig. 4 in
Boothroyd and Sackmann 1999) after both first and second dredge-up
mixing.
Models of the evolution of metal deficient halo stars show
that the \CDC isotopic ratio
should be slightly lower in metal-deficient stars (Charbonnel 1994).

The canonical picture of the evolution of globular cluster stars suggests
that they
should consist of chemically homogeneous populations of stars.
In reality observations indicate that
most galactic globular clusters exhibit star-to-star variations
of atmospheric abundances of C,N,O, and other elements associated
with proton capture reactions.
Gratton et al. (2001) find some evidence that low-mass
clusters are chemically more homogenous;
abundances of Fe and Ca are almost the same among stars
of a given cluster (cf. Kraft et al. 1997; Ivans et al. 2001).
However, there are two exceptions: M22 and $\omega$ Cen (cf.
Vanture et al. 2002, Smith et al. 2002).

Two main explanations have been proposed for the
observed CNO abundance variations (cf Kraft 1994): a) some of the stars
(CN-strong stars) were formed by gas that had been
pre-enriched by nitrogen and other elements during
star formation by accretions of ejecta from more
massive and fast evolving stars; b) observed CN enhancements are due to
deep mixing processes, which have brought freshly synthesized
products from interior regions of active CNO element
nucleosynthesis to the stellar
surface (Sweigart \& Mengel 1979).  Deep mixing elevates
the N-rich and C (and possibly O) deficient matter from
the CN (and possibly ON) cycle regions above the H-burning shell into the
outer convective envelope during RGB ascent, causing corresponding
changes to surface abundances.

The combined model
of primordial abundance variations among the stars and
deep mixing processes during the late stages of evolution
seems to explain the observed abundance inhomonogeneties
(Messenger \& Lattanzio 2002, Ramires \& Cohen 2002, Briley \&
Cohen 2001, Briley et al. 2001).  The solution to this problem is
complicated by the fact that
similar chains of nuclear reactions, i.e. proton
captures by C,N,O, Ne and Mg are associated with both mechanisms
(see Briley et al. 2002 for more details).

Therefore determinations of \CDC can play a key role in improving
our knowledge about the evolution of globular clusters giants.
Specifically, observations of the \CDC isotope ratio should allow us to
pinpoint the evolutionary
phase at which mixing occurs. Isotope ratios in stars with a
range of metallicity can give us information about the role of
metal abundance in suppressing or enhancing mixing. Previous work
concentrated on halo giants, showing that carbon isotope ratios in
metal poor stars decline abruptly to very low values at log g
$\sim$ 2 on the first ascent of the giant branch (Sneden et al.
1986; Pilachowski et al. 1997). Pilachowski et al. (1997) also
found weak evidence for a continuing decline in carbon isotope
ratio with increasing luminosity in metal poor giants from log g
= 2 to 0. Although previous studies benefit from field stars
being relatively brighter there are a couple of advantages to
studying globular cluster giants. Firstly relative to field
giants, globular cluster giants can be expected to have arisen
from similar composition material and thus should  be composed of
more homogeneous material. Secondly, stars in a particular
cluster can be considered to be at a common distance (the
diameter of a globular cluster being negligible in comparison to
the distance from us). Thus their distances are better determined
than the distances of field halo giants.

In this paper, carbon
isotope ratios of 28 relatively well studied red giant stars
taken from globular clusters with a range of
metallicities\footnote{
In this paper the definition of metallicity is related to iron
and other heavy metal abundances though not to those of carbon
and oxygen.} \MU =
[M/H] = [Fe/H]$_{*}$--[Fe/H]$_{\odot}$
= --0.7 (M71) to \MU = --1.6 (M3) were investigated. Section 2
describes the observations, Section 3 presents the model
atmospheres and fits to the observed spectra, Section 4 is
discussion and Section 5 the conclusions.

\section[]{Observations}

Observations were made with the Cooled Grating Spectrometer 4 (CGS4)
on the UK Infrared Telescope (UKIRT) on Mauna
Kea, Hawaii.  The instrument then had a 58 $\times$ 62 InSb array which was
moved in the focal plane in order to three times over-sample the
spectrum.  Sky subtraction was performed by nodding the telescope
approximately 30 arcsec up and down the slit, ensuring that during
alternate `object' and `sky' observations the star remained on the
detector.  The observations presented in this paper were made during
three nights, 1992 April 23 and 24 in a wide
variety of conditions of optical seeing (0.75--2 arcsec) and of
atmospheric humidity (10--75 per cent).

Observations were carried out  for a sample of globular cluster stars,
without detailed analysis of their membership. Our later analysis based
on the literature sources has shown that at least some of the stars are
non-members (see section 4).

The 150 lines mm$^{-1}$ grating was used in third order with the
150 mm focal length camera at a central grating wavelength of
2.34$\mu$m.  This grating position was chosen as it allows
simultaneous coverage of the $^{12}$CO and $^{13}$CO band heads.
This  position also enabled CGS4 to work at high grating
efficiency in a region of relatively high atmospheric transmission
with relatively high resolution.

To remove telluric bands of water, carbon
dioxide and methane, we observed A type standards.
Such stars are not expected to have features in common
with our target stars and are mainly featureless (Malkan et al. 2002).
The airmass difference between object and standard
used never exceeded 0.05 and so we are confident that the spectra have
good cancellation of atmospheric features.
Both the object and the standard were wavelength
calibrated by using arc lines of krypton, argon and xenon and OH
lines. To extract the spectrum from the
sky subtracted signal an Optimal Extraction technique was
used; this combines the rows using weights based on the spatial
profile of the stellar image.  The spectra were reduced following
Jones et al. (1994) using the
$\sc{Figaro}$, $\sc{Specdre}$ and $\sc{Kappa}$ packages provided and
supported by Starlink.

\section{Procedure}

 There are several ways of determining abundances of the CNO elements and
the carbon isotopic ratio in the atmospheres of late type stars. In
studies of CNO abundances, one method is to derive O abundances from
atomic lines or OH bands, followed by CH bands to set [C/Fe] (Cottrell
1978). Other approaches have used CO bands at 2.3 \mum (Smith \&
Suntzeff 1989) or CO bands at 1.6 \mum (Bell \& Briley 1991, Smith et
al. 2002). If C and O are known, CN bands in the blue and red can be
used
to constrain [N/Fe] (Briley et al. 2001, 2002). Note, however, that in
some studies, as in our case, O is not known and some average
value
must be assumed. In some cases it is possible to use NH for a direct
measurement of [N/Fe] (Yakovina \& Pavlenko 1998). A self consistent
approach for determination of log N(C) and \CDC from fits to observed
spectra of OI, molecular C$_2$ and CN was developed by Pavlenko
(1991) and Boyarchuk et al. (1991). In most cases, carbon isotope ratios
can be determined from both red $^{13}$CN bands and $^{13}$CO bands.

\begin{figure*}
\begin{center}
\includegraphics [width=62mm]
{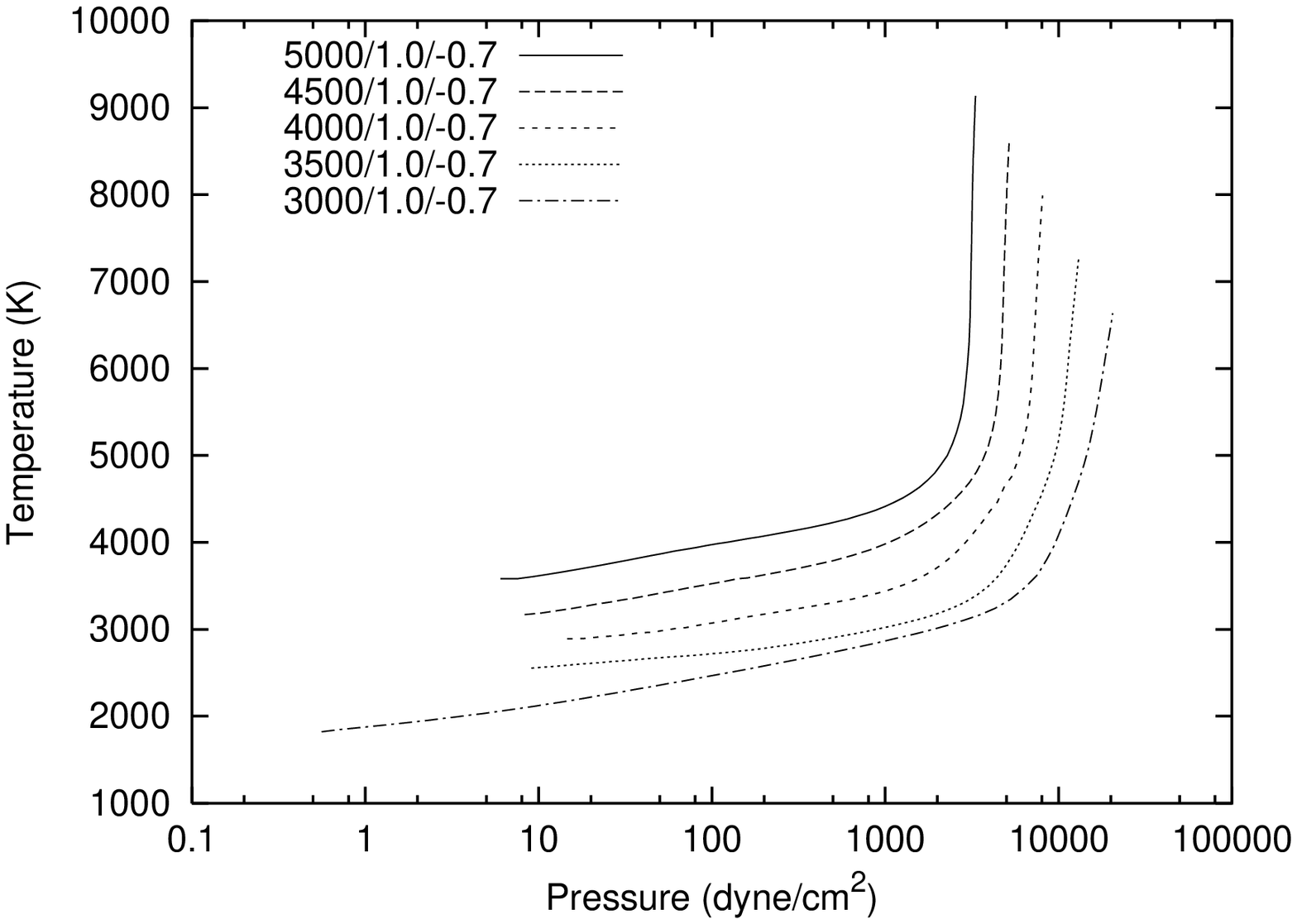}
\includegraphics [width=62mm]
{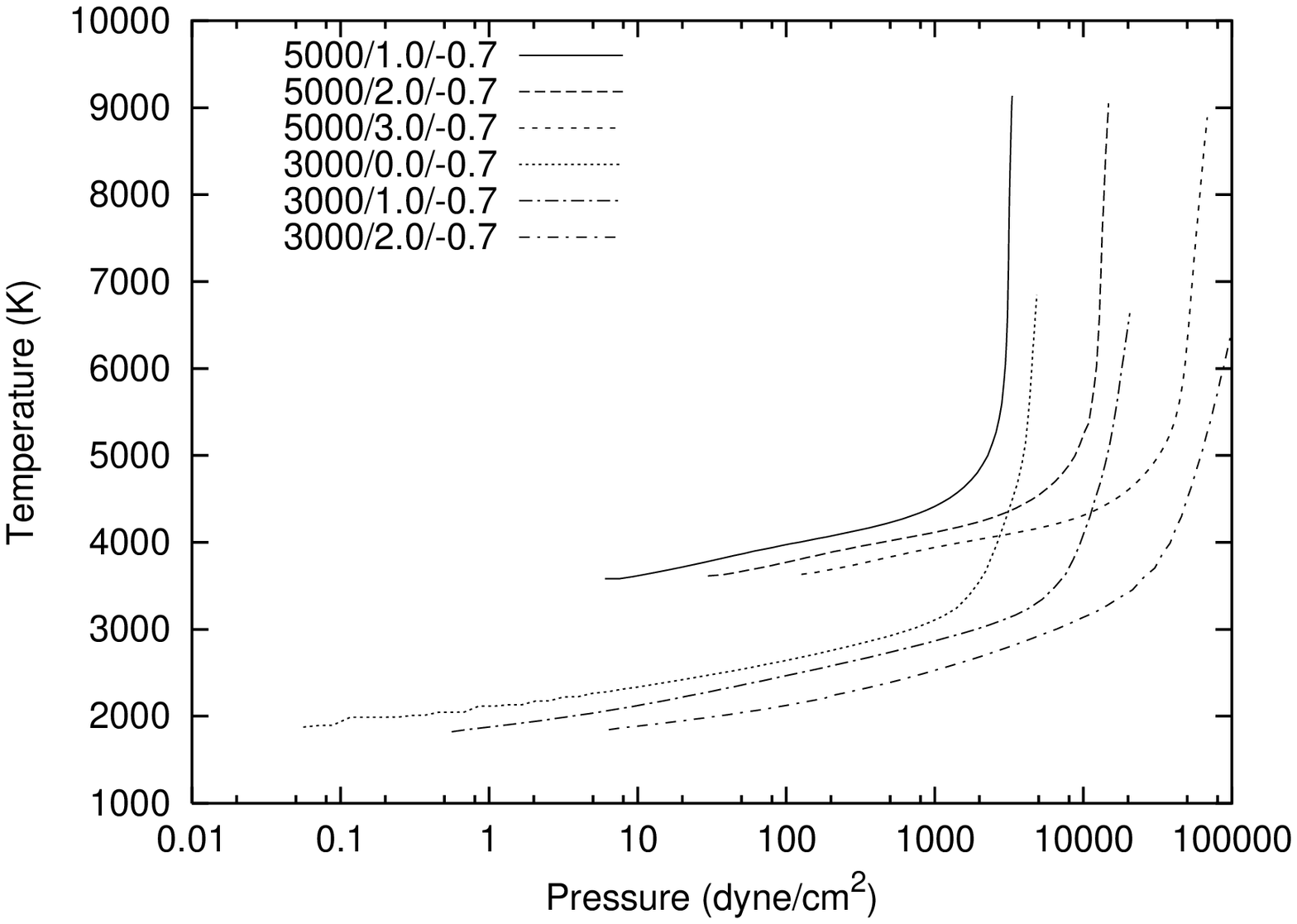}
\includegraphics [width=62mm]
{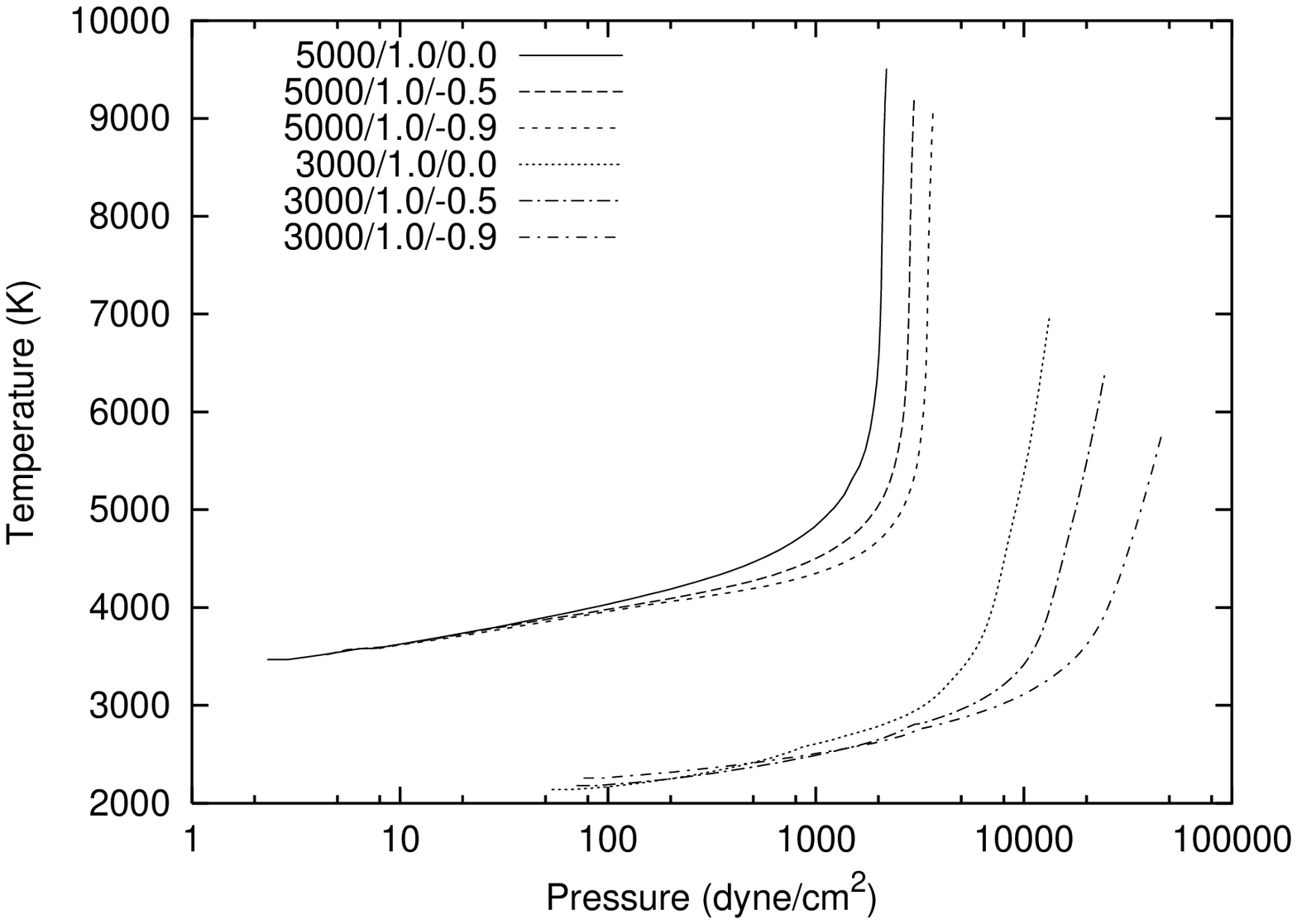}
\includegraphics [width=62mm]
{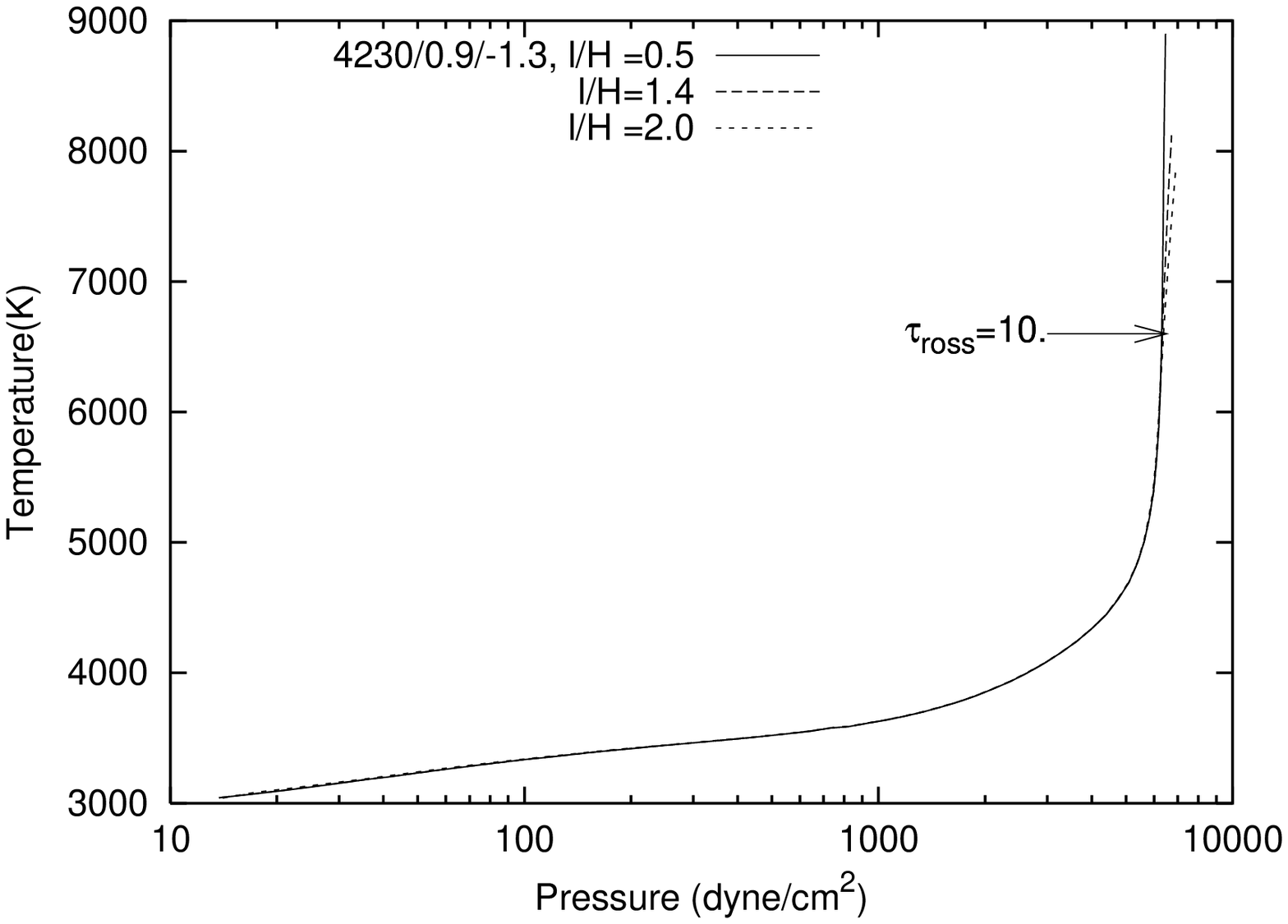}
\end{center}
\caption[]{\label{_001t_} Temperature versus pressure in red giant
model atmospheres of different \Tef, log g, \MU, \lH}
\end{figure*}

\begin{figure*}
\begin{center}
\includegraphics [width=62mm]
{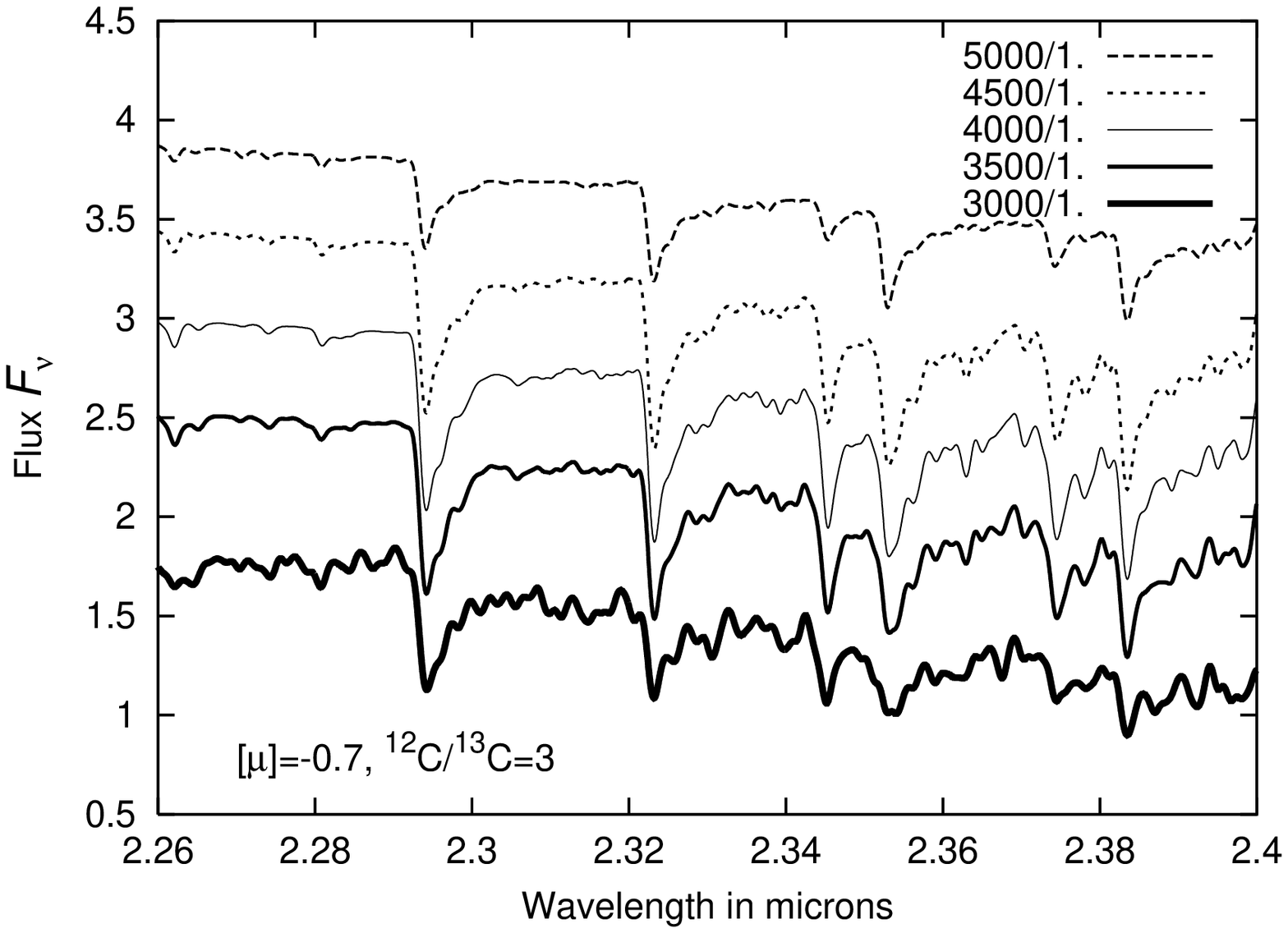}
\includegraphics [width=62mm]
{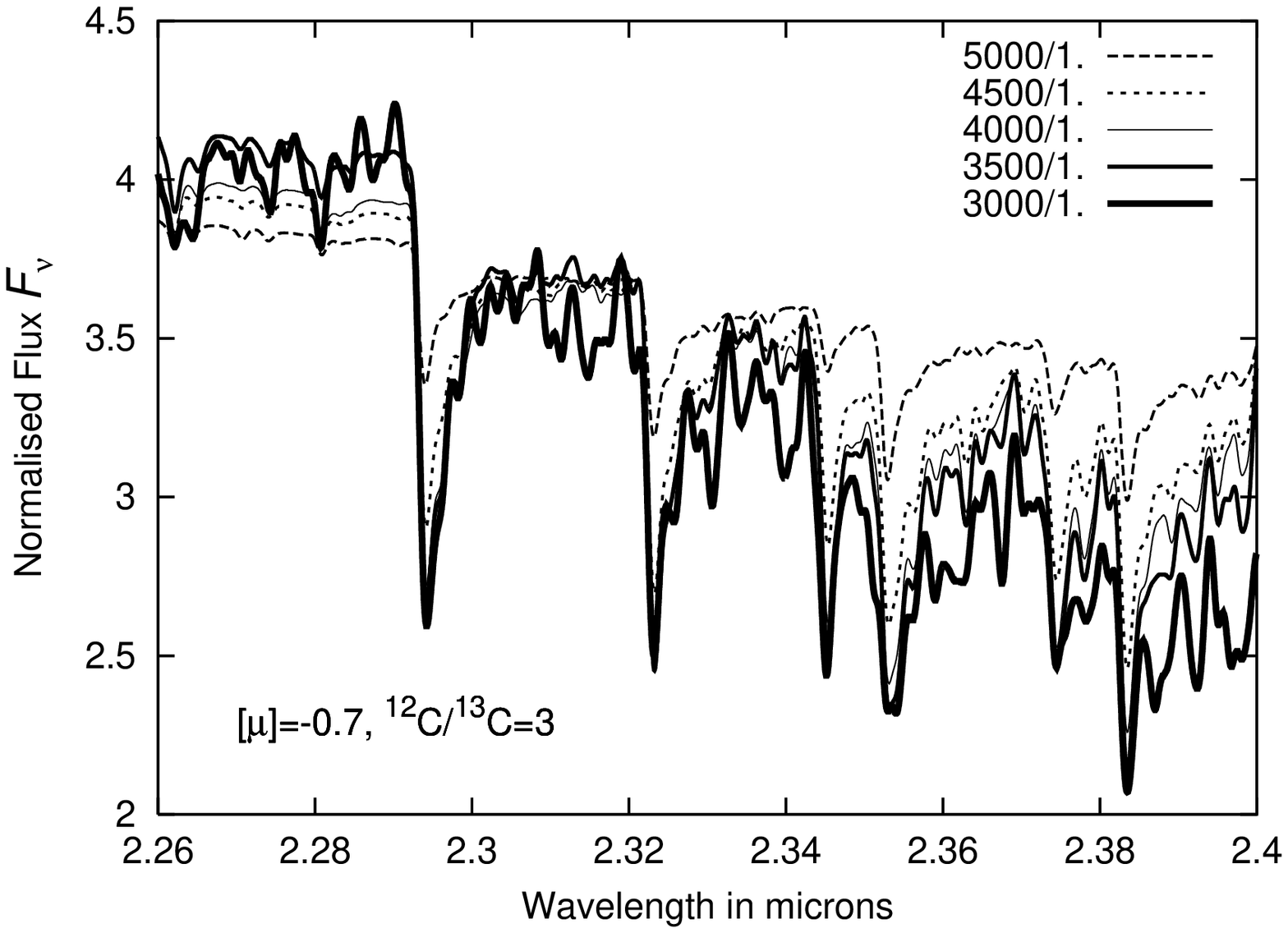}
\end{center}
\caption[]{\label{_001ab_} Left: $F_{\nu}$ fluxes computed for
model atmospheres of \MU = -- 0.7. Right: Fluxes normalised to
equal value at 2.344 $\mu$m. The regular absorption features are
caused by CO bands. The smaller scale features appearing in the
spectra below 4500~K are due to water vapour.}
\end{figure*}


\subsection{Model Atmospheres and Synthetic Spectra}
Model atmospheres were computed using the SAM12 program (Pavlenko
2002, 2003), which is a modification of ATLAS12 (Kurucz 1999).
 The standard set of continuum opacities included in ATLAS12
by Kurucz (1999) is used. Computations of a few opacities were added or
upgraded: the absorption of C$^-$ (Myerscough \& McDowell
1966) and H$_{2}^-$ Doyle (1968). To compute opacities due to
bound-free absorption of C, N, O atoms we used TOPBASE (Seaton et
al. 1992) and cross-sections from Pavlenko \& Zhukovska (2003). Our
bound-free
transition opacity tables are available on the Web (Pavlenko 2003b).

Opacities due to atomic and molecular lines absorption were taken
into account using the opacity sampling approach (Sneden et al.
1976). Line list data were taken from different sources: atomic
lines from VALD (Kupka et al. 1999); lines of diatomic molecules
CN, C$_2$, CO, SiH, CO from CDROM 18 of Kurucz (1993); H$_2$O line
list from the AMES database (Partridge \& Schwenke 1998); OH line list
from Schwenke (1997). TiO and VO opacities were taken into account
using the JOLA approach. Algorithms for molecular band opacity
computations allow reproduction of the relative strength and
shapes of the spectral energy distributions for late type stars
(Pavlenko 1997, 2000, 2003).

The shape of every line was determined using the Voigt function $H(a,v)$.
Damping constants were taken from line databases or computed using
Unsold's approach (Unsold 1955). In model atmosphere computations
we adopt a microturbulent velocity, \Vt, of
2 km/s. For some spectra we investigated the effect of varying \Vt
(see Table \ref{__M5_1__} and section 4).

 The photospheres of K- and M-giants lie on the upper boundary of their
convective envelopes. In fact convection determines the lower boundary
conditions for the equations of energy transfer in model atmosphere
computations. The mixing length theory of 1D convection modified by
Kurucz(1999) is used in ATLAS12 and SAM12.

In the frame of this approach the mixing length parameter \lH should
be defined. Heiter et al., (2002) find that spectroscopic measurements
require 0.5, Asida (2000) derives \lH =1.4, Kurucz (1993, 1999) used \lH
= 1.25. Generally speaking, the choice of \lH is rather arbitrary --- it
is actually a combination of different parameters often constructed with
different formalisms. In that sense different headline values of mixing
lengths from different models may in fact be equivalent (Salaris,
Cassisi \& Weiss 2002).

Two grids of model atmospheres used in this paper were computed for
\lH = 2.0 and 1.4. \lH = 2.0 represents an approximate upper limit for
\lH. However,
our computations indicate that only the inner region of our globular
cluster photospheres is sensitive to \lH (Fig. \ref{_001t_}). Our
synthetic spectra show very weak dependence: variations in $S$ used to
quantify the minimisation procedure are  marginal ($<$ 1 \%) when  \lH
changes from 0.5 to 2.0 in the model atmospheres (Fig.
\ref{_001t_}); however, also see section 3.4.

The model atmospheres were computed in the range of   T$_{\rm
eff}$ = 3000 to 5000 K, log g = 0.0 to 2.0, \MU = --0.5 to --2.0, and
different carbon abundances log N(C). For every giant we computed 10
model atmospheres of different log N(C), with 0.2
dex as a step for log N(C).   Then, for each value of log N(C) we
computed a grid of synthetic spectra for the wavelength range from
2.26--2.39 \mum. Synthetic   spectra were computed for a grid
of \CDC = 90, 40, 20, 10, 9, 8, 7, 6, 5, 4, 3.

In our computations the abundances of oxygen and nitrogen were fixed,
i.e. log N(N)= log N$_{\odot}$(N) + \MU, log N(O) = log
N$_{\odot}$(O) + \MU.
To study the dependence of our results on log N(O) we
carried out some numerical experiments with variable oxygen abundances
(see section 3.4)

Numerical computations of theoretical radiative fluxes $F_{\nu}$
in spectra of red giants were carried out within the classical approach:
LTE, plane-parallel model atmosphere and no energy divergence
by WITA6 (Pavlenko 2000).

The spectra are calculated using a self-consistent approach:
model atmospheres and synthetic spectra are computed for the same
opacity source lists and sets of abundances. This approach allows us to
treat the possible impact of abundance changes on the temperature
structure of the model atmosphere as well as on the emitted fluxes
in a direct way.

We will use  a definition of
`synthetic spectra' as an alias for ``theoretical spectral
energy distributions of the spectrum of red giants''. Computations
were carried out with wavelength steps $\Delta\lambda$ = 0.05 nm
and 0.01 nm (see section 3.4).

Our spectra of the second overtone bands of CO are
contaminated by water lines formed inside the M-giants
atmospheres.
In our computations we always took into account
absorption using AMES line lists of CO (Goorvitch 1994), \HHO
(Partrige \& Schwenke 1997) and atomic line list from VALD (Kupka et al. 1999).
The reliability of the AMES lists for astrophysical
computations was shown by Jones et al. 2002 and
Pavlenko \& Jones 2002.

\subsection{Dependence on input parameters}

\subsubsection{Dependence on T$_{\rm eff}$}

In Fig. \ref{_001t_} we show the temperature structure T =
f(P$_g$) for some of our model atmospheres with
different \Tef, log g, and \MU. Photospheres of cooler
giants move downwards to more dense layers due to the overall
reduction of opacity. In general, for lower T$_{\rm eff}$ the
curves T = f(P$_g$) in Fig. \ref{_001t_} are systematically shifted toward
lower temperature/higher pressure regions. A comparison of
computed fluxes is shown in Fig. \ref{_001ab_}.
It is worth noting: \\

$\bullet$ for T$_{\rm eff}$ = 3500--4500
K the overall shape of spectral energy distributions is governed
by CO; larger changes are seen for T$_{\rm eff}$ =
4500--5000 K, these are primarily caused by differences in the
temperature structure of the models  (see right panel of
Fig. \ref{_001ab_}).\\

$\bullet$ if T$_{\rm eff}$ drops below 4500 K bands of H$_2$O
appear in the region (see Pavlenko \& Jones 2002 for more details).\\

\begin{figure*}
\begin{center}
\includegraphics [width=62mm]
{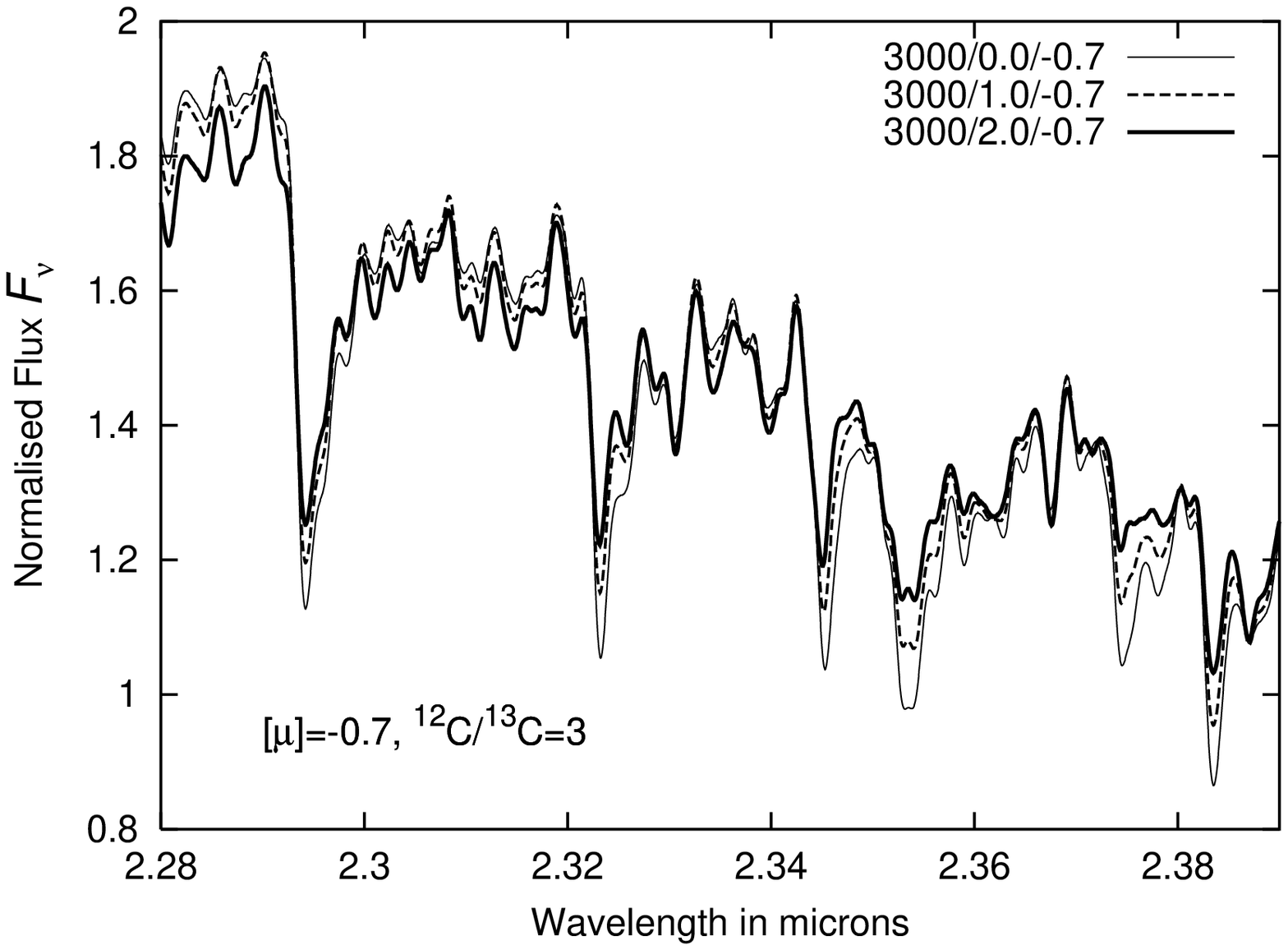}
\includegraphics [width=62mm]
{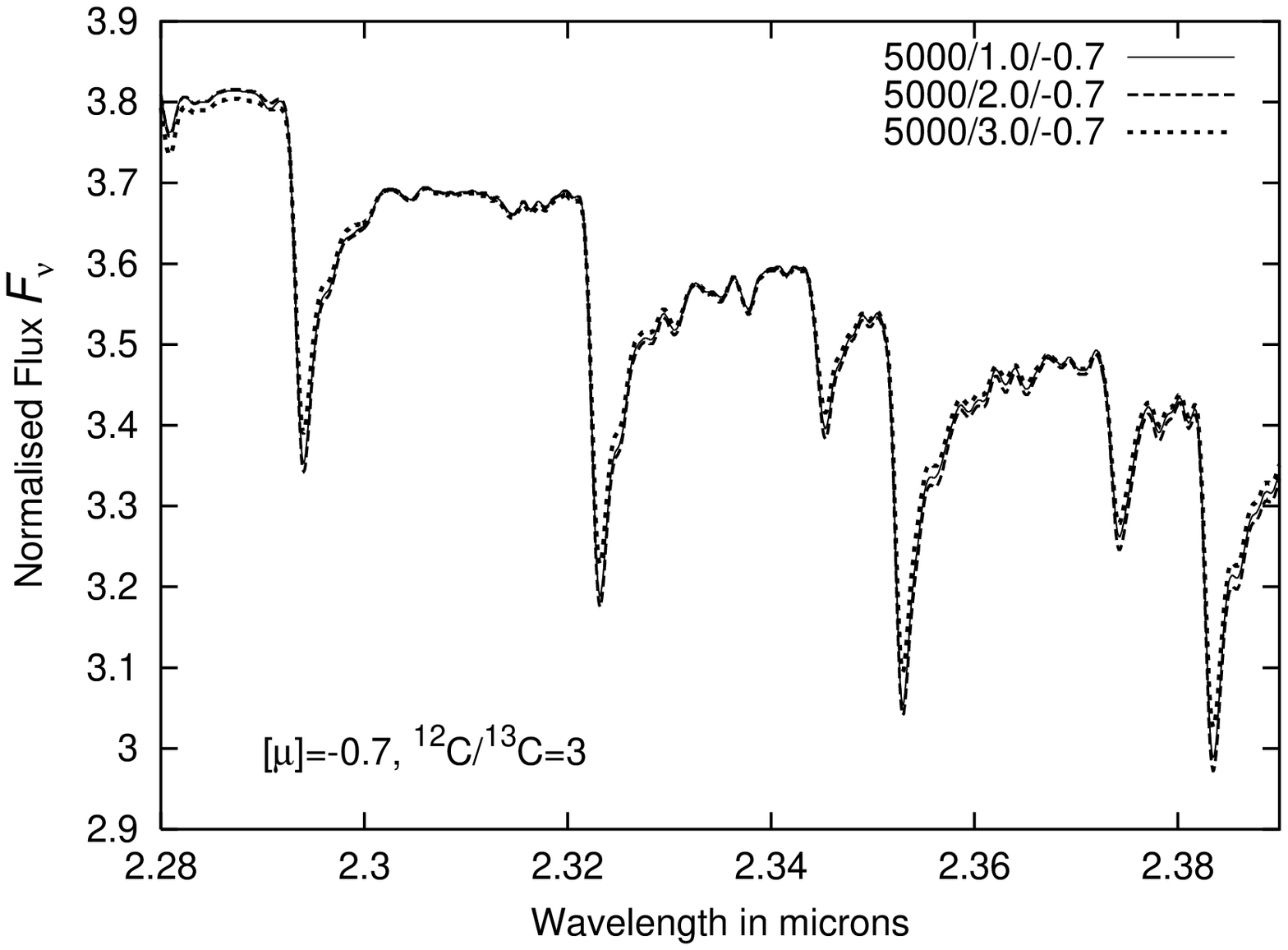}
\end{center}
\caption[]{\label{_fnu-logg_} Synthetic spectra computed for model
atmospheres of different log g --- left: T$_{\rm eff}$ = 3000 K,
right: T$_{\rm eff}$ = 5000 K.}
\end{figure*}

\begin{figure*}
\begin{center}
\includegraphics [width=62mm]
{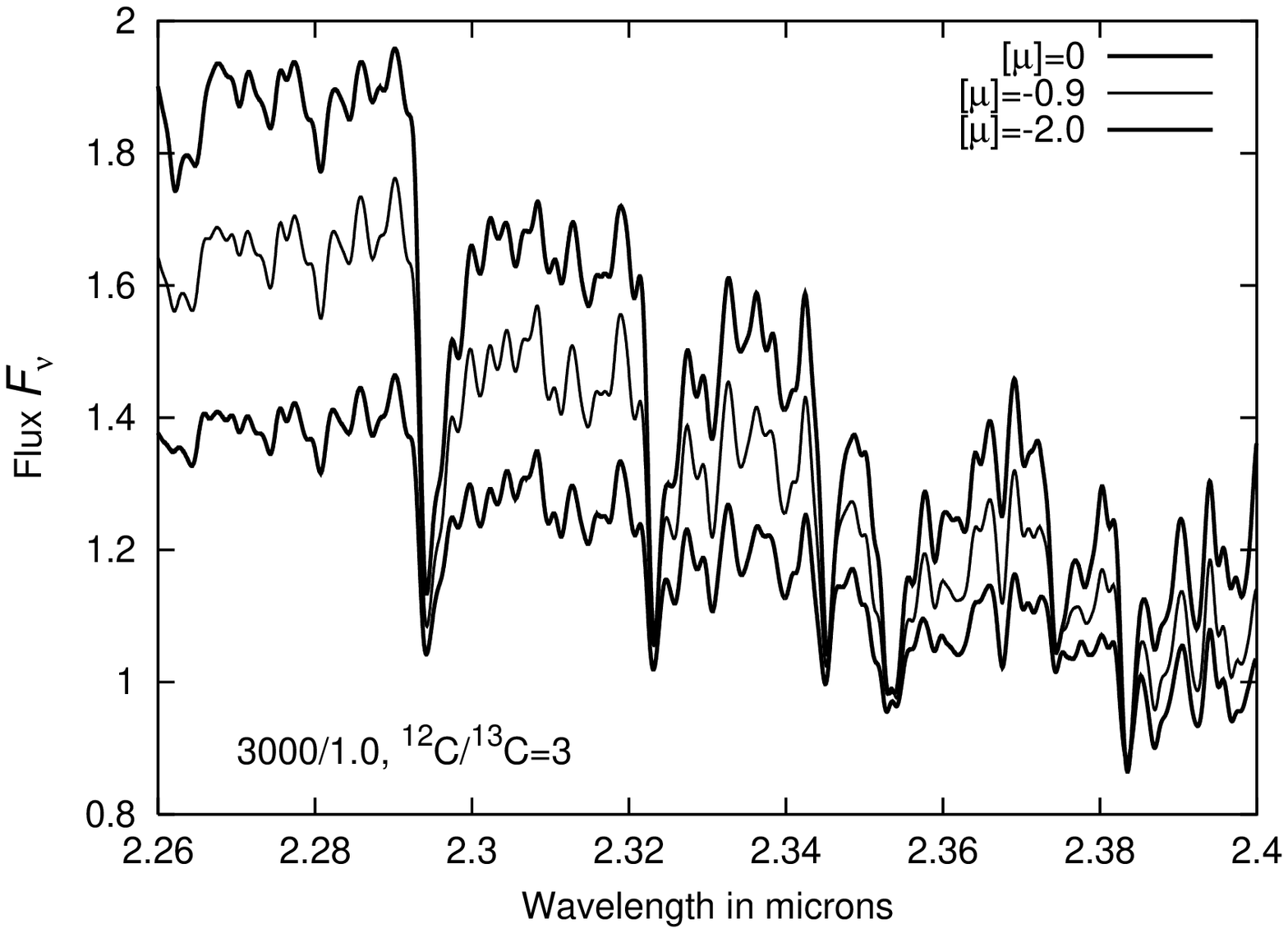}
\includegraphics [width=62mm]
{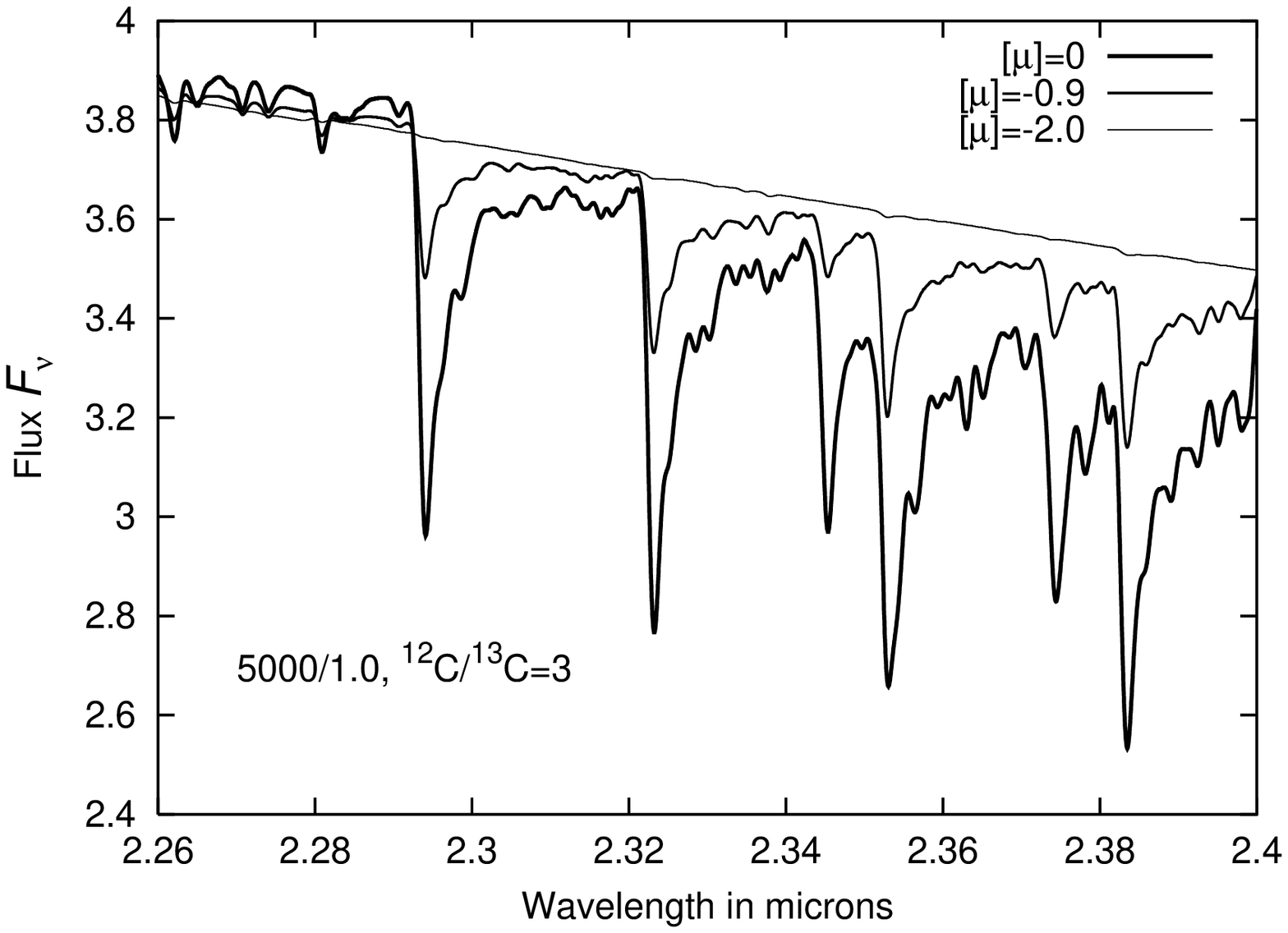}
\end{center}
\caption[]{\label{_fnumu_} Synthetic spectra computed for model
atmospheres of different \MU --- left: T$_{\rm eff}$ = 3000 K,
right: T$_{\rm eff}$ = 5000 K.
}
\end{figure*}

\subsubsection{Dependence on log g}

Computed temperature structures T =f(P$_g$) of some model atmospheres
of T$_{\rm eff}$ = 3000 and 5000 K and different log $g$ are
shown in Fig. \ref{_001t_}.
There is a wide range of characteristic pressure for the photospheric
layers across all the models we calculated.
On the other hand, the dependence on log g of our
computed spectra for model atmospheres of T$_{\rm eff}$ = 5000 and 3000 is
quite different (Fig. \ref{_fnu-logg_} ). Specifically,
spectra computed for T$_{\rm eff}$ = 5000 K show a rather
weak dependence on log g.

\subsubsection{Dependence on metallicity}

Small variations of \MU ($\pm$ 0.2 dex) have little influence on the
temperature structure of model
atmospheres (Fig. \ref{_001t_}). Continuum absorption governs
the opacity in metal deficient atmospheres. Both CO
and H$_2$O molecular densities are very sensitive to the \MU parameter
(Fig. \ref{_fnumu_}).

\subsubsection{Dependence on log N(O) and log N(C)}

\begin{figure*}
\begin{center}
\includegraphics [width=62mm]
{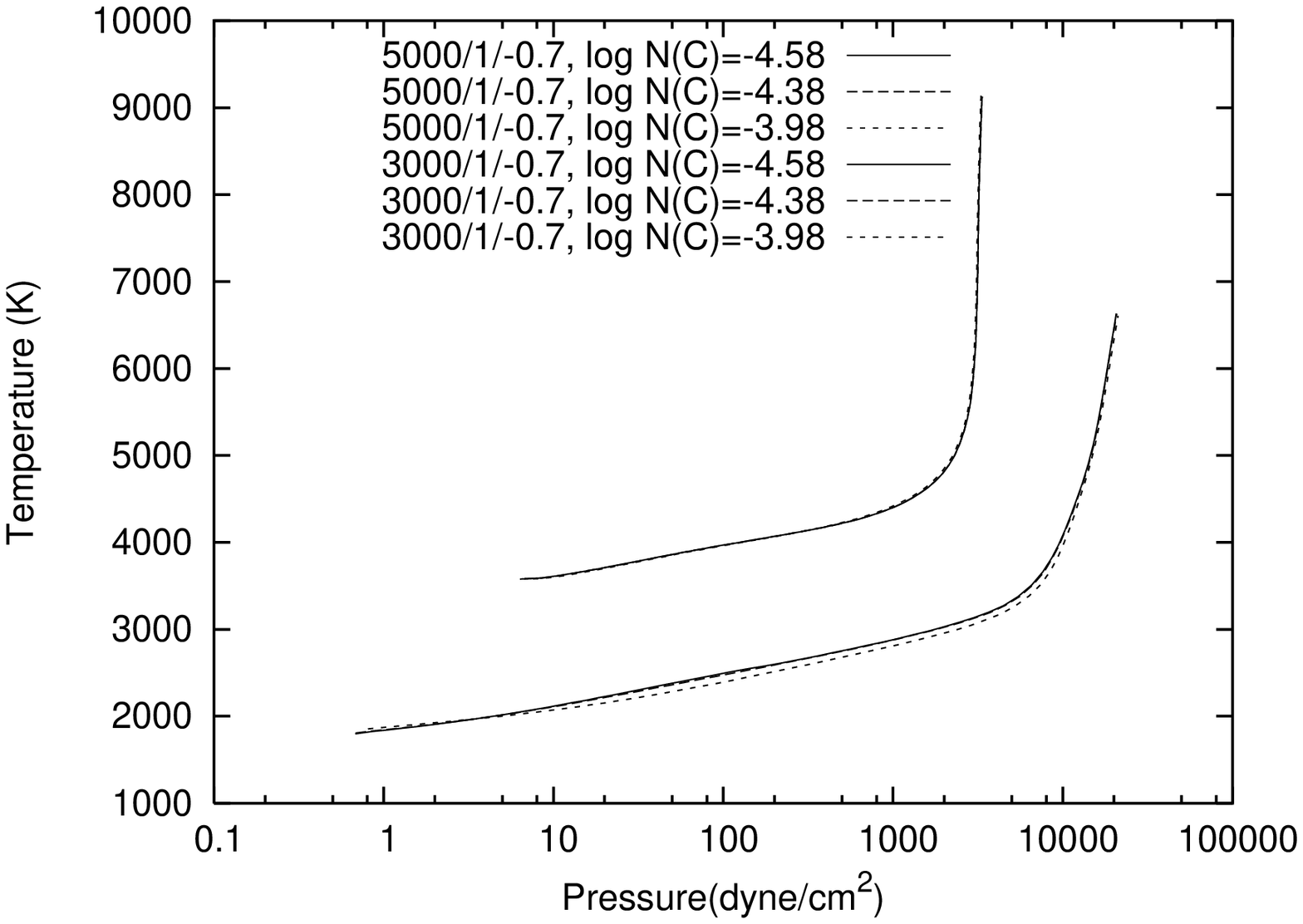}
\includegraphics [width=62mm]
{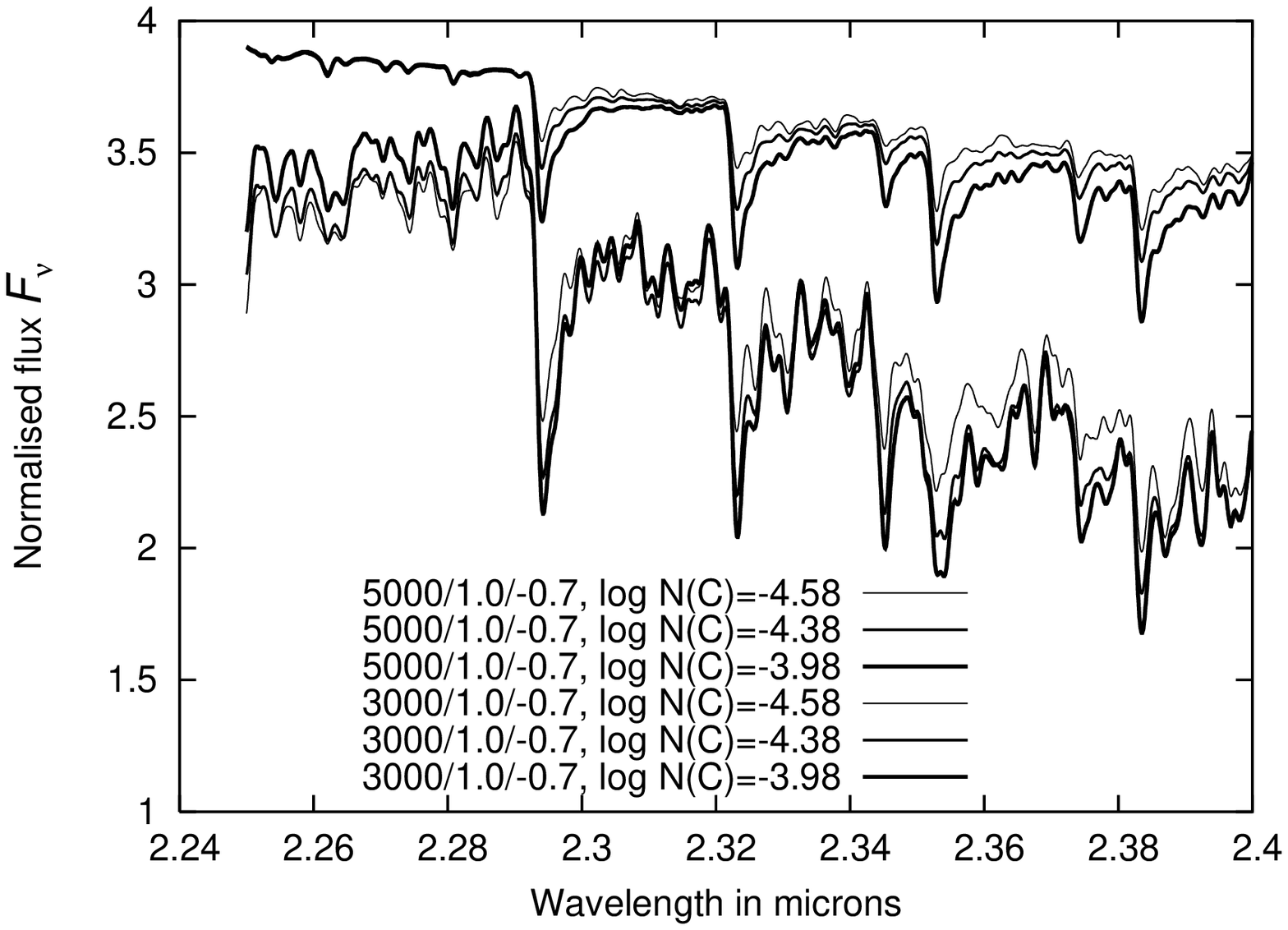}
\end{center}
\caption[]{\label{_tc4_} Computed temperature structures of model
atmospheres with different log N(C) and the synthetic spectra for
these model atmospheres.}
\end{figure*}

\begin{figure*}
\begin{center}
\includegraphics [width=62mm]
{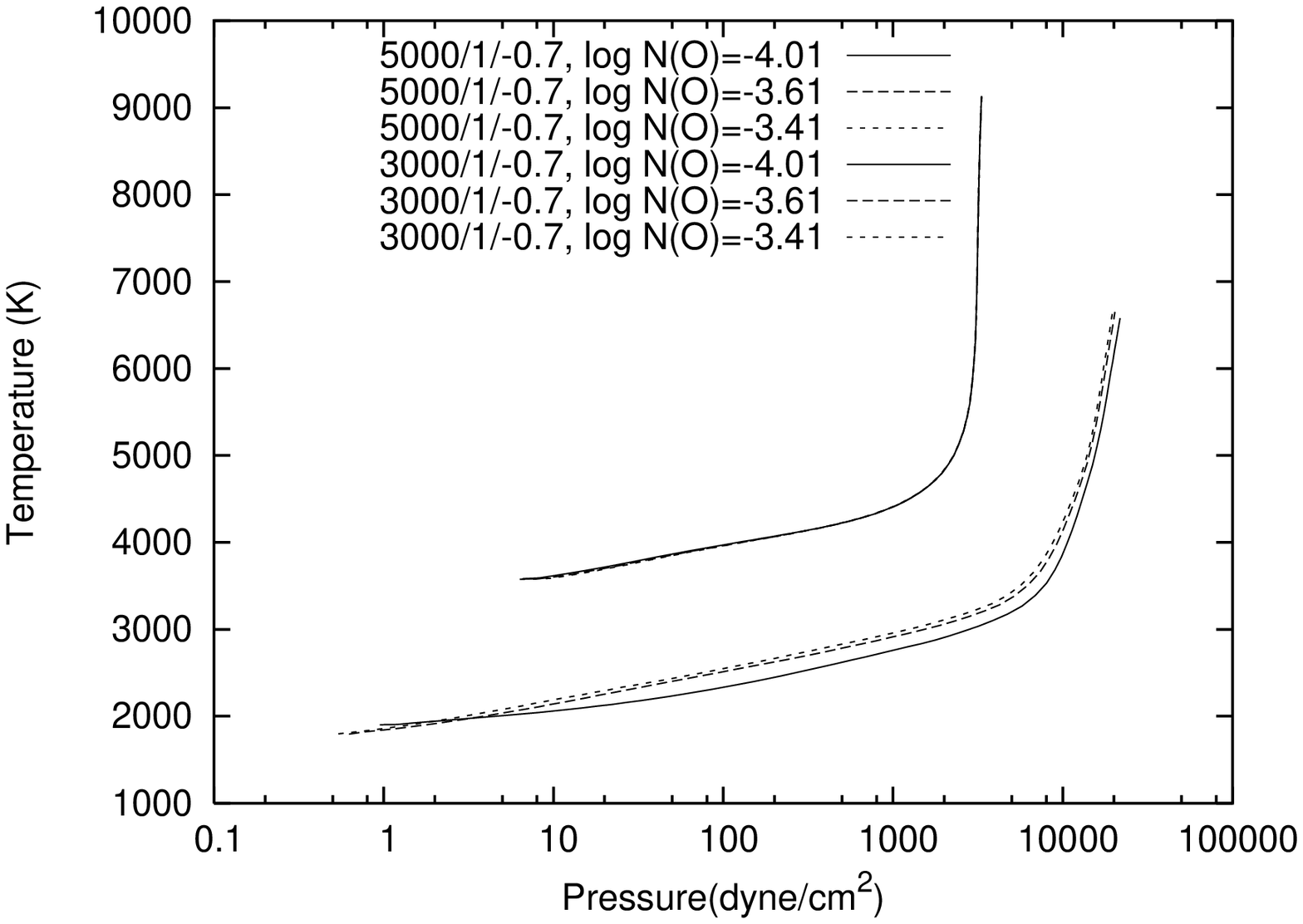}
\includegraphics [width=62mm]
{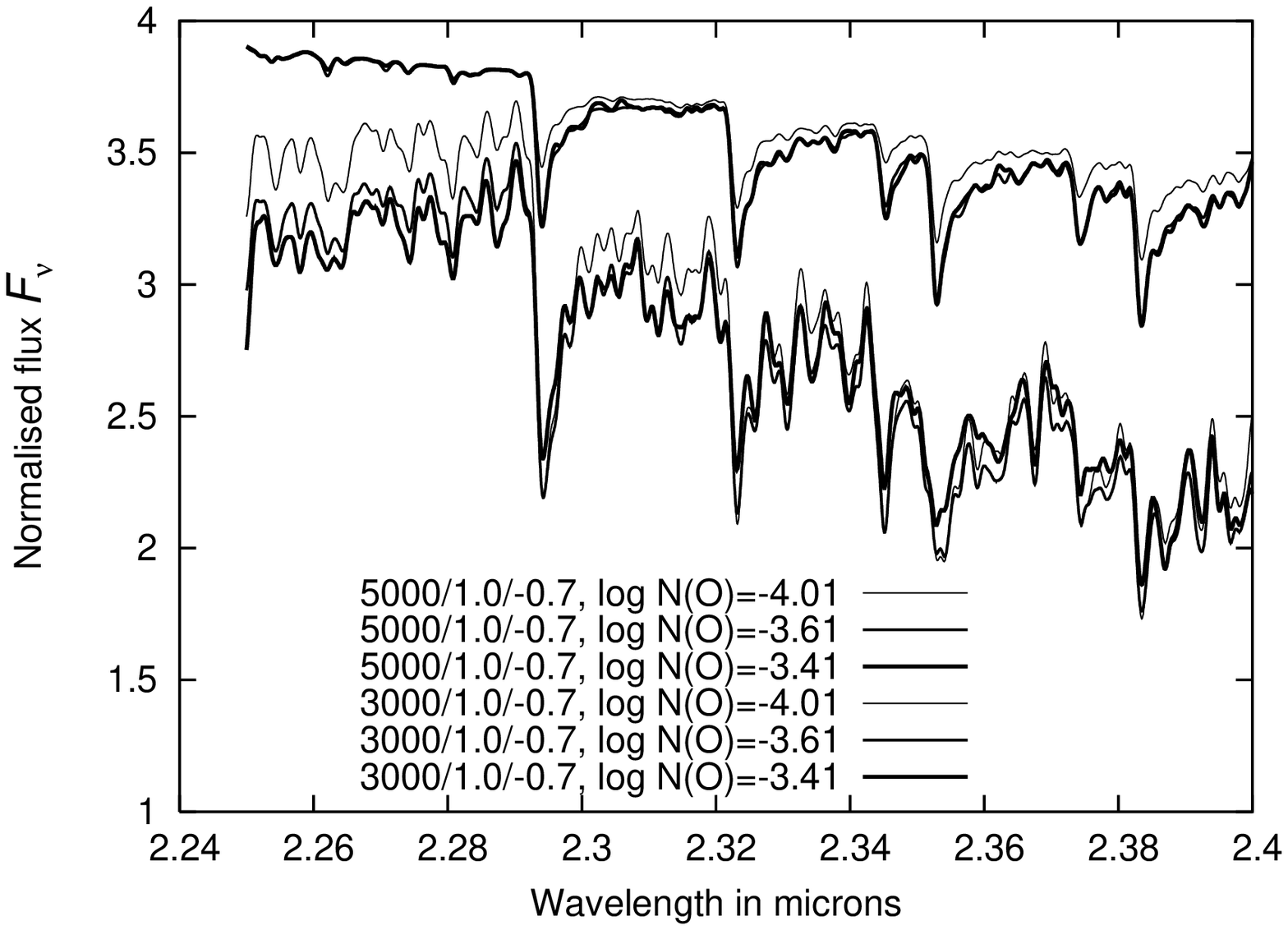}
\end{center}
\caption[]{\label{_to4_} Computed temperature structures of model
atmospheres with different log N(O) and the
synthetic spectra for these model atmospheres.}
\end{figure*}

The chemical balance in red giant atmospheres is governed by the
C/O ratio through the formation of CO and H$_2$O molecules (Tsuji
1973). In the atmospheres of M-giants C/O $<$ 1 and the majority
of carbon atoms are bound into CO molecules. For globular cluster
giants, C/O $<<$ 1. As a result, the profiles of CO bands in IR
spectra of the M giants we modelled show strong sensitivity to log
N(C),  see Fig. \ref{_tc4_}. We used this dependence for finding log N(C) for our M giant
spectra. On the other hand, the temperature structure of model
atmospheres as well as the profiles and relative strengths of the
computed CO bands show rather weak dependence on log N(O) (see
Figs. \ref{_to4_}), especially if log N(C) and log N(O) differ
significantly, as in atmospheres of our giants. Similar results
were found by Smith \& Suntzeff (1989).

\subsubsection{Dependence on \Vt} 
\begin{figure*}
\begin{center}
\includegraphics [width=62mm]
{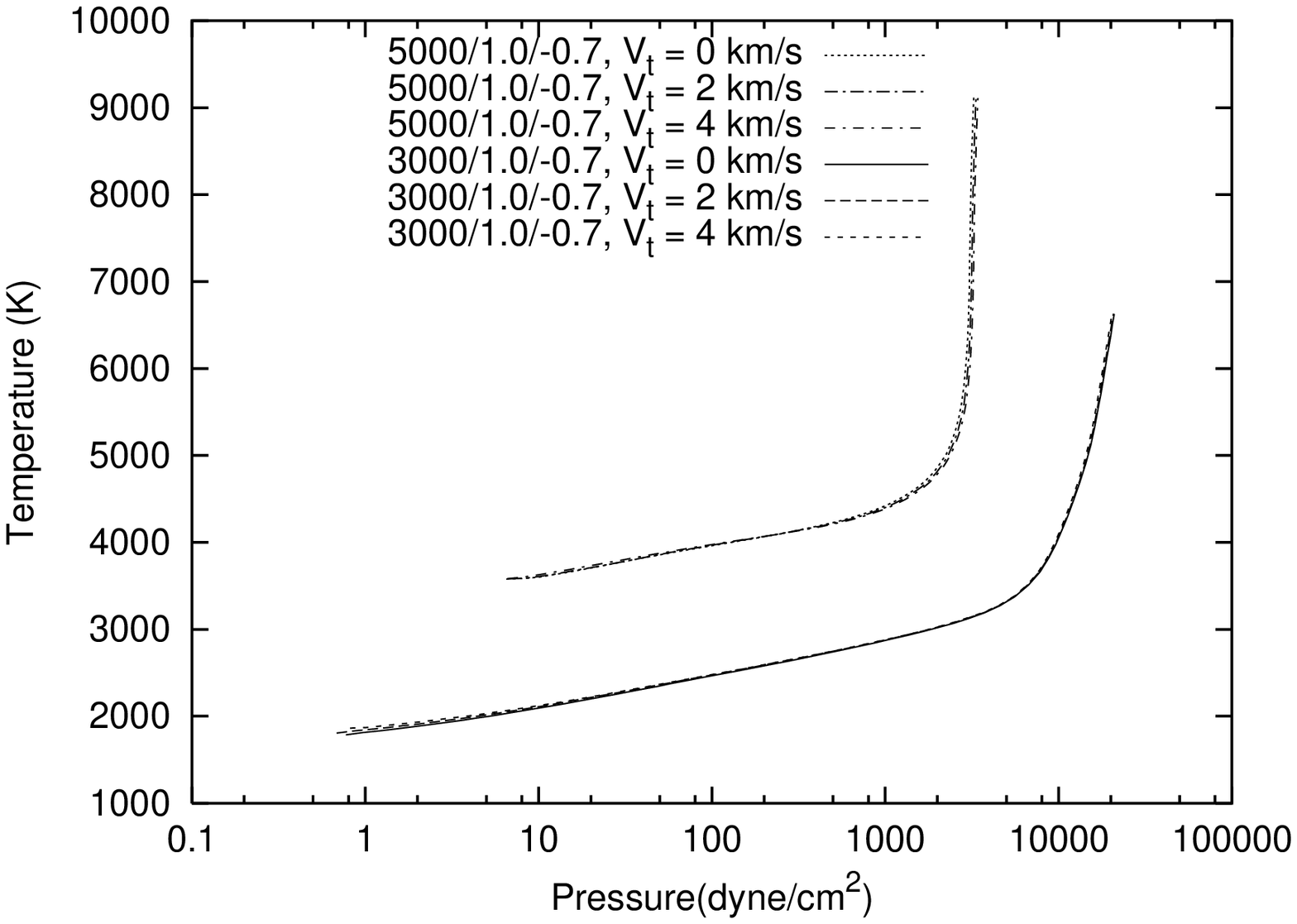}
\includegraphics [width=62mm]
{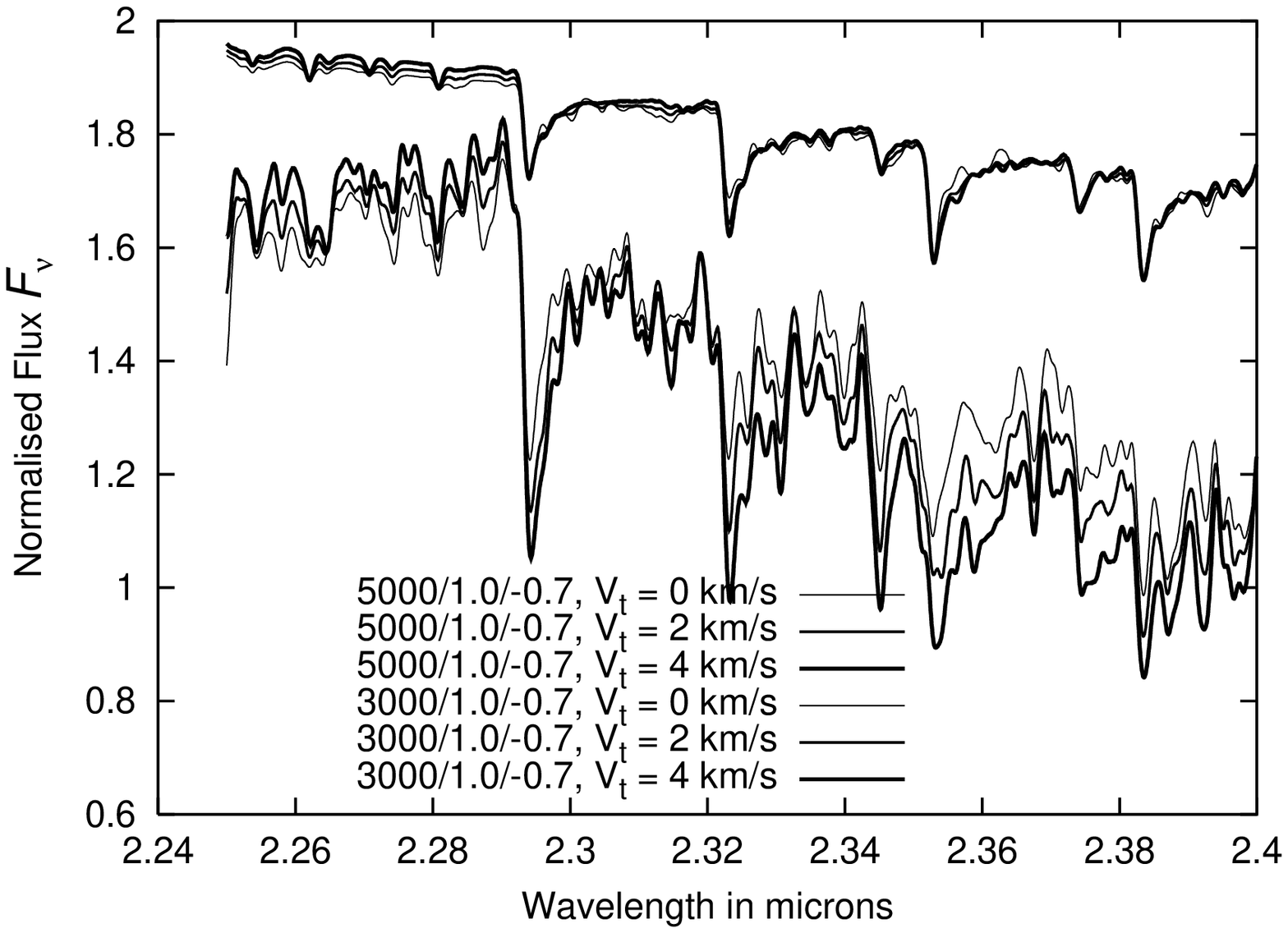}
\end{center}
\caption[]{\label{_Vt_} Computed temperature structures
 and the synthetic spectra for different \Vt
(\CDC = 3 for all models).}
\end{figure*}

In our analysis we use absorption
lines of intermediate strength. Therefore our results should show
a dependence on microturbulent velocity (cf. Smith \& Suntzeff
1989). Microturbulent velocities in the atmospheres of giants of
globular clusters lie in the range 1--3 km/s (see Kraft et al.
1997, Ivans et al. 2001, Ramirez and Cohen (2002, 2002a) and
references therein, as well as our Table \ref{__M5_1__}). We computed a few
model atmospheres and synthetic spectra for metal deficient giants
with different \Vt (Fig. \ref{_Vt_}). In general, the dependence
on \Vt increases towards lower \Tef. This arises because:

a) In the low temperature regime the relative contribution of microturbulent
velocities to the total velocity field $V_{total} = (2*R*T/\mu +
V_t^2)^{1/2}$ increases.

b) Lower temperatures lead to the formation of many absorption
lines of intermediate strength which have a strong dependence on
\Vt.

\subsection{Choice of $T_{\rm eff}$ and log g parameters}

In order to investigate the carbon abundance and isotopic ratio in
the target stars it is important that other parameters can be
fixed in a reliable and consistent manner. Effective temperatures
and gravities for nearly all of the targets
are taken from the
study of Alonso et al. (1999, 2000). The empirical nature of
the study, the large sample size and the use of interferometric
and Hipparcos data gives us confidence that our results will not
be undermined by uncertainties in radii and temperature.

\subsection{Fits to observed spectra}
In fitting the observational data to synthetic spectra we
followed a scheme described by Jones et al. (2002) and Pavlenko \&
Jones (2002). In order
to determine the best fit parameters for our targets in a
systematic fashion, we performed a least-squared minimisation on
the observed spectra within a grid of synthetic spectra smoothed
to the resolution of the data using a triangular function to mimic
the square pixels of the detector. Then, for every
spectrum we minimised a 3D function $S=f(x_s,
x_f, x_w) = 1/N \times \sum(1-F_{obs}/F_{synt})^2$, where
$F_{obs}, F_{synt}$ are observed and computed fluxes,  $N$ is
the number of points in the observed spectrum to be fitted.
 $x_s, x_f, x_w$ are respectively the
relative shift in wavelength scale, a normalisation factor
used to overlay observed and computed spectra and an instrumental
broadening parameter.

Our ``best fits'' for every giant of our sample are shown in the
next section of the paper. All these fits were found from
minimisation of  the $S$ function. In the observed region there
are four band heads of $^{12}$CO at 2.29, 2.32, 2.35, 2.38 $\mu$m
and two of $^{13}$CO at 2.345, 2.37 $\mu$m. After a number of
trials we found that our most reliable solutions could be found
from the CO band heads in the spectral region from 2.285 -- 2.36
$\mu$m. Unfortunately only one $^{13}$CO band head, at 2.345
$\mu$m, lies in this region. In principle, the other band of \CCC
at 2.375 $\mu$m provides a more sensitive tool for \CDC
determination in spectra of M-stars (see Pavlenko \& Jones 2002)
but our data do not cover properly the whole feature. The
spectra have a slight curvature, which we correct for but which
has greatest error at the edge of the array.

\begin{figure*}
\begin{center}
\includegraphics [width=62mm]
{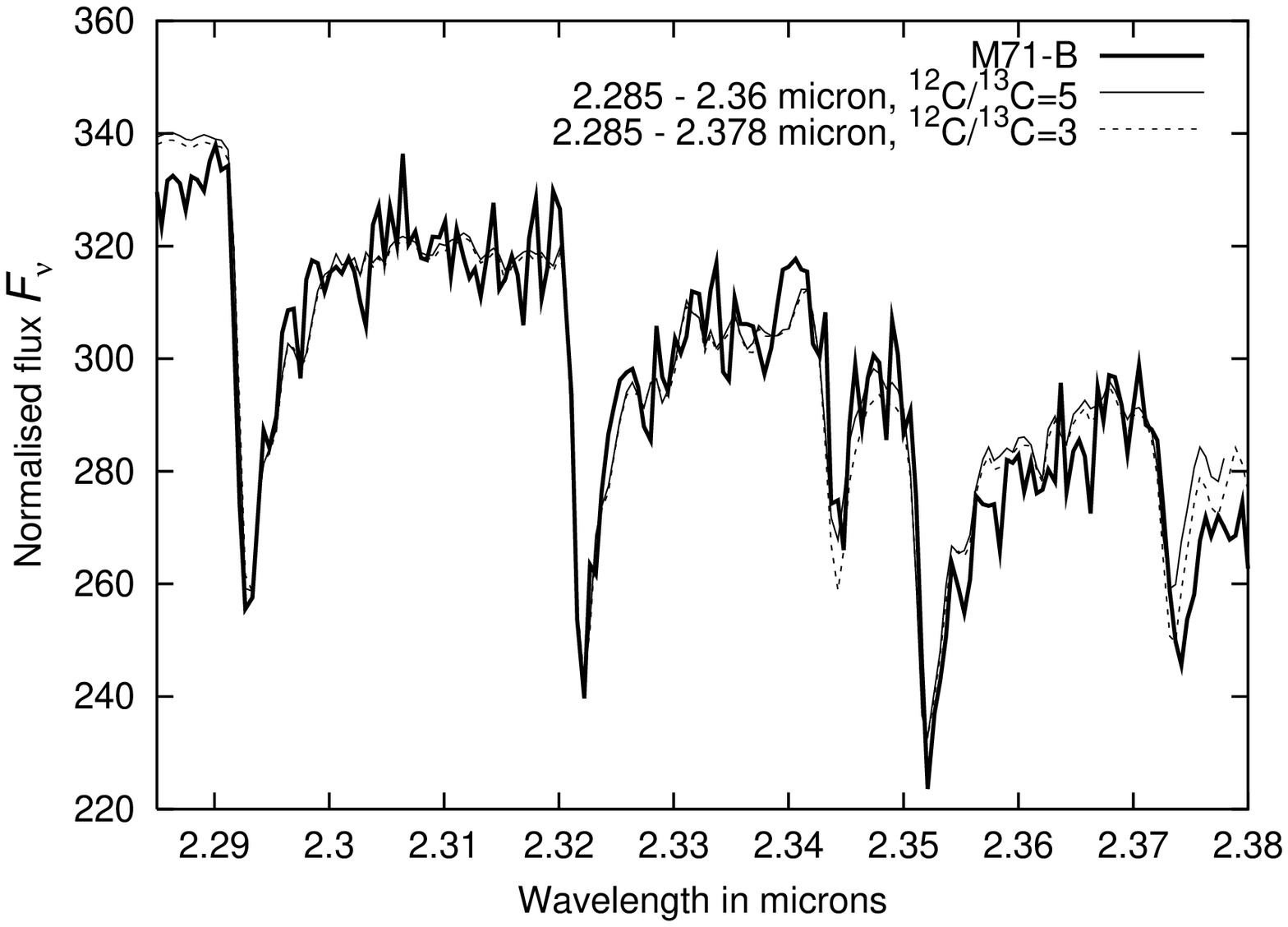}
\includegraphics [width=62mm]
{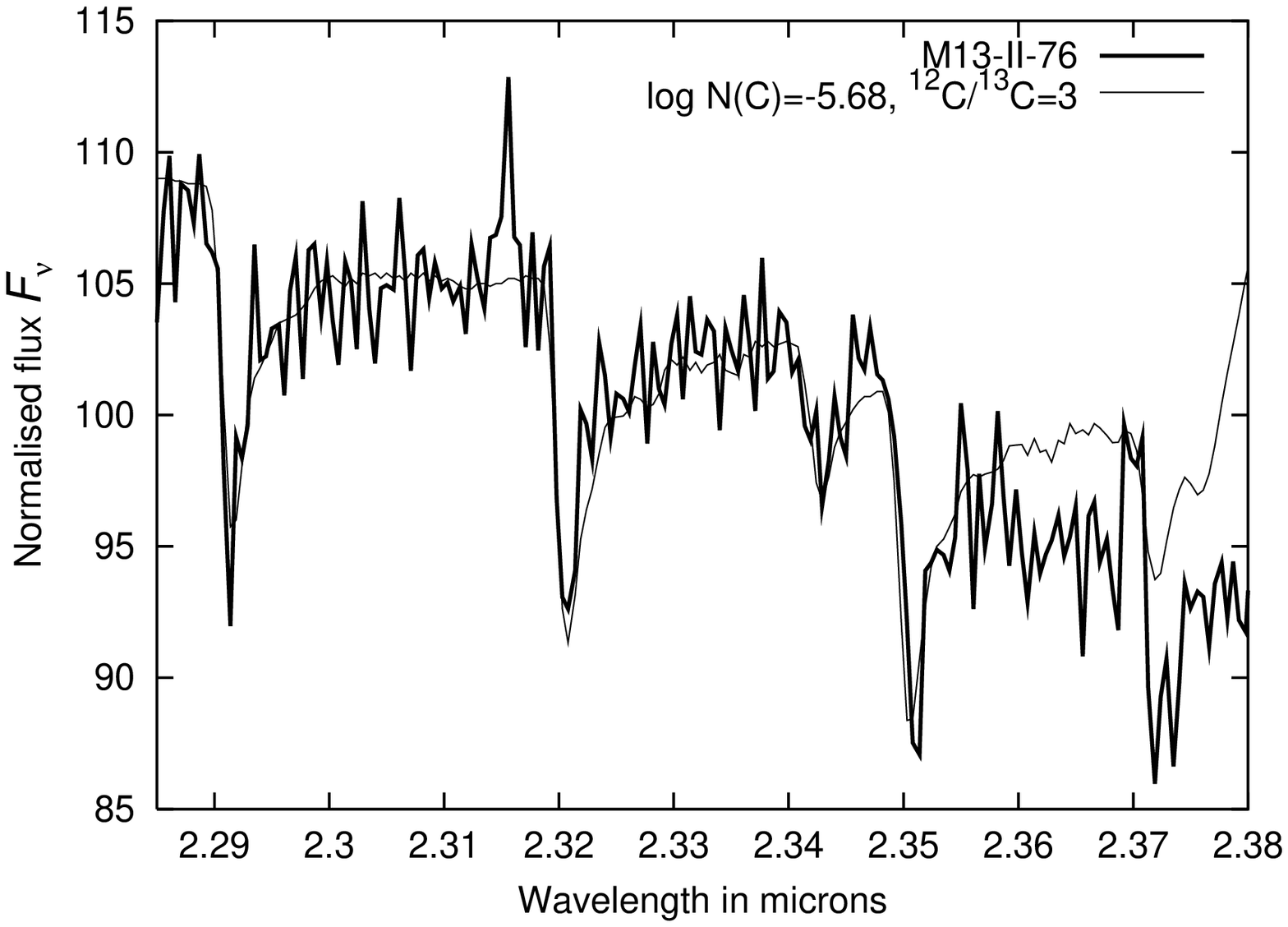}
\end{center}
\caption[]{\label{__bCC__}  Left: Fits to observed spectra
Fits to observed spectra for the full spectral region for
M71-B (left) and M13-II-76 (right).
} 
\end{figure*}

Fig. \ref{__bCC__} shows two examples of fits to observed spectra
using a narrow (2.285 - 2.36 $\mu$m) and  wider wavelength region
(2.285 - 2.38 \mum). The latter contains also the 2.375 $\mu$m \CCC
band head. We note a number of features:

(1) Our computations of synthetic spectra were carried out with
wavelength resolution of $\Delta \lambda$ = 0.01 nm. 
We found that the lower wavelength resolution introduced extra
    computational noise, although the value of the min ($S$)
    for $\Delta \lambda$ = 0.01 nm and 0.05 nm remained 
    the same in most cases.

(2) Since we believe the quality of the theoretical spectra is expected
to be excellent throughout the modelled region we may cautiously use $S$
to estimate the quality of the observed spectra. Our minimisation value
$S$ increases when we use the 2.375 $\mu$m band in our
analysis: from our ``standard''  fit over 2.285 --- 2.36$\mu$m to
M71-B  (3600/0.33/-0.7) we got $S$ = 583 (log N(C)= -5.38, \CDC=5),
whereas for the extended region 2.285 --- 2.378$\mu$m we obtained a less
good minimisation of $S$ = 664 (log N(C) = -5.38, \CDC=3), see left
panel of Fig. \ref{__bCC__}.

(3) We obtained the same log N(C) from the fit to the extended region
because the main contribution to $S$ was provided by strong $^{12}$CO
bands which are common in both cases. When this is not the case or when
edge effects are evident our results for the \CDC ratio become very
sensitive to the quality of spectrum on the redward side of the
2.375 \mum band.

(4) Carbon isotope ratios are systematically lowered, if we use
2.36-2.38$\mu$m for our analysis (see table \ref{_tablelong_} for
three M71 giants).

(5)  In the right panel of Fig. \ref{__bCC__} we show another example of
a fit to M13-II-76 (4202/1.10/-1.4). In that case the fit was
performed to the narrower wavelength region. The measured flux at
wavelengths greater than 2.36 $\mu$m is significantly lower than the
theoretical values.  Furthermore, without a
sophisticated normalisation we get a non-physical \CDC $<$ 3 if we fit
to the 2.285-2.38$\mu$m region.

We conclude that the use of the second band-head of $^{13}$C$^{16}$O
at
2.375 $\mu$m in our analysis degrades rather than improves our results.
Due to its poorer signal-to-noise ratio and the possibility of
non-optimal extraction at the end of the spectrum, we excluded it from
our analysis. 

\begin{table*}
 \centering
 \begin{minipage}{140mm}
  \caption{Carbon abundances and \CDC in giants of M71
  from the fits to 2.285 --- 2.36 and 2.285 --- 2.38 \mum}
\label{_tablelong_}
  \begin{tabular}{@{}ccccccccccc@{}}
\hline
    &        & & \multicolumn{3}{c}{2.285 --- 2.36 \mum}&
    \multicolumn{3}{c} {2.285 --- 2.38 \mum} \\
\hline Cluster & Object &\Tef/log g &log N(C)& \CDC & $S$ & log
N(C)& \CDC & $S$ \\

\hline M71 & b  & 3600/0.33 & -5.38  & 5  &583 & -5.38 & 3  & 664
\\
& 30 & 3925/0.88 & -4.38  & 7 & 420 & -4.38 & 4 & 688 \\ & 21
&4349/1.65 & -4.38  & 5 & 467 & -4.38 & 3  & 642 \\ \hline \hline
\end{tabular}
\end{minipage}
\end{table*}

\begin{table*}
\centering
\begin{minipage}{140mm}
\caption{Derived values for [C] and $^{12}$C/$^{13}$C are given.
Values of metallicity, temperature and gravity  are taken from
Alonso et al. (1999, 2000). Data was also taken of
i61 in M5 and I-21 and IV-59 in M13, however, the S/N of the data is
rather poor and so data for these objects is not included.
Cluster non-members as determined by proper motions studies are
given in italics.}
\label{__table1__}
\begin{tabular}{@{}llllllllllll@{}}
\hline Cluster & [Fe/H]$\pm$0.1 & Object & log (L/L$_{\odot}$) &
\Tef & log $g$ & log N(C) & [C] $\pm$0.1& \CDC \\ \hline
&   &   &       &            &      &       &       &         \\
M71 members    &  --0.71     &21 & 2.33  &4349$\pm$48 & 1.65 & --4.38& -0.19 & 5$\pm$2 \\
   &      &30 & 2.84  &3925$\pm$56 & 0.88 & --4.38& -0.19 & 7$\pm$2 \\
   &      &B  & 3.08  &3600$\pm$72 & 0.33 & --5.38& -1.19 & 5$\pm$3 \\
   &      &29 & 3.19  &3574$\pm$50 & 0.09 & --4.58& -0.39 & 9$\pm$2 \\
&  & C, noise  & 2.08 & 4856$\pm$60 & 2.51  --3.78& +0.41
& 95$\pm$10 \\
\\
M71 non-member?   &      &79? & 1.80  &4556$\pm$56 & 2.20 & --3.98& +0.36 & 5$\pm$2 \\
\\
M71 non-members   
  &--0.71&{\em N}, noise  &  {\em 1.99} &{\em 4840$\pm$60} & {\em 2.12} & {\em --2.82}& {\em +1.37} &{\em 3$\pm$1} \\
   &      &{\em A5} & {\em 1.86}  &{\em 4531$\pm$57} & {\em 2.10} & {\em --4.18}& {\em +0.01} &{\em 10$\pm$3} \\
   &      &{\em A7} &{\em 2.04}  & {\em 4411$\pm$53} & {\em
   1.92} & {\em  --4.38}&{\em -0.19} & {\em 20$\pm$5} \\
   &      &{\em A6} & {\em  2.70}  & {\em 3897$\pm$72} & {\em
   1.02} & {\em  --4.18}&{\em -0.01} & {\em 20$\pm$5} \\
      &      &   &       &            &      &          &         \\
M5 & --1.3 & IV-59 & 2.90 &4243$\pm$45 & 1.0 & -5.38 &-0.6 & 7$\pm$3 \\
   &       & II-9  & 3.03 &4230$\pm$48 & 0.9 & -5.58 &-0.8 & 3$\pm$1 \\
   &       & IV-81 & 3.23 &3963$\pm$53 & 0.6 & -5.58 &-0.8 & 4$\pm$1 \\
\\
M13&       &I-24, noise   & 2.61 &4374$\pm$57  & 1.10  &  -5.58& -0.5 & 5$\pm$5 \\
   & --1.4 &II-76  & 2.87 &4202$\pm$52  & 1.10       &  -5.58& -0.7 & 3$\pm$1 \\
   &       &III-73 & 2.99 &4164$\pm$52  & 0.90       &  -5.58& -0.7 & 3$\pm$1 \\
   &       &III-56, noise &3.13  &4013$\pm$41  & 0.70       &  -6.08& -1.0 & 7$\pm$2 \\
   &       &I-48, noise   & 3.21 &3929$\pm$45  & 0.60       &  -6.68& -1.8 & 90$\pm$20 \\ 
   &       & II-67 & 3.28 & 3894$\pm$45 & 0.50       &  -6.38& -1.5 & 4$\pm$1 \\
\\
M3 & --1.6 &  I-21 &2.91  & 4124$\pm$52 & 1.0 & -5.58 &-0.5 & 4$\pm$1 \\
   &       & III-28&3.03  & 4092$\pm$43 & 0.8 & -5.78 &-0.7 & 4$\pm$1 \\
   &       & AA   &3.12  & 3977$\pm$50 & 0.7 & -5.78 &-0.7 & 3$\pm$1 \\
   &       & II-46 &3.12  & 3951$\pm$52 & 0.7 & -5.58 &-0.5 & 4$\pm$1 \\
\\
M3 non-member? & & 1397? &3.15  &3916       & 2.5 & -4.72 &+1.3 & 5$\pm$1 \\
\\
\hline \hline

\end{tabular}
\end{minipage}
\end{table*}

\begin{table*}
 \centering
 \begin{minipage}{140mm}
  \caption{Comparison with literature values for target objects:
M5 from Ivans (2001), M3-IV-59 Smith et al.(1997),  M13 from
Smith et al. (1996), M13 from Kraft et al. (1997).}
 \label{__abund}
  \begin{tabular}{@{}lllrrrrlrlr@{}}
\hline Cluster & Object & Fe/H & lit [O]   & lit [C] & our [C] &
\\ \hline
M5    & IV-59  & -1.25--1.40 & +0.37 & -0.6$^{sm}$ & -0.5   & \\
M3  & AA     & -1.45       & -0.22 & -1.15$^s$           & -1.0
\\
    & III-28 &-1.49        &  0.31 & -0.88$^s$           & -0.8  \\
    & II-46  &-1.46        &  0.29 & -0.84$^s$           & -0.7  \\
M13 & III-56 & -1.51       &-0.02  & -1.19$^s$           & -1.2
\\
    & III-73 &-1.51        &0.34   & -0.86$^s$           & -0.8  \\
    & II-67  & -1.51       &-0.64, -0.79$^k$ & -1.34$^s$ & -1.45 \\
    & II-76  & -1.49       &0.46   & -0.82$^s$           & -0.9  \\
M13 & I-48   & -1.57       &-0.75  &                     & -1.8 \\
    & II-67  & -1.52       &-0.79  &                     & -1.6 \\
\hline \hline
\end{tabular}
\end{minipage}\\
$^s$ --- Smith et al. (1996) \\ $^{sm}$ --- Smith et al. (1997) \\
$^k$ --- Kraft et al. (1997)    \\
\end{table*}

\section{Results}

Results of log N(C) and  \CDC determinations for our giants are
shown in  Table \ref{__table1__}. Before addressing each cluster
we mention some other general results. First of all, in Table
\ref{__abund} we list objects for which there are C and \CDC
abundance values in the astronomical literature, along with our
log N(C) and \CDC determinations. In general, within the errors of
our spectra and procedure our carbon abundances are consistent
with those of giants with known carbon and oxygen abundances.
Then, to study the impact of variations in some of our other input
parameters we carried out some numerical experiments:

\begin{itemize}
\item To test the dependence of our
results on \Vt we carried out some additional modelling of
our M5 giant spectra. We determined log N(C) and \CDC for
them adopting different \Vt. Our results are shown in Table
\ref{__M5_1__}. They show that the models are very sensitive to the
value of \Vt . We are able to determine individual values of
\Vt for given giants:
namely, \Vt = 2 km/s for M5-II-9 and M5-IV-81, \Vt = 3 km/s for M5-IV-59.

\item In the last two columns of Table \ref{__M5_1__} we show the
dependence of our results on \Tef. Effective temperatures of all
stars were increased by 100 K and the fitting procedure
was repeated. It can be seen that our
results, i.e. log N(C) and \CDC, remain in the frame of our error
bars, i.e. $\pm$ 0.2 dex and $\pm$ 1, respectively.

\end{itemize}

\begin{table*}
 \centering
  \caption{Dependence of log N(C) and \CDC on metallicity,
  microturbulent velocities and temperatures for M5
  giants.
   Below the pairs log N(C)/\CDC the ($S$) values are
  given(see Section 3)}
\label {__M5_1__}
  \begin{tabular}{||c|c|c|c|c|c|c|c||}
   \hline
      &          &           &  [$\mu$]/$V_t$/$\Delta$T   \\
Giant &-1.13/1/0 & -1.13/2/0 & -1.13/3/0 & -1.13/4/0 & 1.11/2/0 & -1.11/2/+100 \\ \hline
      &          &          &          &          &         &               \\
II-9  & -5.36/5  & -5.58/3 & -5.58/3 & -5.58/3  & -5.58/3 & -5.38/3     \\
      & (937) & (912) & (929) & (957) & (926)& (916)    \\
      &          &          &          &          &         &               \\
IV-59 & -4.98/20 & -5.38/7  & -5.38/5  & -5.38/2  & -5.18/8 & -5.18/8       \\
      & (538) & (509) & (500) & (529) & (515)& (503)     \\
      &          &          &          &          &         &               \\
IV-81 & -4.98/10 & -5.58/5  & -5.78/4  & -5.79/3  & -5.58/5 & -5.38         \\
      & (414) & (379) & (408) & (410) & (377)& (394)   \\
      &          &          &          &          &         &               \\
\hline
\hline
\end{tabular}
\end{table*}

\subsection{M71 (NGC 6838)}

Globular cluster M71  is a nearby metal-rich cluster. Sneden et
al. (1994) found for M71 some evidence of star-to star differences
in O (and Na) abundances. Briley et al. (1994, 1997) carried out
isotopic carbon abundance analysis using fits of weak $^{12}$CN
and $^{13}$CN bands in spectra at 800.2-800.6 nm across a narrow
range of effective temperatures (4000-4350 K). Ramirez \& Cohen
(2002a) performed an extensive analysis of abundances in
atmospheres of red giants and subgiants of M71 based on the
optical and near-infrared spectra. There is no overlap in objects
or in spectral region between these studies and ours.

\begin{figure*}
\begin{center}
\includegraphics [width=62mm]
{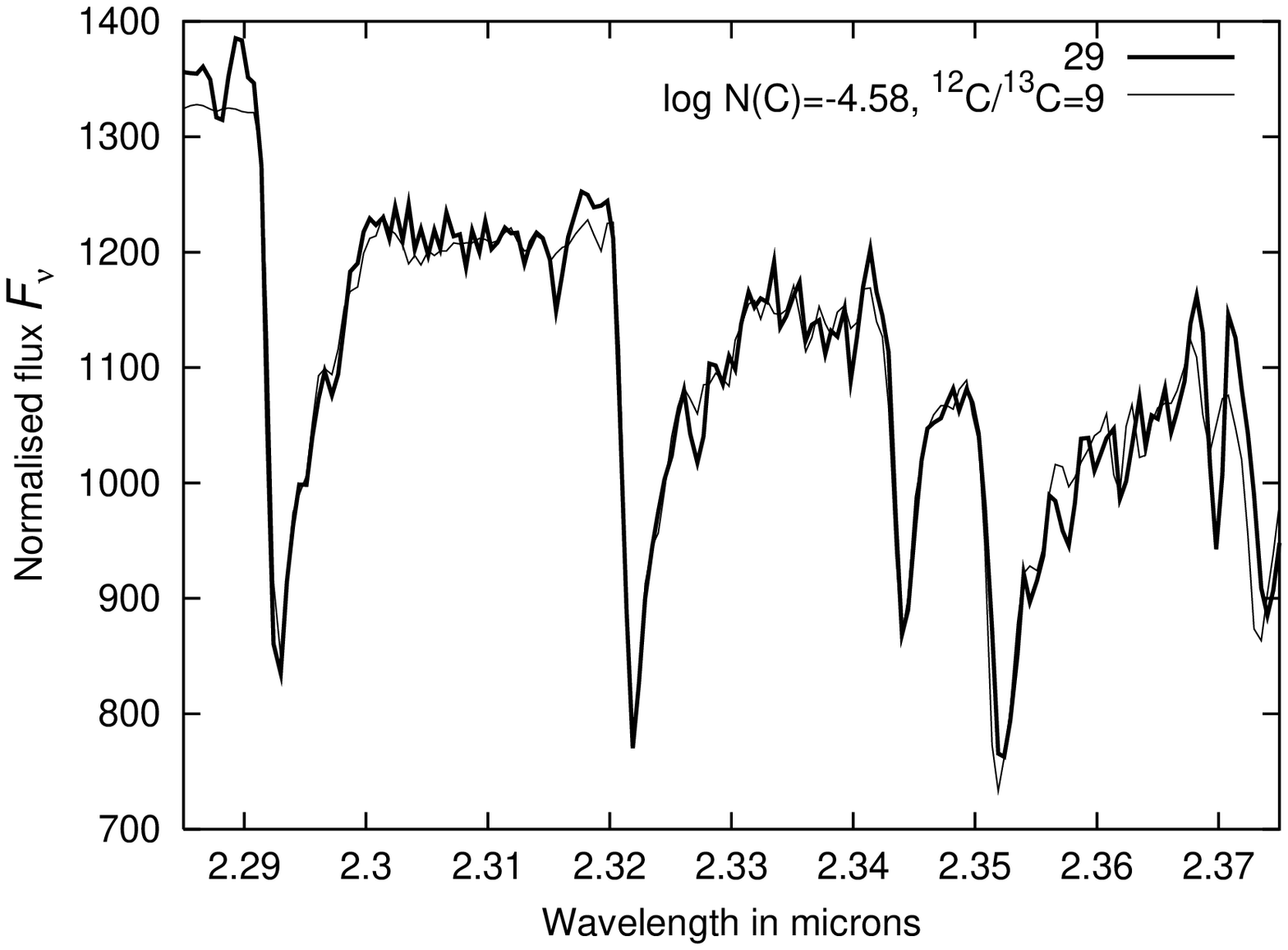}
\includegraphics [width=62mm]
{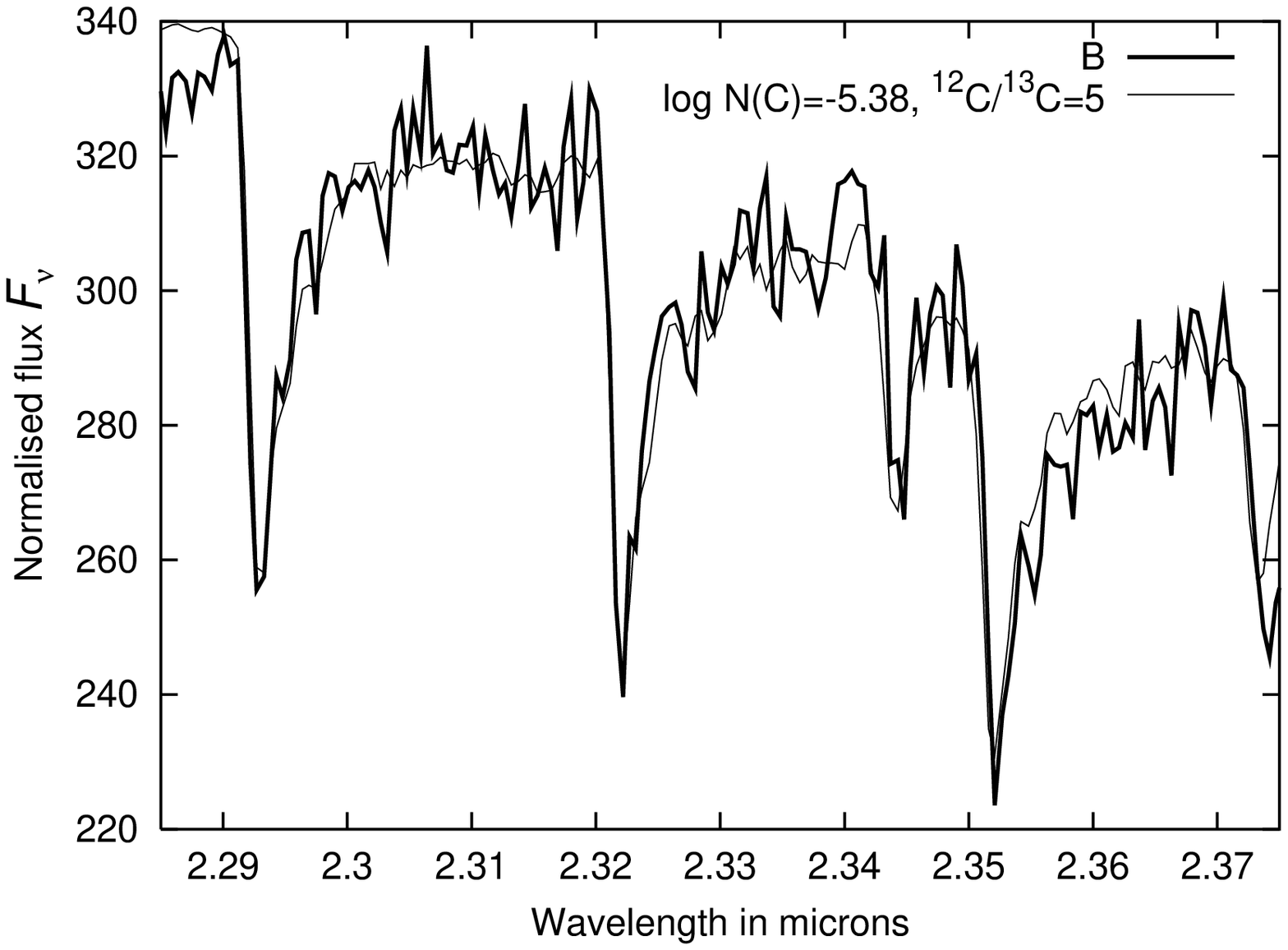}
\includegraphics [width=62mm]
{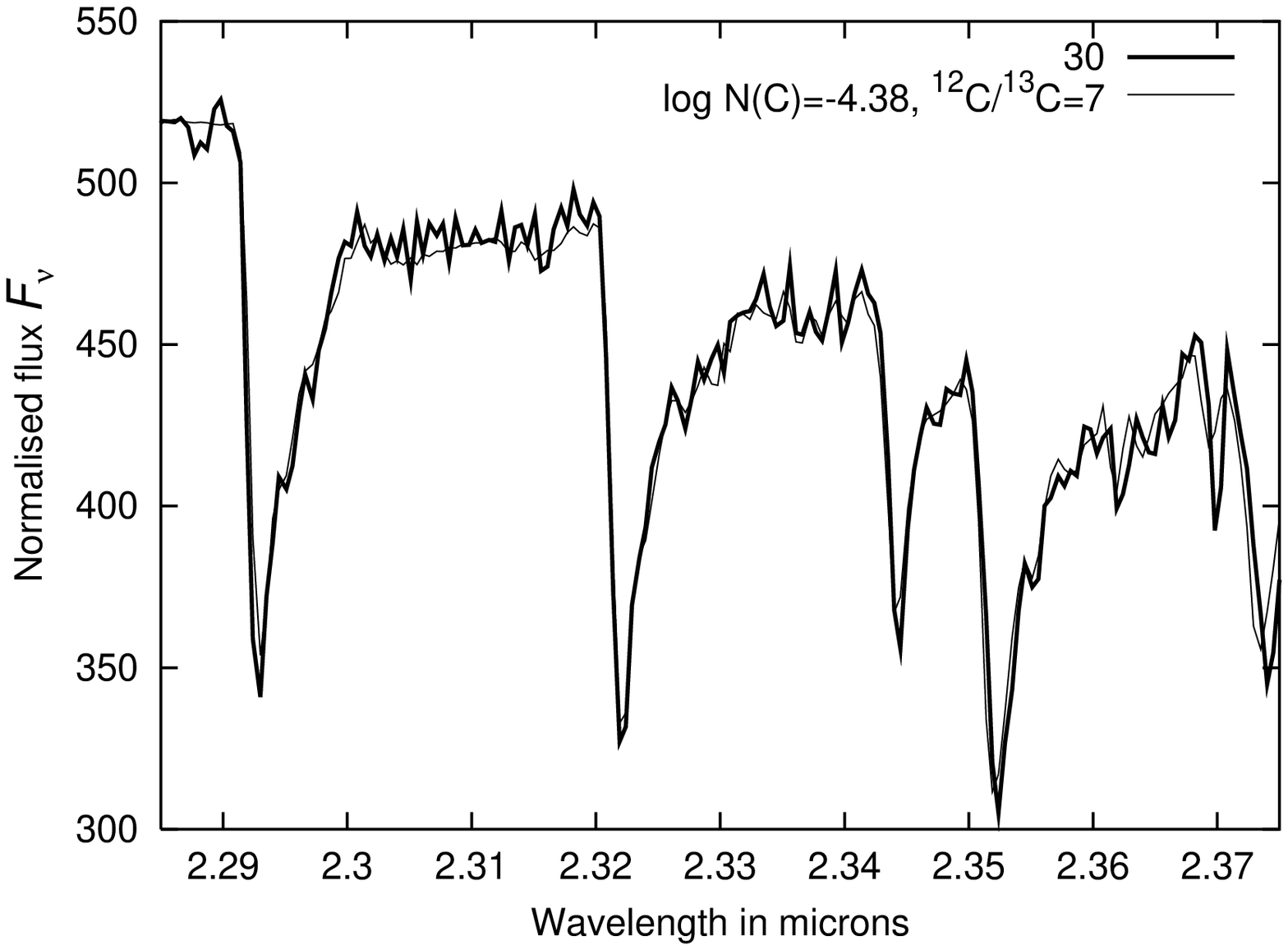}
\includegraphics [width=62mm]
{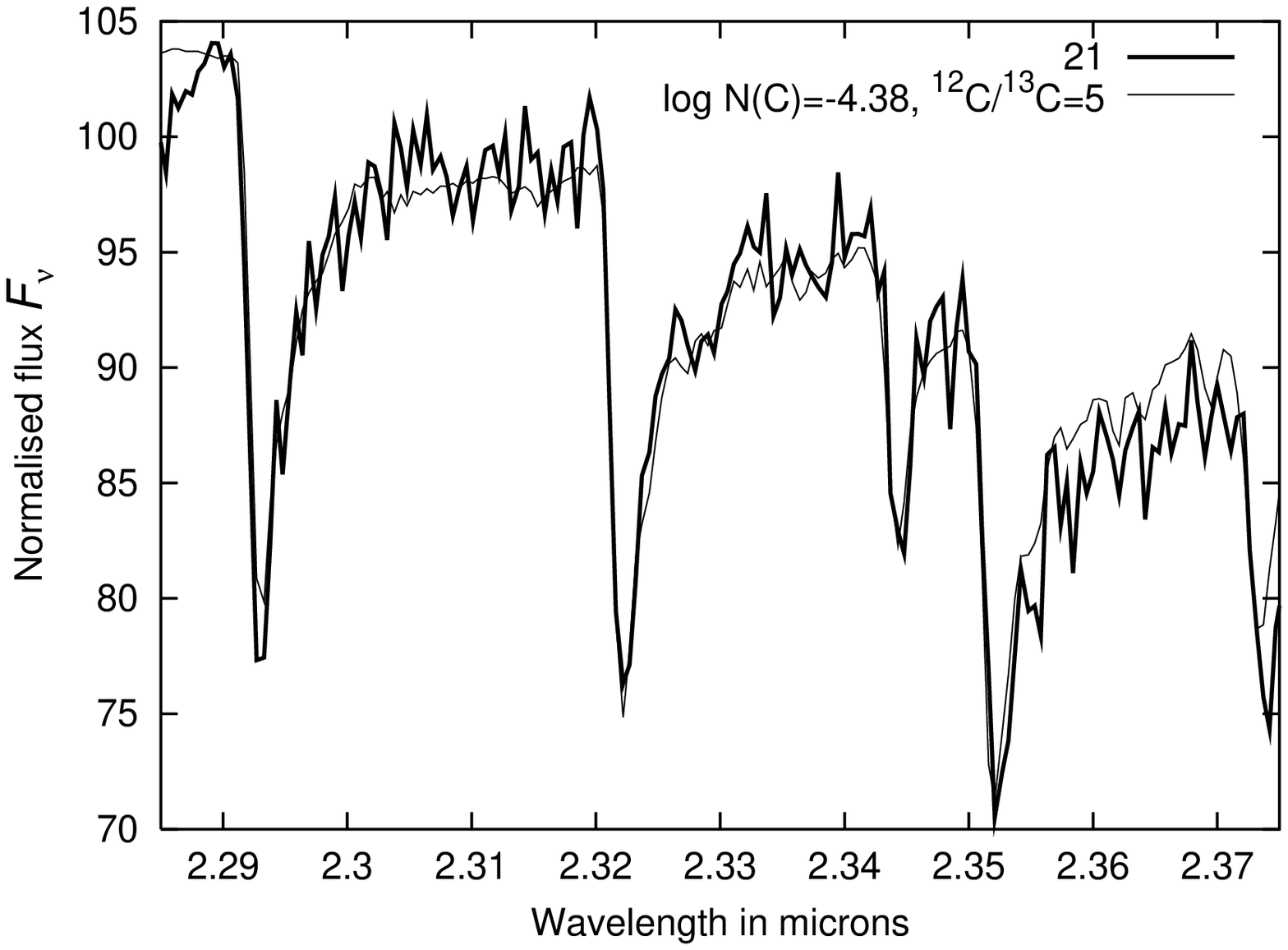}

\end{center}
\caption[]{\label{_M71_1}Fits to observed spectra of M71 giants
arranged by effective temperatures.}
\end{figure*}

\begin{figure*}
\begin{center}
\includegraphics [width=62mm]
{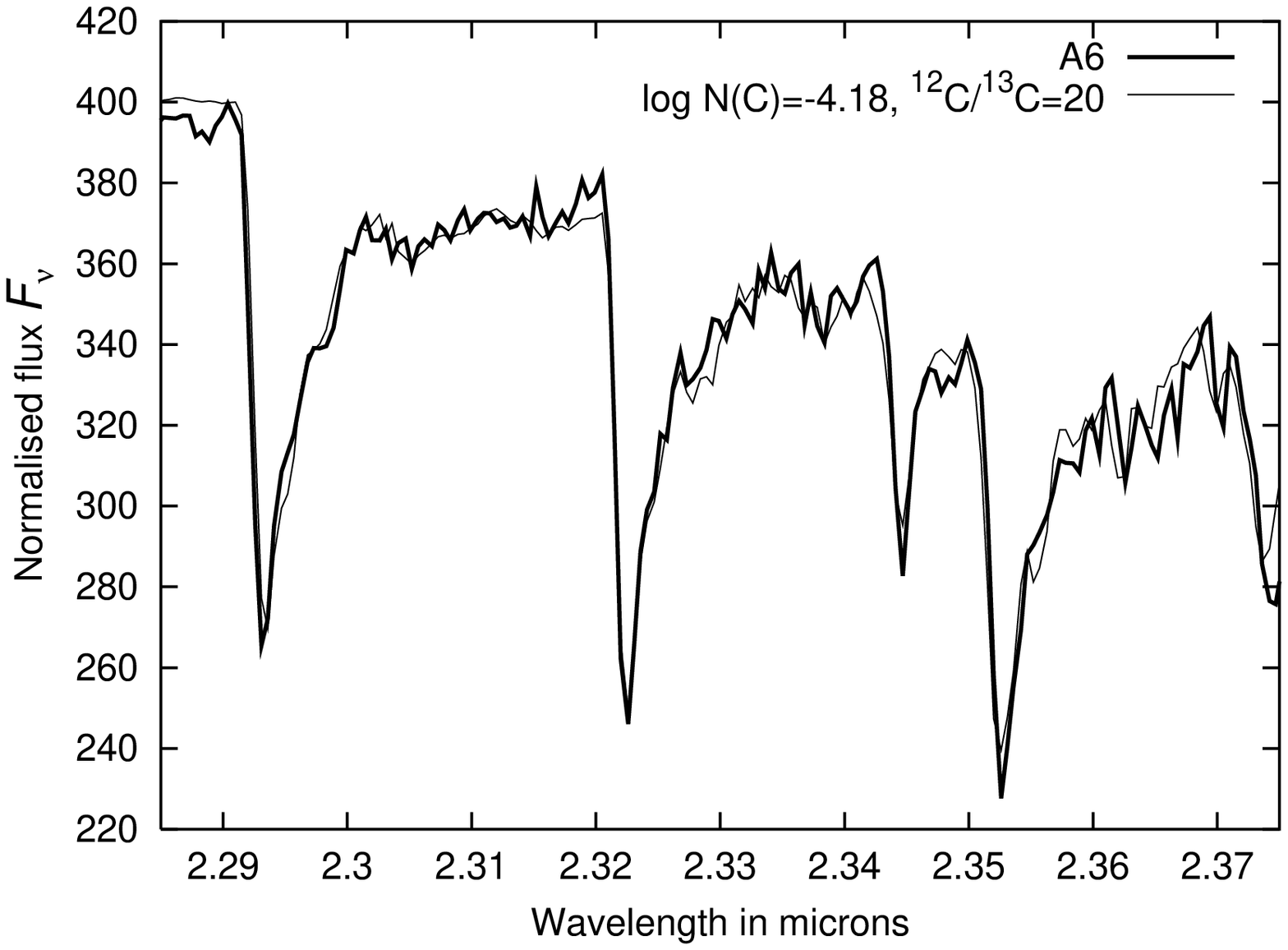}
\includegraphics [width=62mm]
{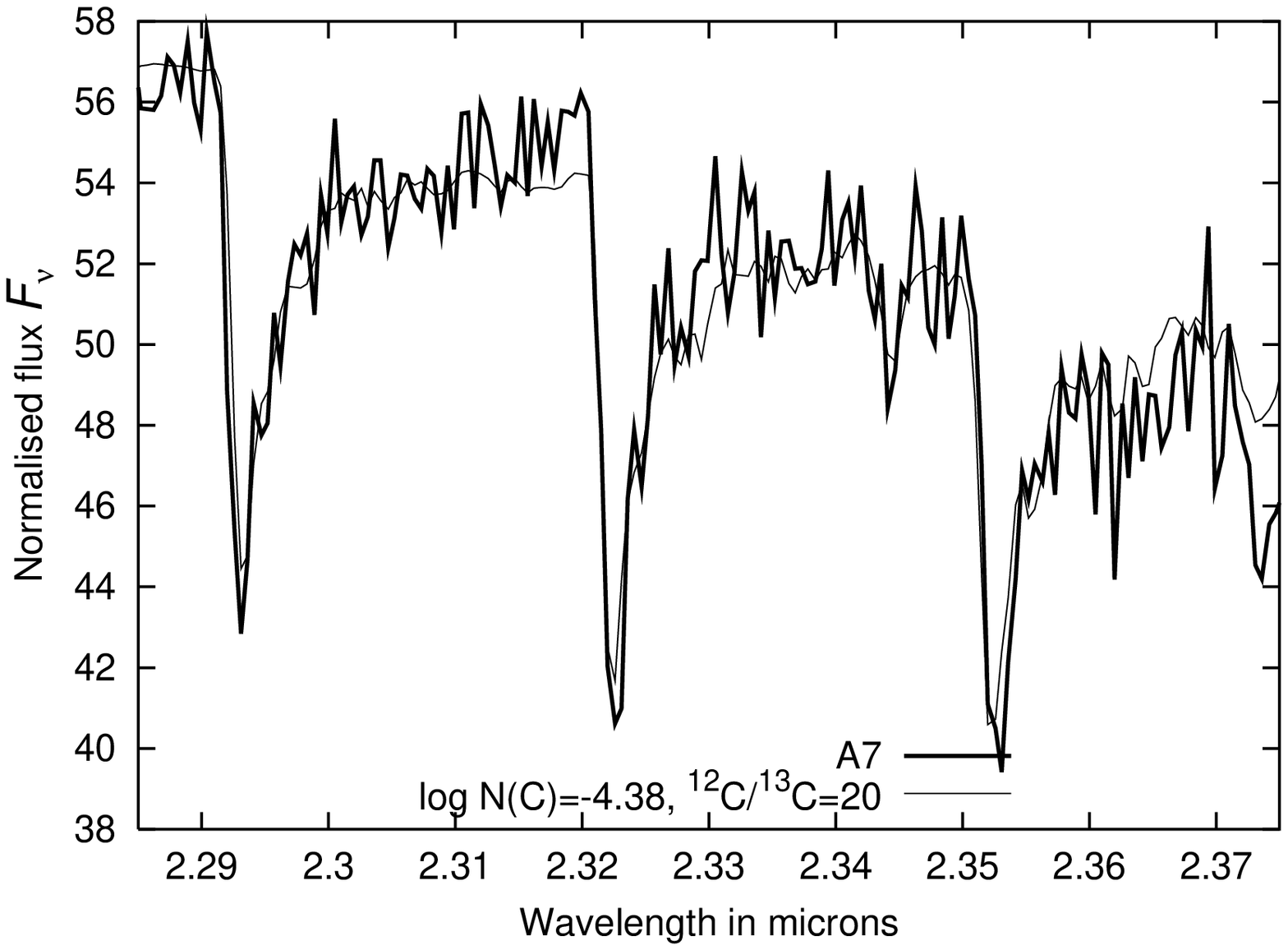}
\includegraphics [width=62mm]
{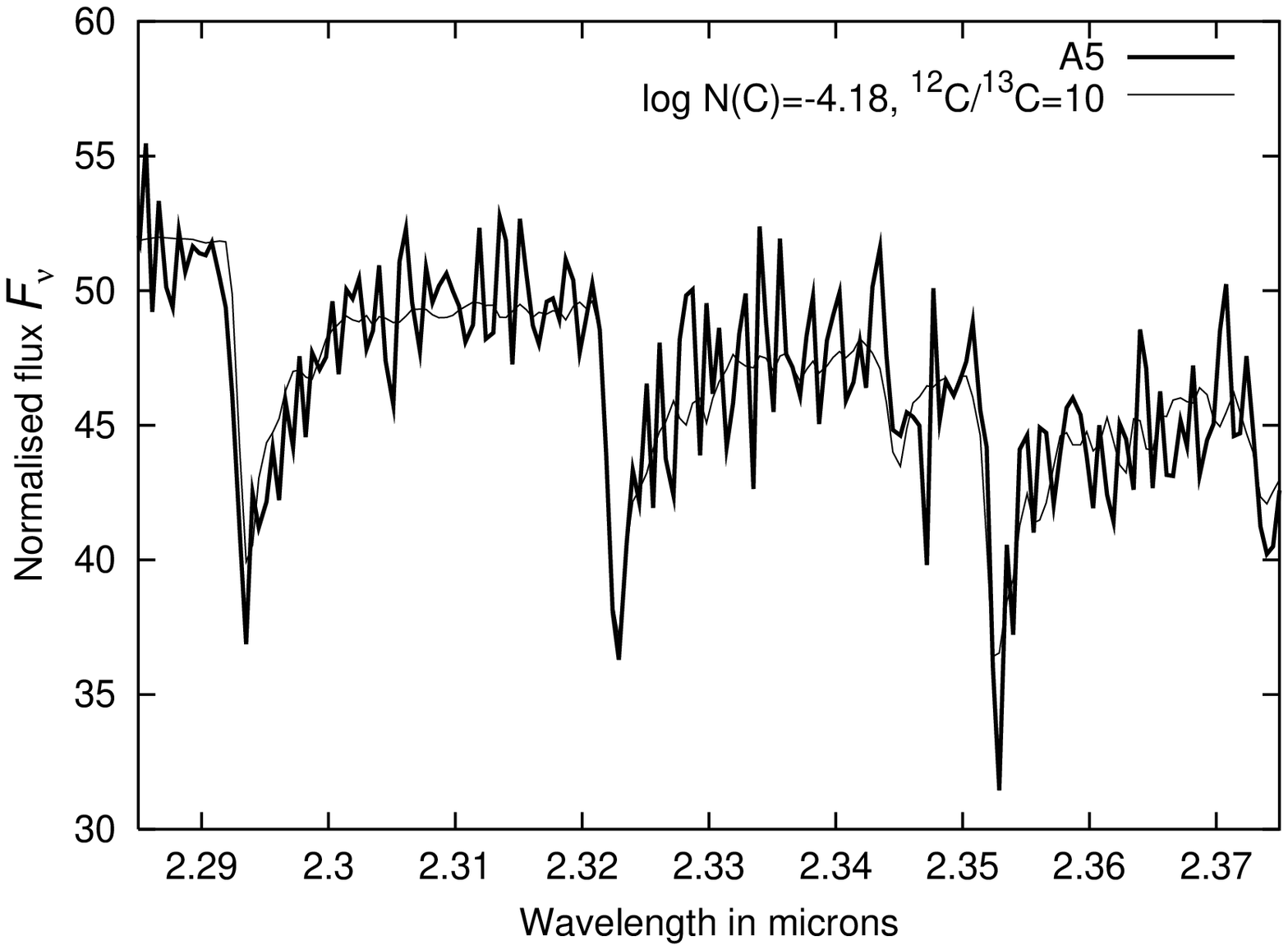}
\includegraphics [width=62mm]
{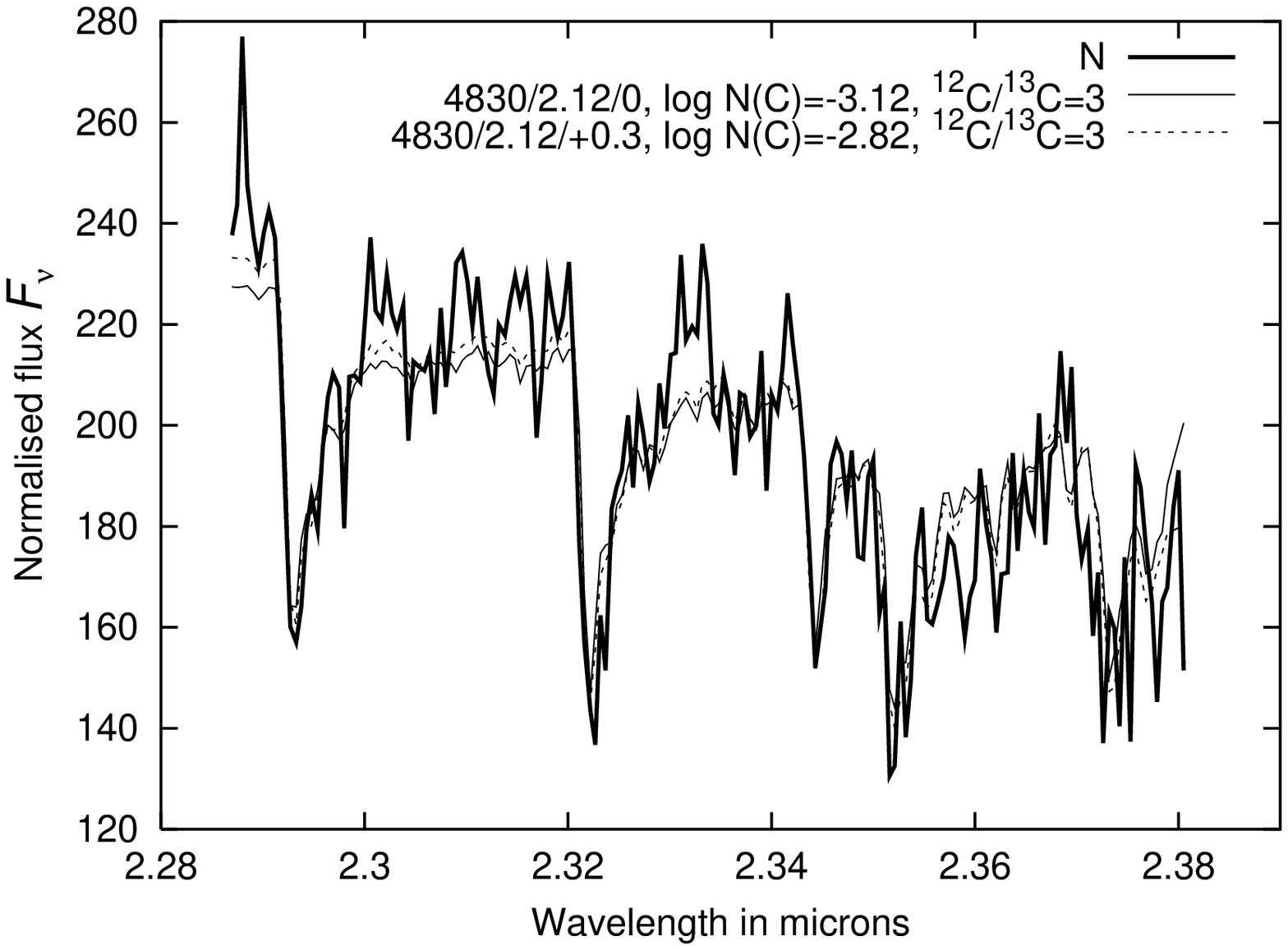}
\includegraphics [width=62mm]
{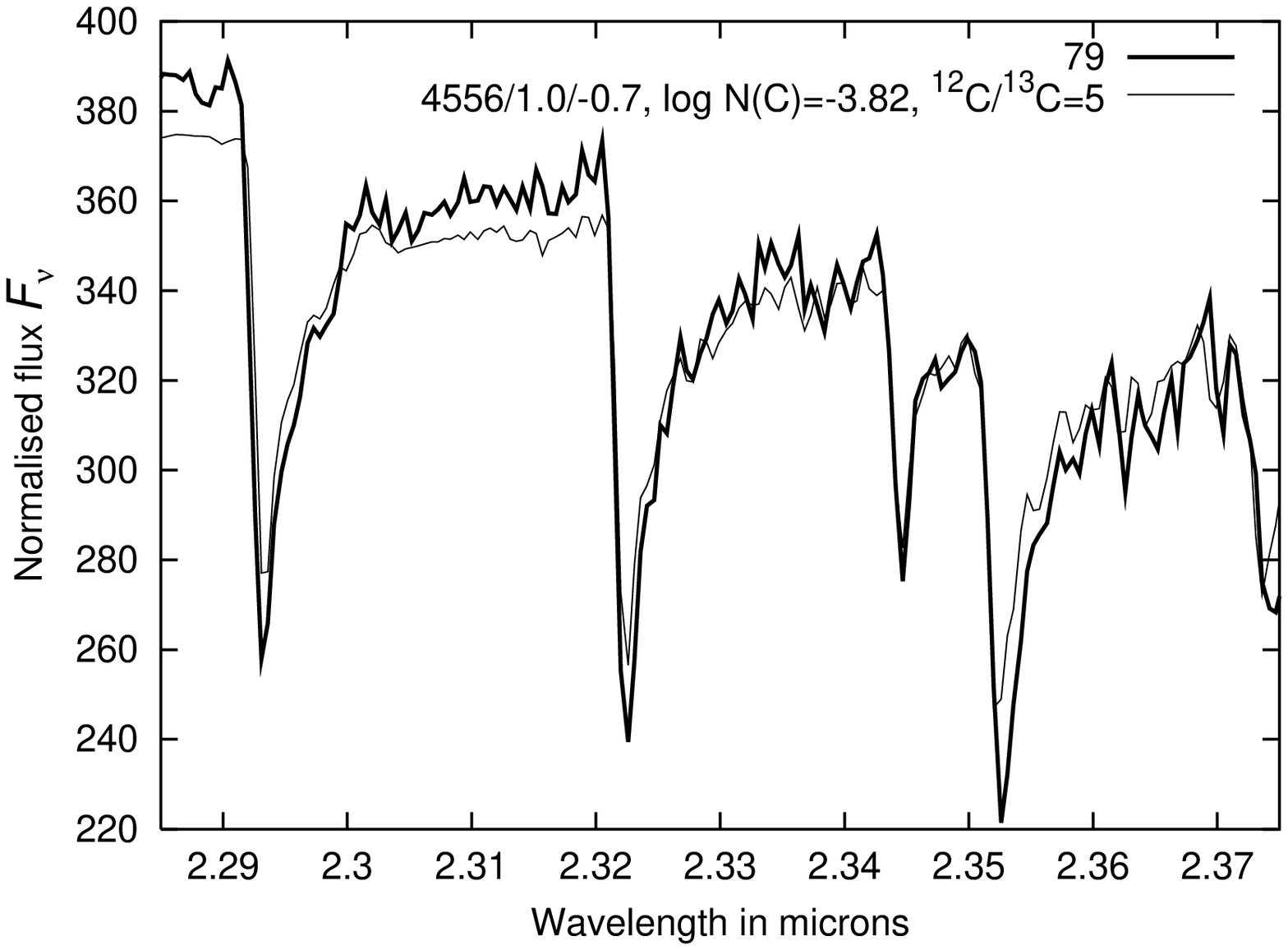}
\includegraphics [width=62mm]
{ff-M71-n.eps}
\end{center}
\caption[]{\label{_M71_0}Fits to observed spectra of field stars
of our ``M71 sample'' arranged by effective temperatures.}
\end{figure*}

\begin{figure*}
\begin{center}
\includegraphics [width=62mm, height=32mm]
{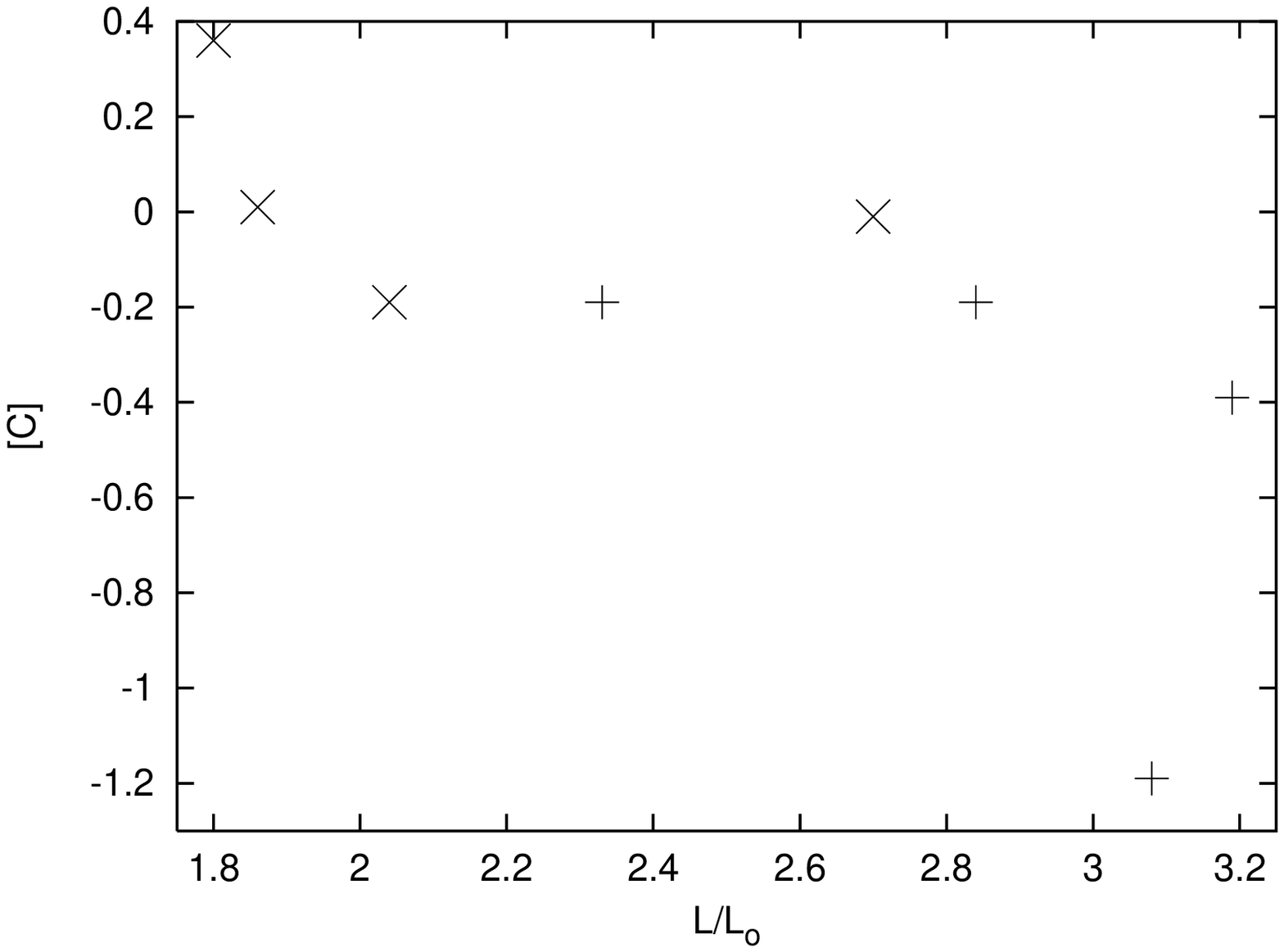}
\includegraphics [width=62mm, height=32mm]
{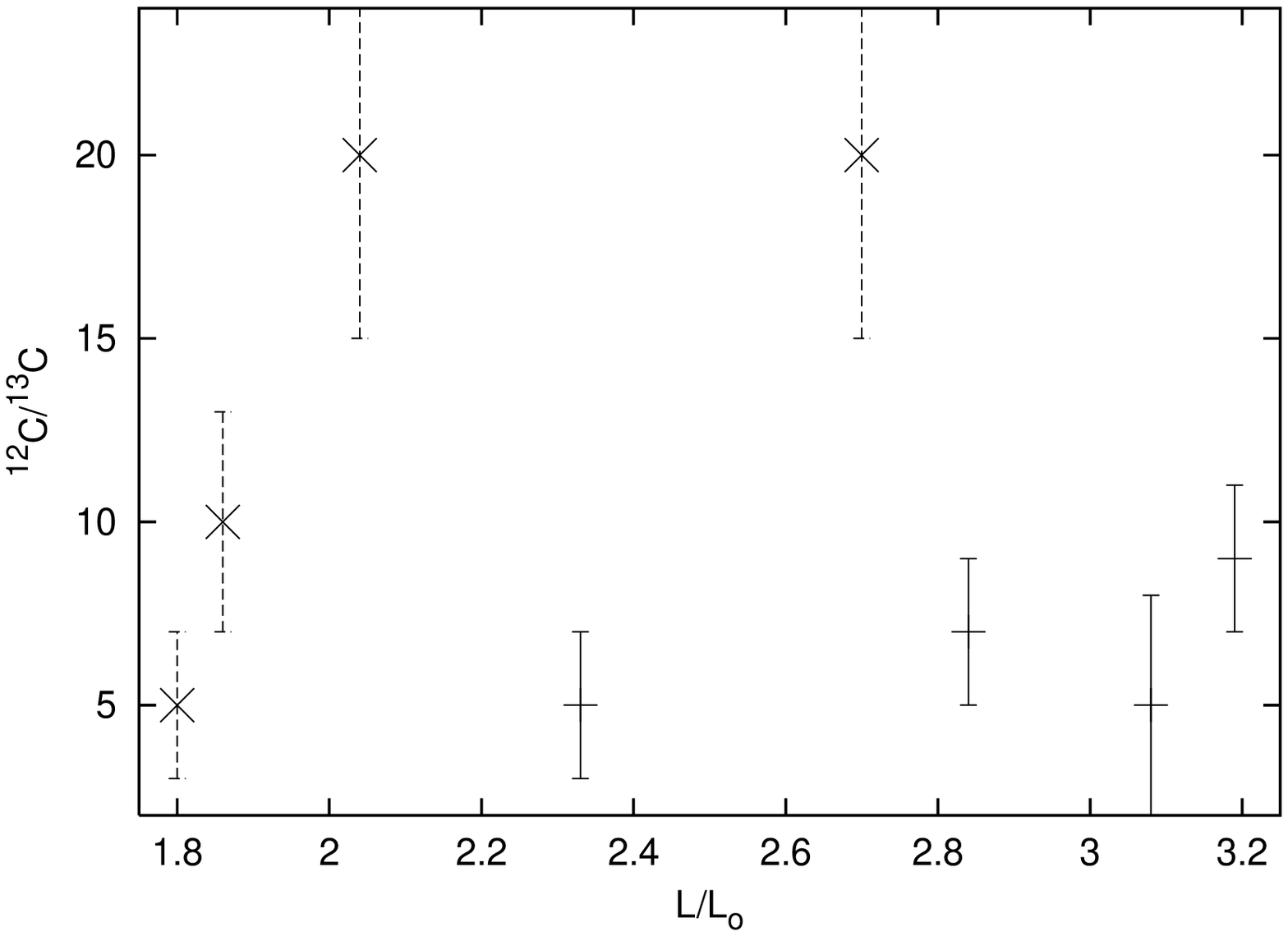}
\includegraphics [width=62mm, height=32mm]
{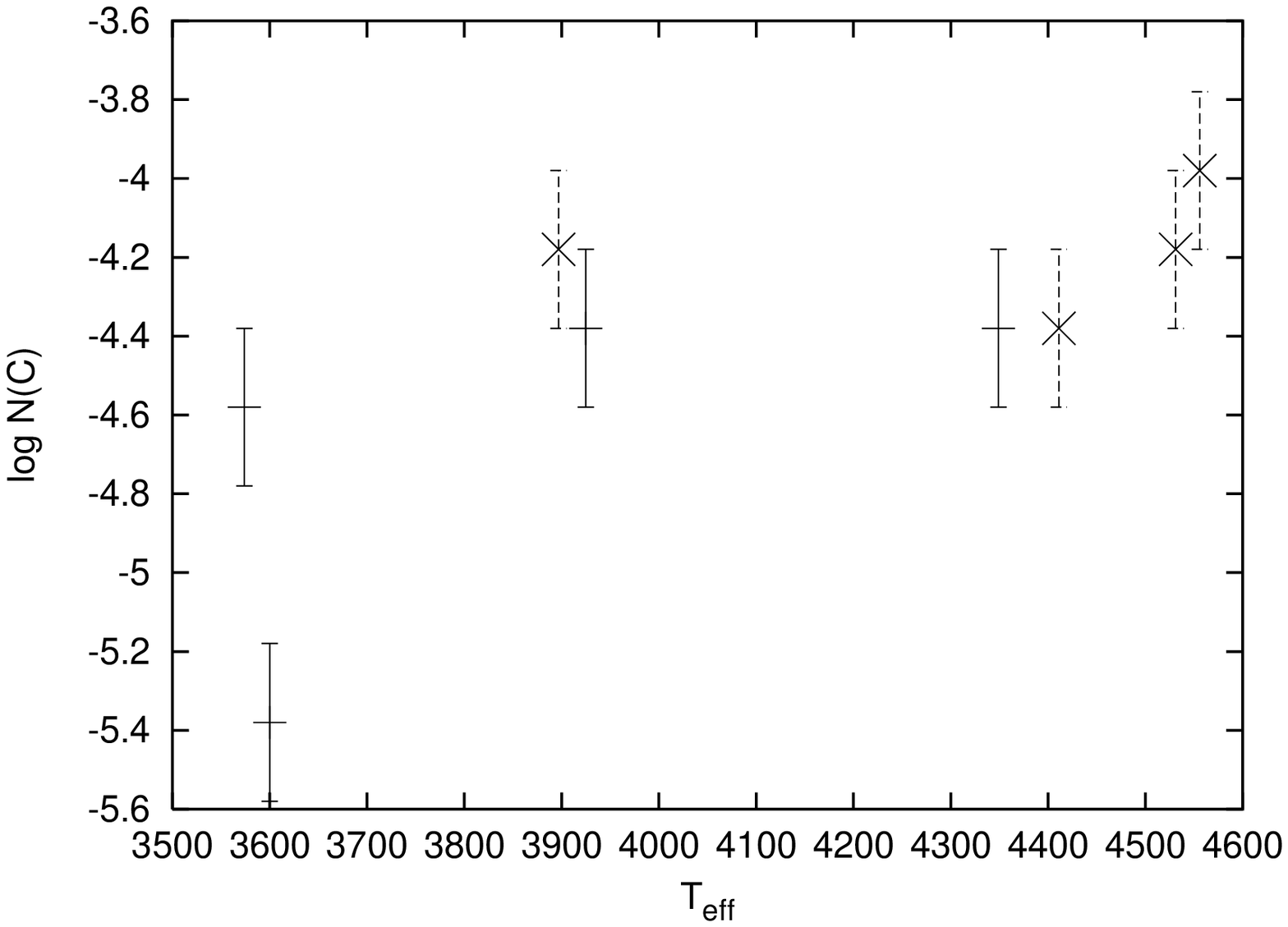}
\includegraphics [width=62mm, height=32mm]
{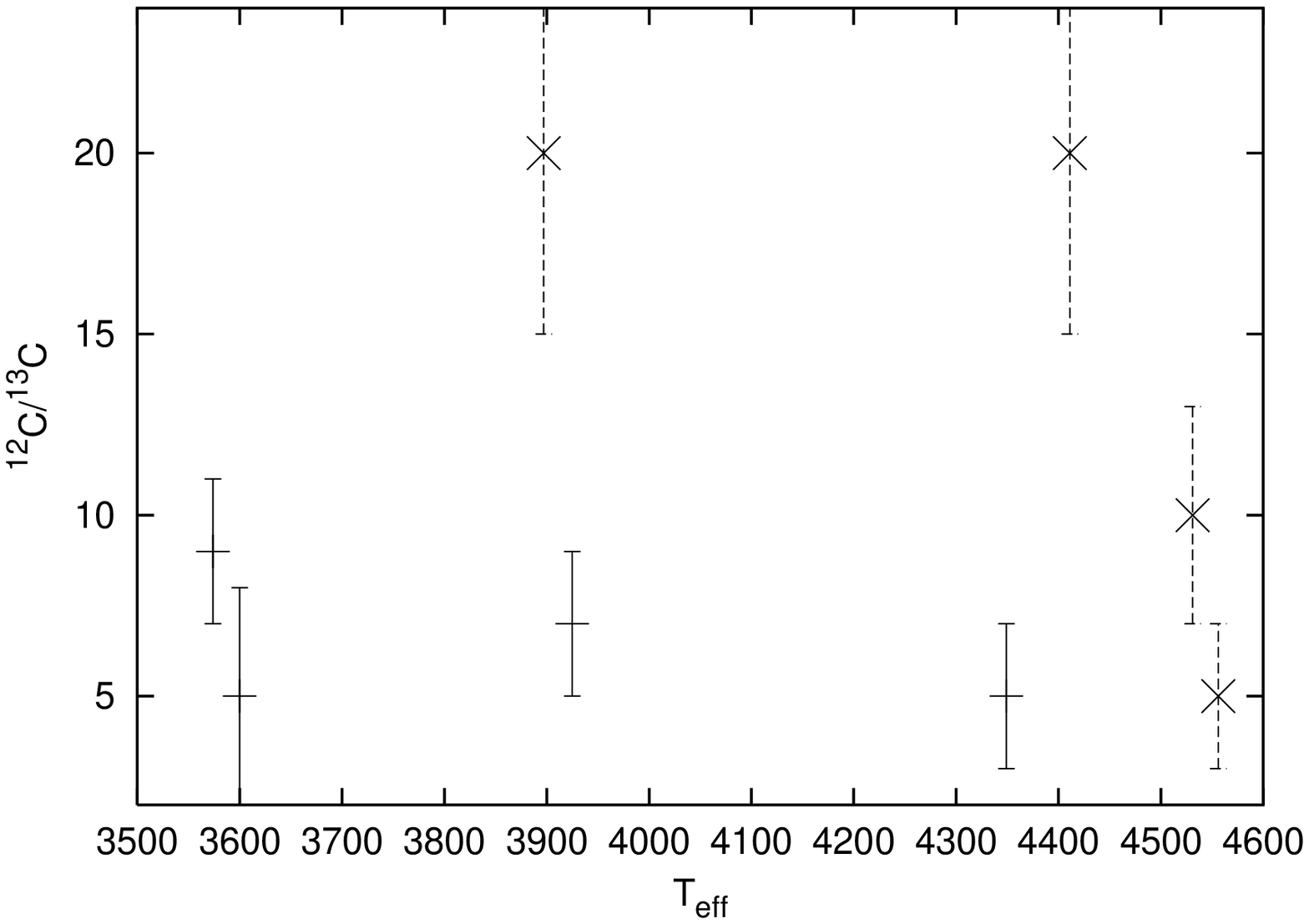} \caption[]{\label{_M71__} The interdependence of
carbon abundance (log N(C)), isotopic ratio (\CDC), effective
temperature (\Tef) and luminosities (L\L$_o$) for ``M71 sample''
giants: cluster members (+), field stars($\times$)}
\end{center}
\end{figure*}

The cluster lies in a relatively crowded field at low
galactic latitude.  From the proper motion study results of Cudworth
(1985) we found that at least four stars of our sample are non members
of M71. The membership probabilities for stars N, A5, A7, A6  are of 0
\%, 6 \%, 0\%, 10 \%, respectively. M71-79 has a 51 \% probable
membership. Our determination of log N(C) and \CDC for the field stars
should be considered as indicative not qualitative results, in part
because we fix their metallicity as \MU = --0.7.
Nevertheless,  analysis of the whole sample of stars enabled us to
confirm that we can indeed see differences between cluster and field
stars.

Our fits to the observed spectra of M71 giants are given in Fig.
\ref{_M71_1}, fits to the field star spectra are shown in
Fig.\ref{_M71_0}.  
In general, the fits are of good
quality. Only fits to spectra with the poorest signal-to-noise ratio,
the hottest giants `N' and `C', cause problems. The membership
probability of M71-C
is 95 \% (Cudworth 1995) and the `best' solutions
suggest too a high value of \CDC =90.
Perhaps we are seeing some peculiarities in the chemical abundance of
this star. This relatively hot (and faint) giant should be studied in
more detail.

Non-member M71-N is explained more simply. We obtained a fit for
a carbon-rich atmosphere with low \CDC, though this probably
results from incorrect input data --- at least the spectrum belongs to a
cooler star with low \CDC.

We found a similar picture for M71-79 --- even with [C/Fe] = 0.36, we
do not reach a minimum in $S$. Most probably, star 79 is not member of
M71.

The values of log N(C) and \CDC determined for our M71 sample are
given in Table \ref{__table1__}. We find:

\begin{itemize}

\item Giants of M71 show \CDC =6 $\pm$ 3; this dispersion is
generally
consistent with observational errors.

\item The dispersion of \CDC in the atmospheres of M71 giants is
smaller than in the M71 non-members sample.


\end{itemize}

\subsection{M5 (NGC 5904)}

M5  is a mildly metal-poor cluster ([Fe/H]= -1.4, Zinn
\& West 1984, [Fe/H]=-1.11, Sneden et al. 1992; [Fe/H]=-1.1
Caretta \& Gratton 1997). From analysis of 36 giants in M5 Ivans
et al. (2001) found no significant abundance variations for Mg,
Si, Ca, CS, Ti, V, Ni, Ba and Eu. However, large
variations were found for abundances of O, Al, Na, which are
sensitive to the proton-capture nucleosynthesis. Recently Cohen
et al. (2002) carried out abundance analysis for stars at the base
of the red giant branch (RGB) of M5. Star-to-star variations with a
significant range in both [C/Fe] and [N/Fe] are found at all
luminosities extending to the bottom of RGB at M$_v \sim$ 3$^m$.

\begin{figure*}
\begin{center}
\includegraphics [width=62mm]
{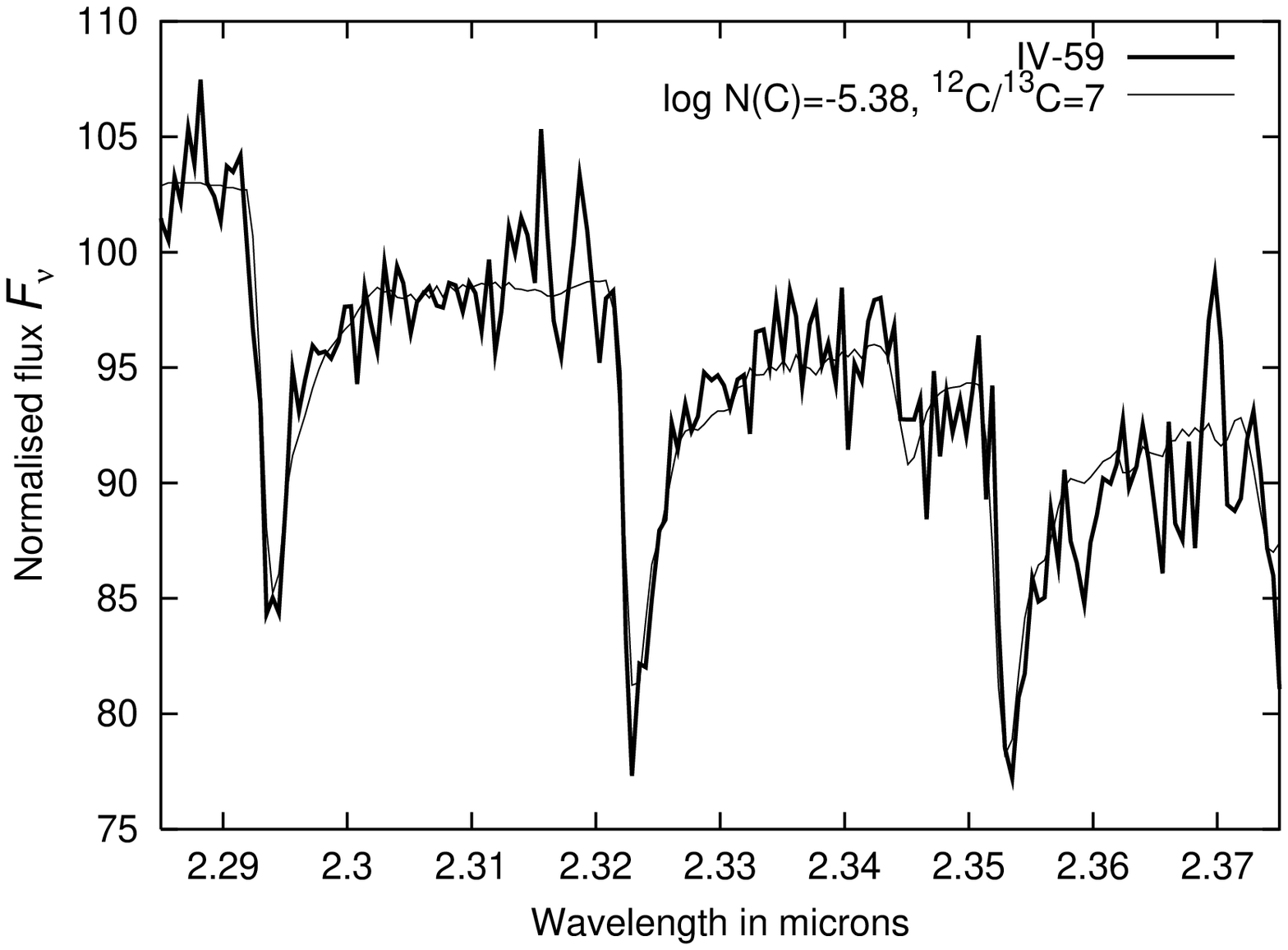}
\includegraphics [width=62mm]
{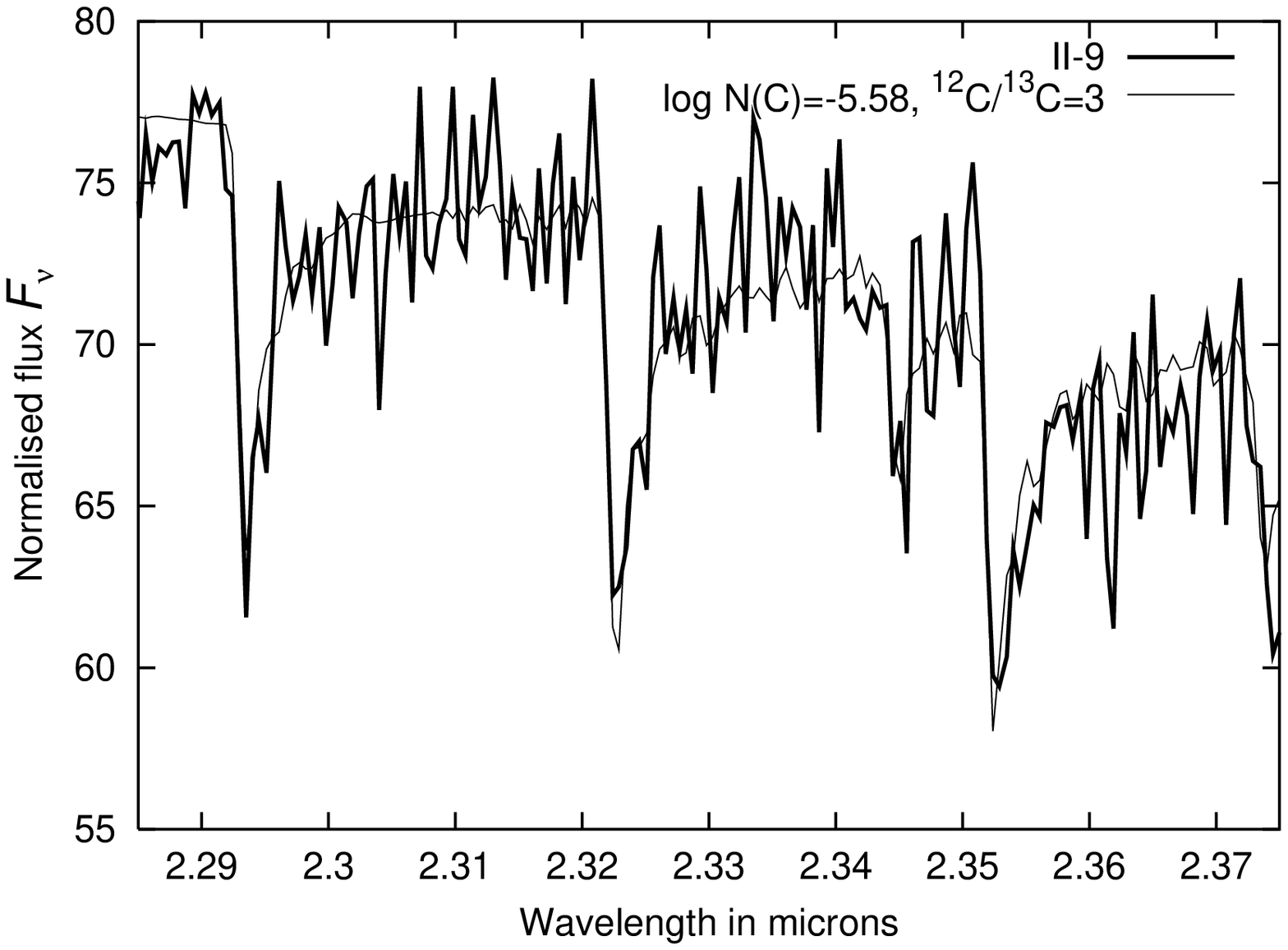}
\includegraphics [width=62mm]
{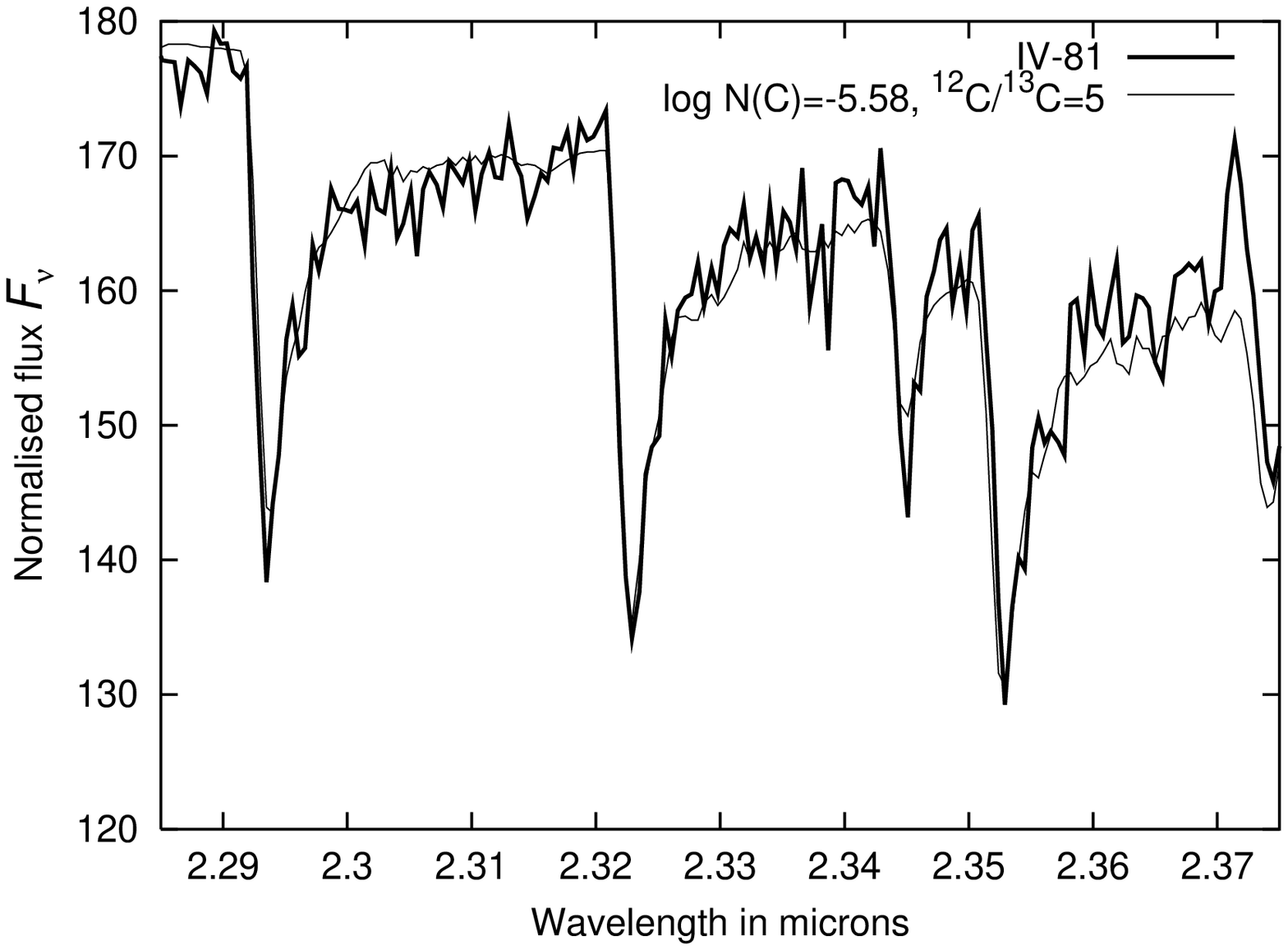}
\end{center}
\caption[]{\label{_M5_} Fits to observed spectra in M5.}
\end{figure*}

Giants IV-59 and II-9 are members of M5
 (99 \% probability, Cudworth
1979); for IV-89 membership is not yet established from proper
motion measurement. However the results of log N(C) and \CDC
determinations for  M5-IV-89 agree well with those for the
established members IV-59 and II-9 (see table \ref{__table1__}).
We conclude that we have confirmed the membership of M5-IV-89
spectroscopically.

Our fits to the observed spectra of M5 giants are shown in Fig
\ref{_M5_}. In all cases we can find reasonable fits. [C/Fe] $<$ --
0.6 in all the giants we measured.

We used the data for the M5 giants to carry out an additional
test, on the sensitivity to metallicity of the carbon abundance we
derive. We performed sets of computations for two metallicities,
[$\mu$] = -1.1 and  [$\mu$] = -1.3. All model atmospheres and
synthetic spectra were re-computed for these abundance grids.
Results are given in Table \ref{__M5__mu}. The changes in isotopic
ratio \CDC and carbon abundance log N(C) deduced indicated that
the uncertainty in metallicity introduced errors of only $\pm$ 1
and $\pm$ 0.2 dex, respectively.

We found a comparatively high \CDC = 7 ratio
 in giant IV-59. Its atmosphere
is moderately deficient in carbon compared to the other giants in
M5.   However, the spectrum of IV-59 is relatively noisy. 
Also, our experiments
with \Vt
show that \CDC =5 $\pm$ 2 for \Vt = 3 km/s 
for this star (see table \ref{__M5_1__}).

\begin{table*}
 \centering
 \begin{minipage}{140mm}

  \caption{Carbon abundances in giants of M5 computed for model atmospheres of
  [$\mu$] = 1.1 and -1.3.}
  \label{__M5__mu}
  \begin{tabular}{@{}ccccccccccc@{}}
\hline
    &        & & \multicolumn{3}{c}{[$\mu$]=-1.3}&
    \multicolumn{3}{c} {[$\mu$]=-1.1} \\
\hline
Cluster & Object &\Tef/log g &log N(C)&[C/Fe]& \CDC& log N(C)&[C/Fe] & \CDC \\
\hline
M5 & IV-59 & 4343/1.0  & -5.38  & -0.6 & 7 & -5.18 & -0.6  & 8 \\
   & II-9   & 4320/0.9  & -5.58  & -0.8 & 3 & -5.58 & -1.0 & 3 \\
   & IV-81  & 3963/0.6 & -5.58   & -0.8 & 5 & -5.58 & -1.0 & 5 \\
\hline
\hline
\end{tabular}
\end{minipage}
\end{table*}

\subsection{M13 (NGC 6205)}

Kraft et al. (1997) studied abundances of O, Na, Mg and Al in
giants of M13 ([Fe/H] = -1.5).
 Briley et al. (2002) found that  M13 is distinguished by rather
large star-to-star variations in carbon (and possibly other light
elements).

\begin{figure*}
\begin{center}
\includegraphics [width=62mm]
{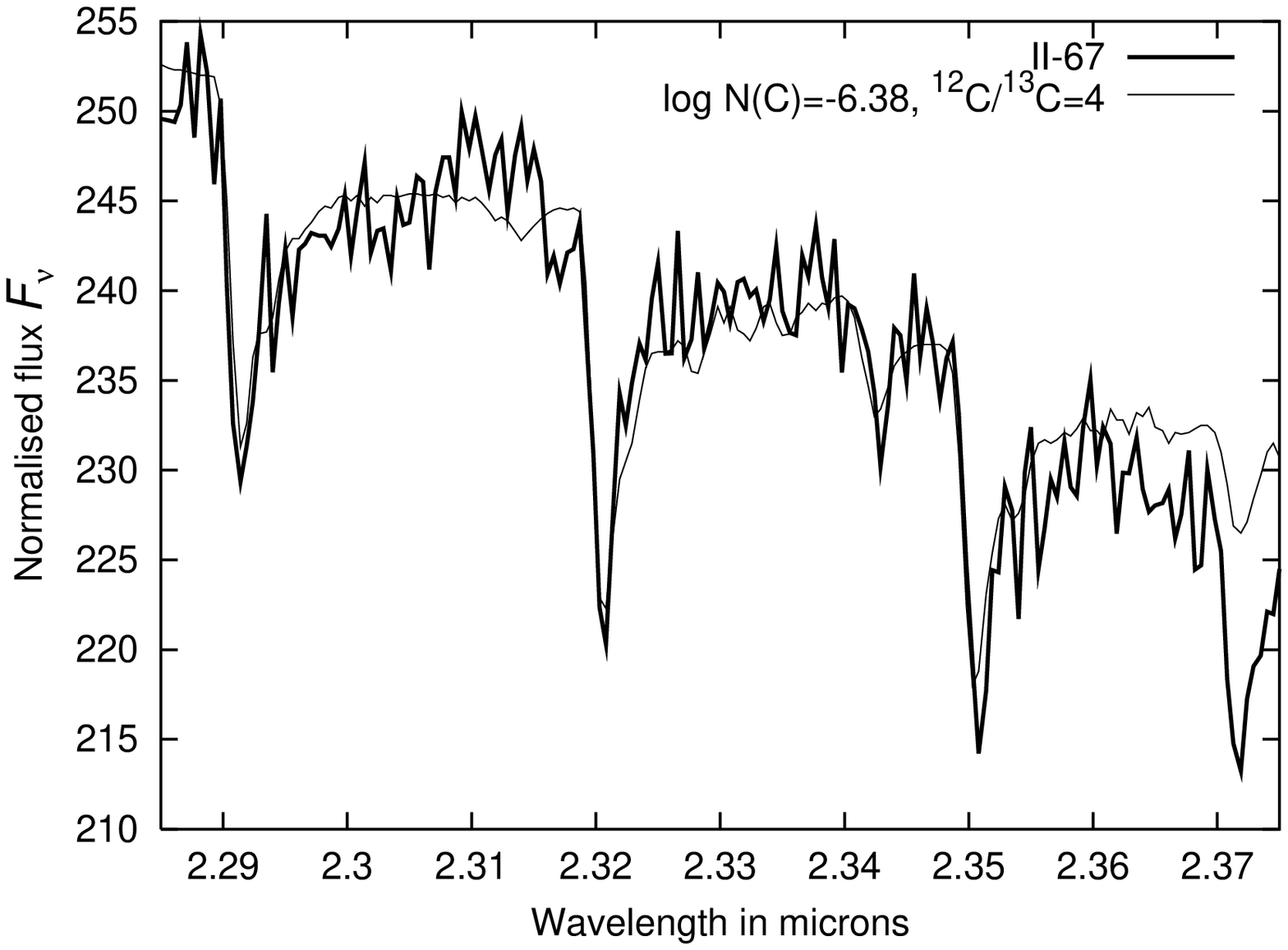}
\includegraphics [width=62mm]
{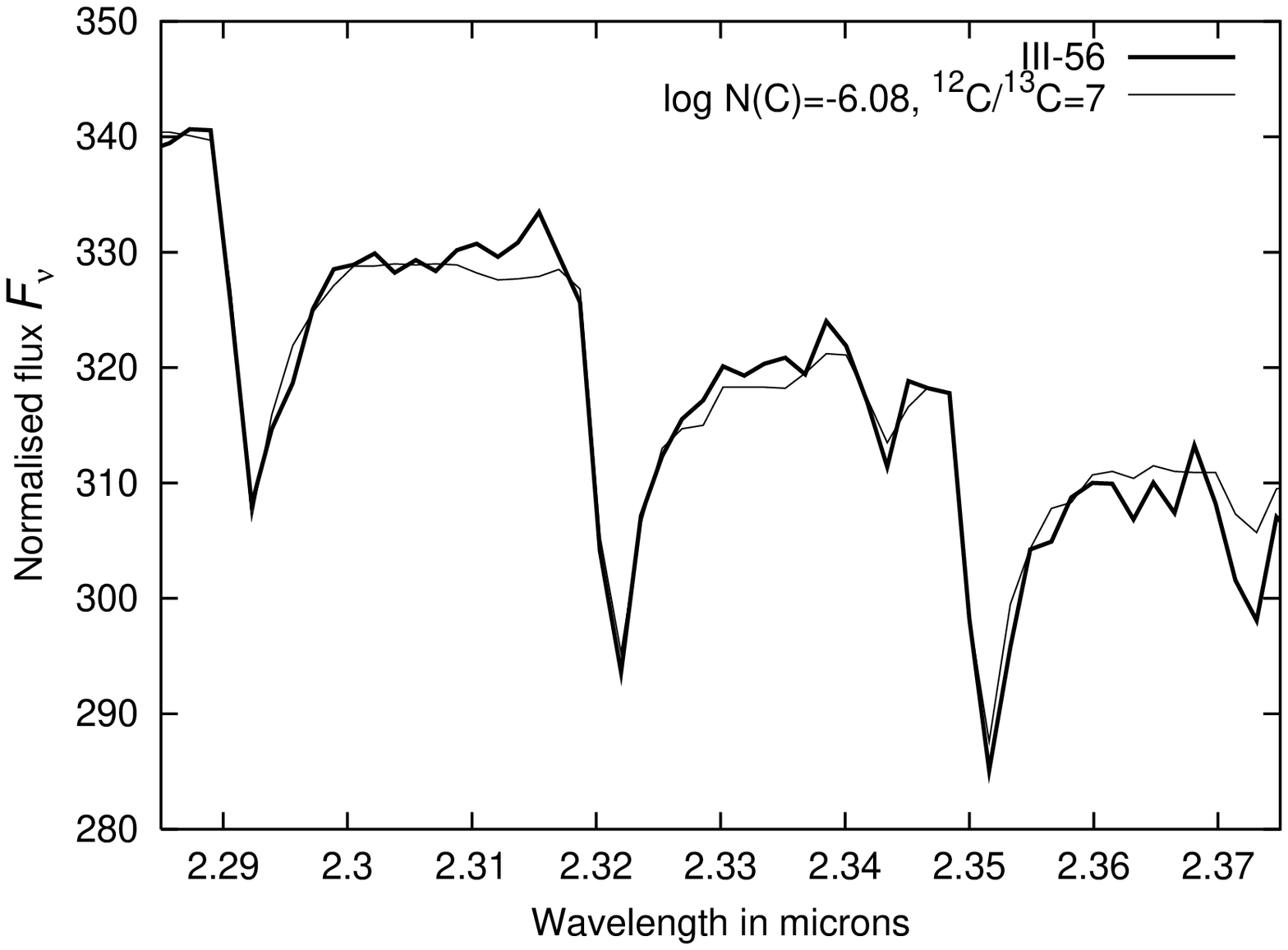}
\includegraphics [width=62mm]
{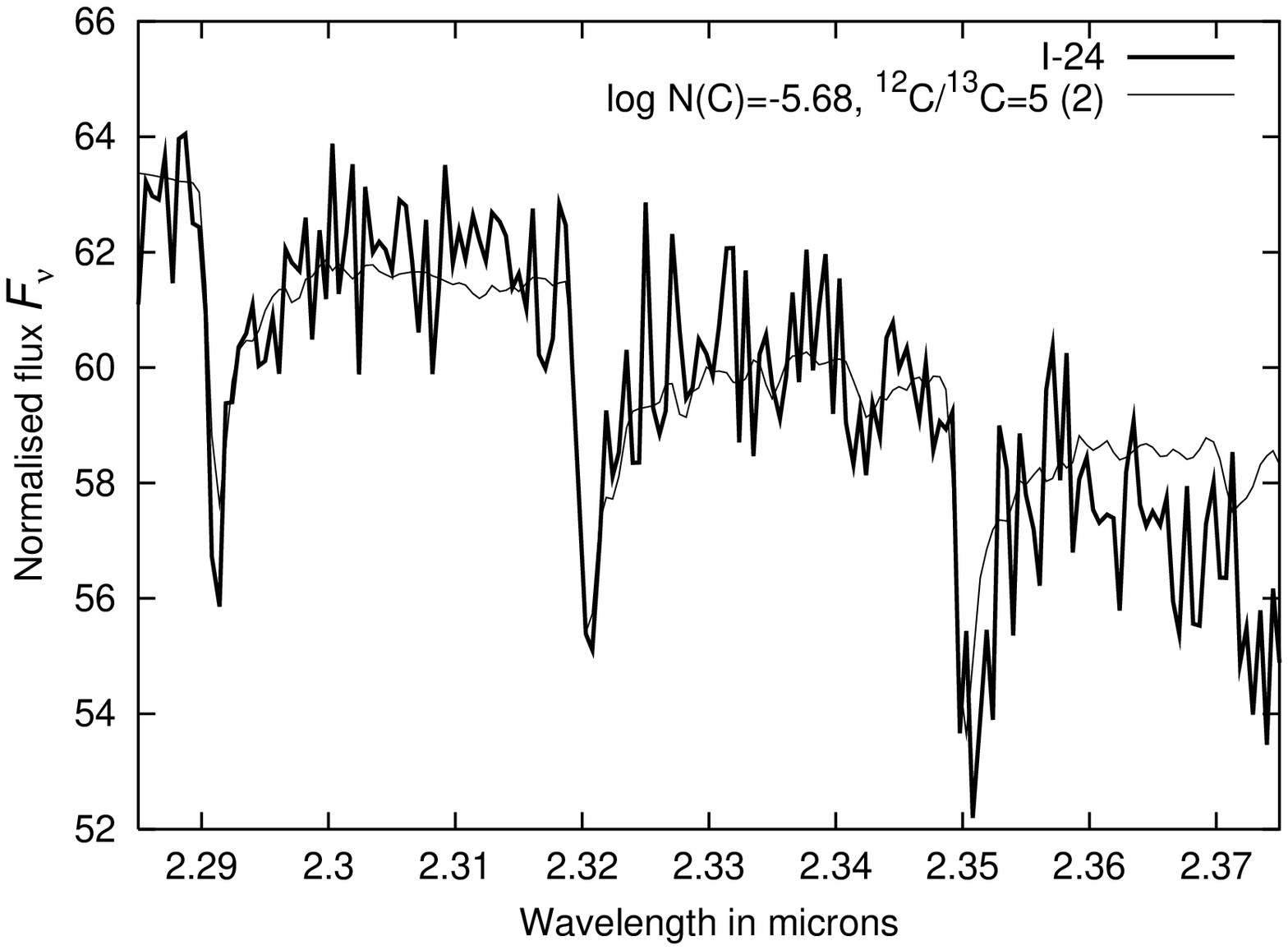}
\includegraphics [width=62mm]
{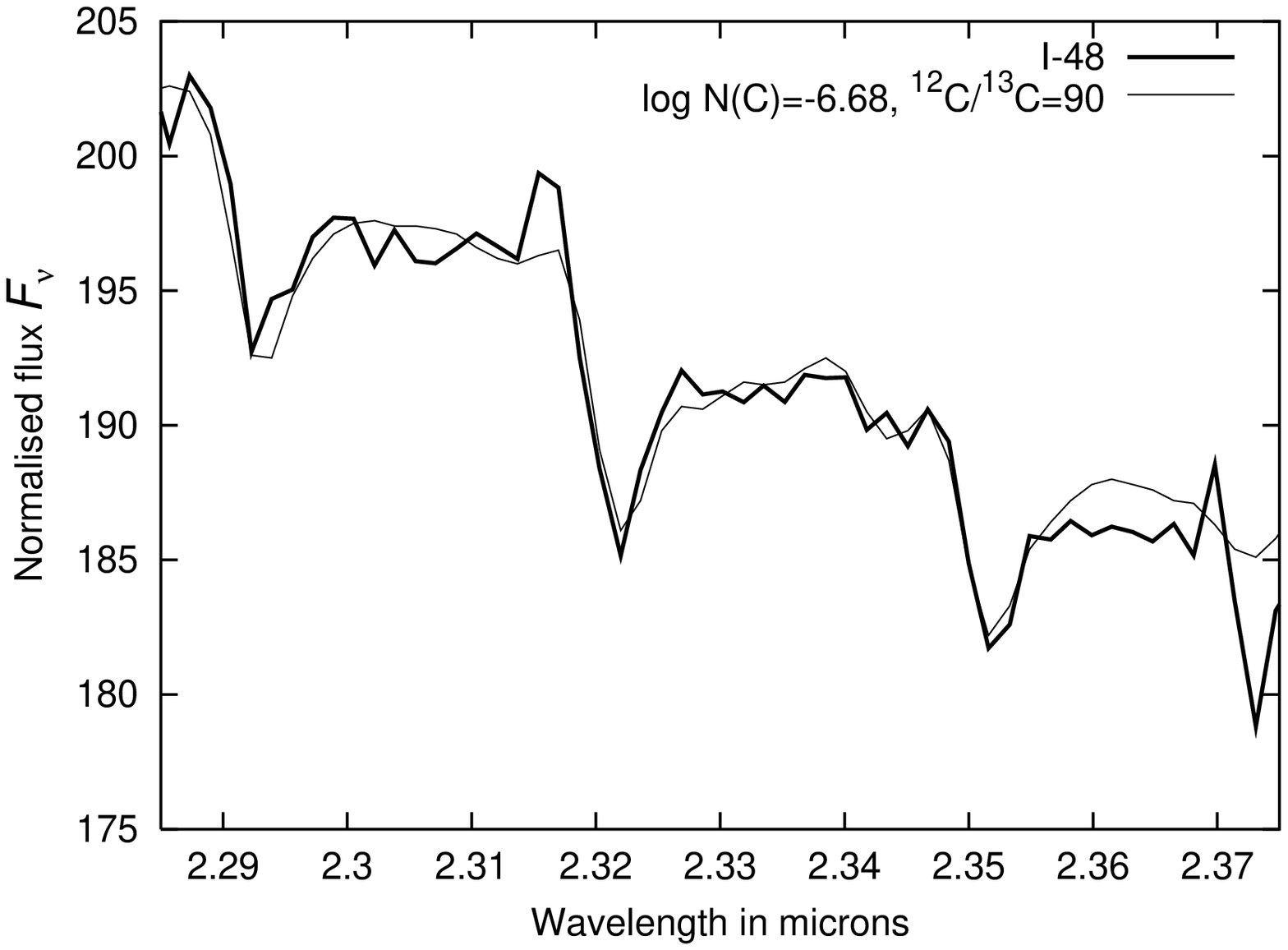}
\includegraphics [width=62mm]
{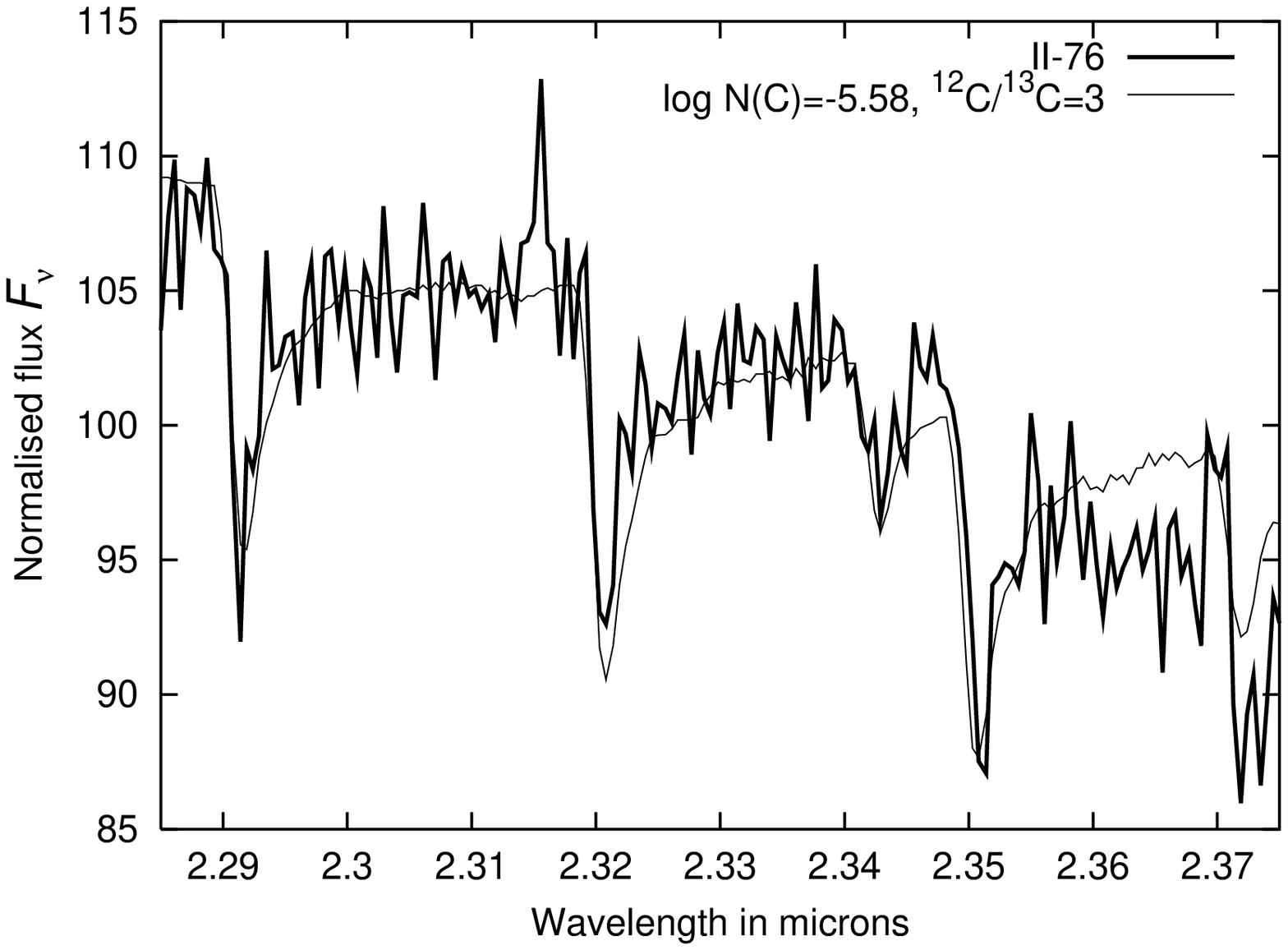}
\includegraphics [width=62mm]
{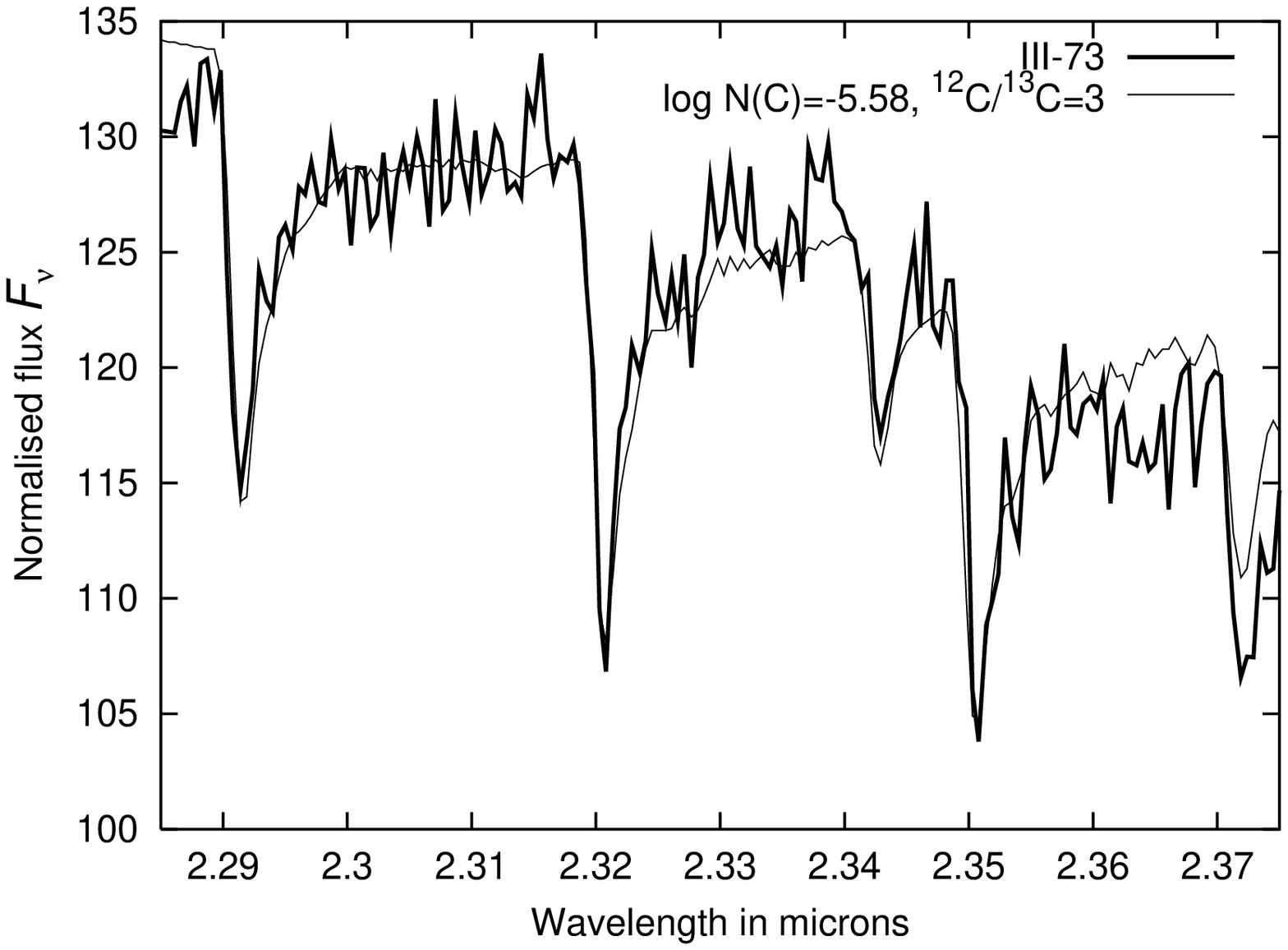}
\end{center}
\caption[]{\label{_M13_} Fits to the observed spectra in M13.}
\end{figure*}

 All stars of our sample are members of M13 with probability
99 \% (Cudwotrh 1979a).
The fits to M-giants spectra of
M13 are presented in Fig \ref{_M13_}. The spectrum of M13-I-18 is too
noisy to permit any reliable analysis and spectra of
M13-III-56, M13-I-48, M13-I-24 are also noisier than for our other giants.

\begin{figure*}
\begin{center}
\includegraphics [width=62mm, height=32mm]
{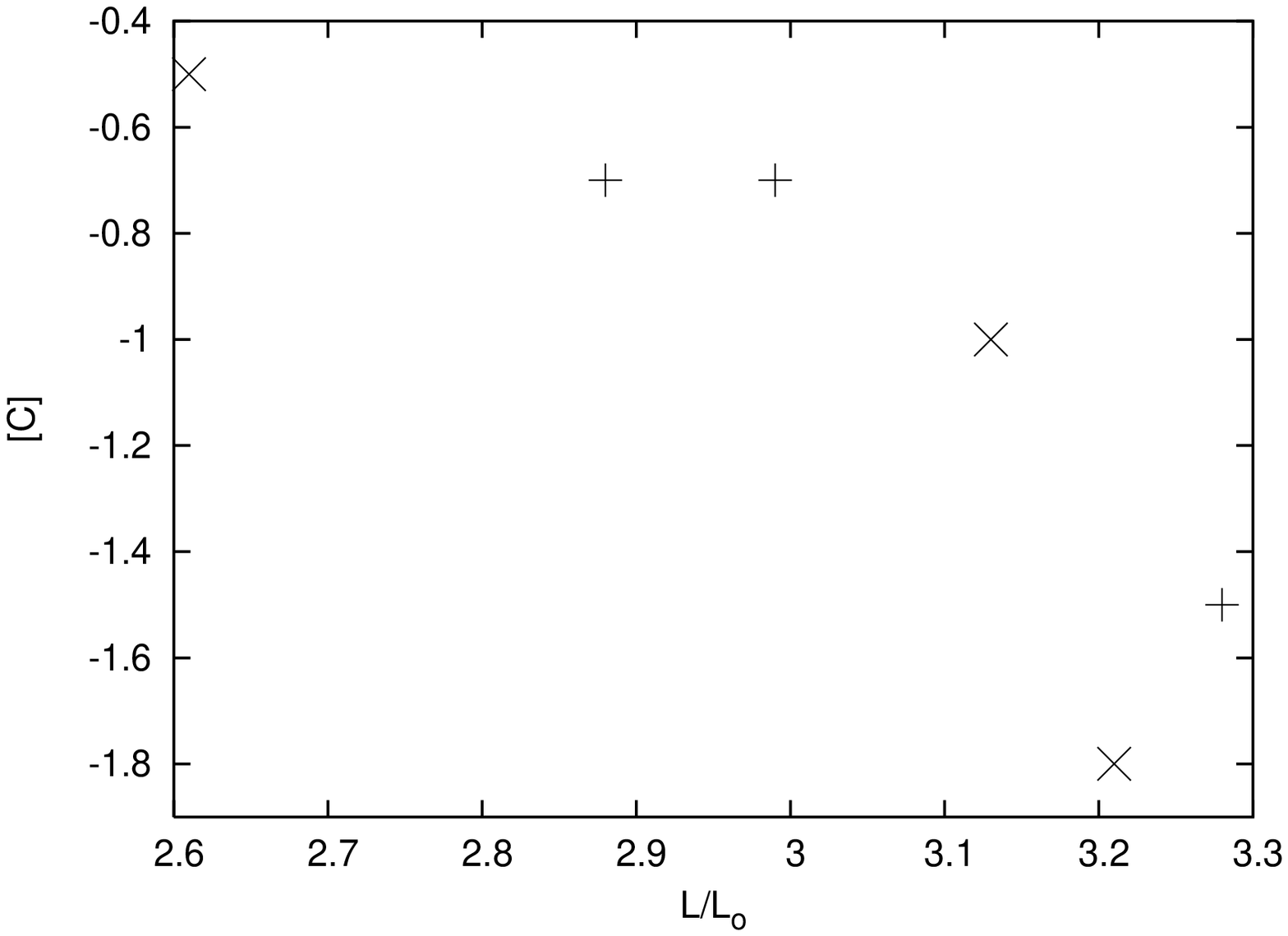}
\includegraphics [width=62mm, height=32mm]
{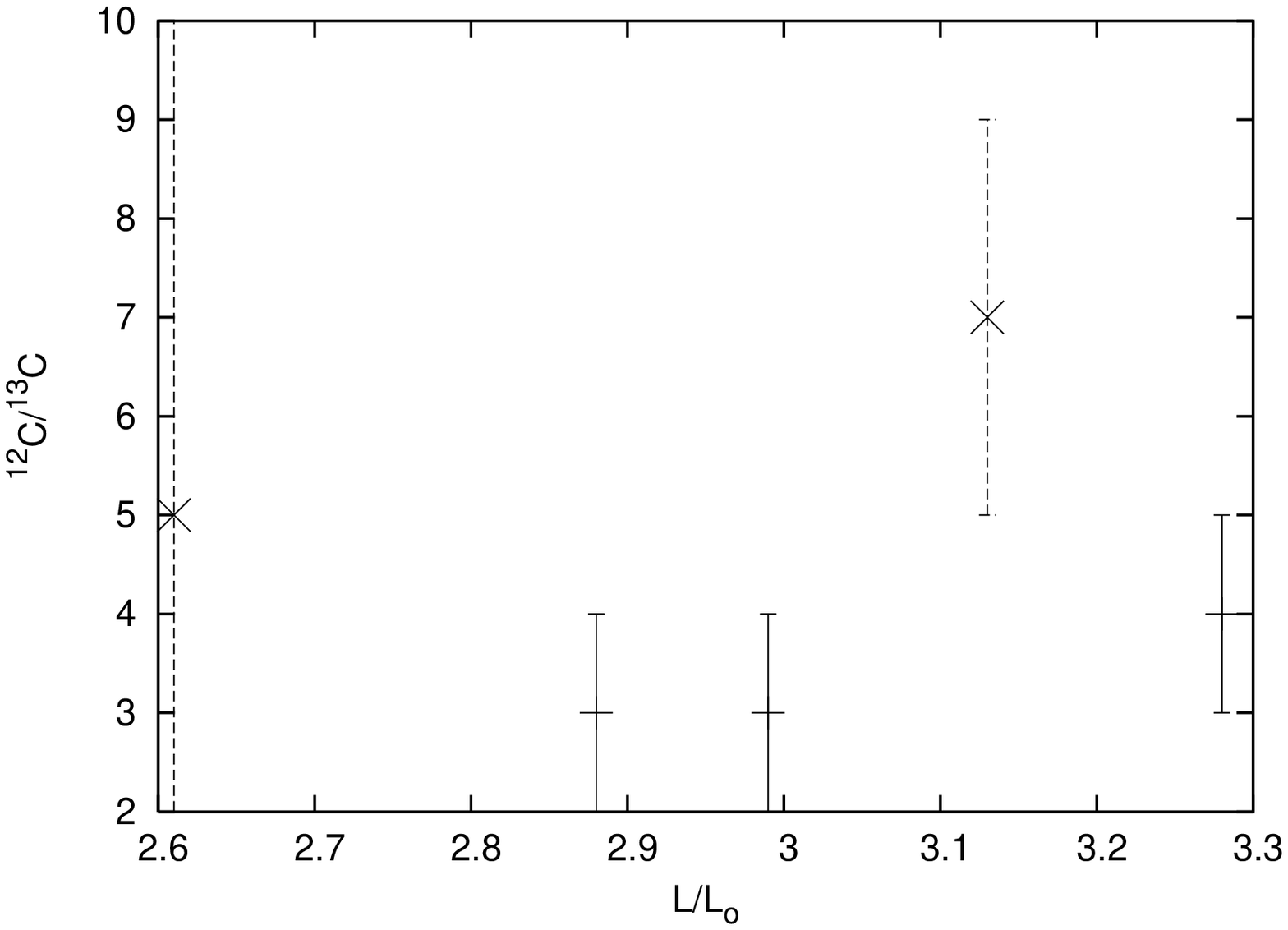}
\includegraphics [width=62mm, height=32mm]
{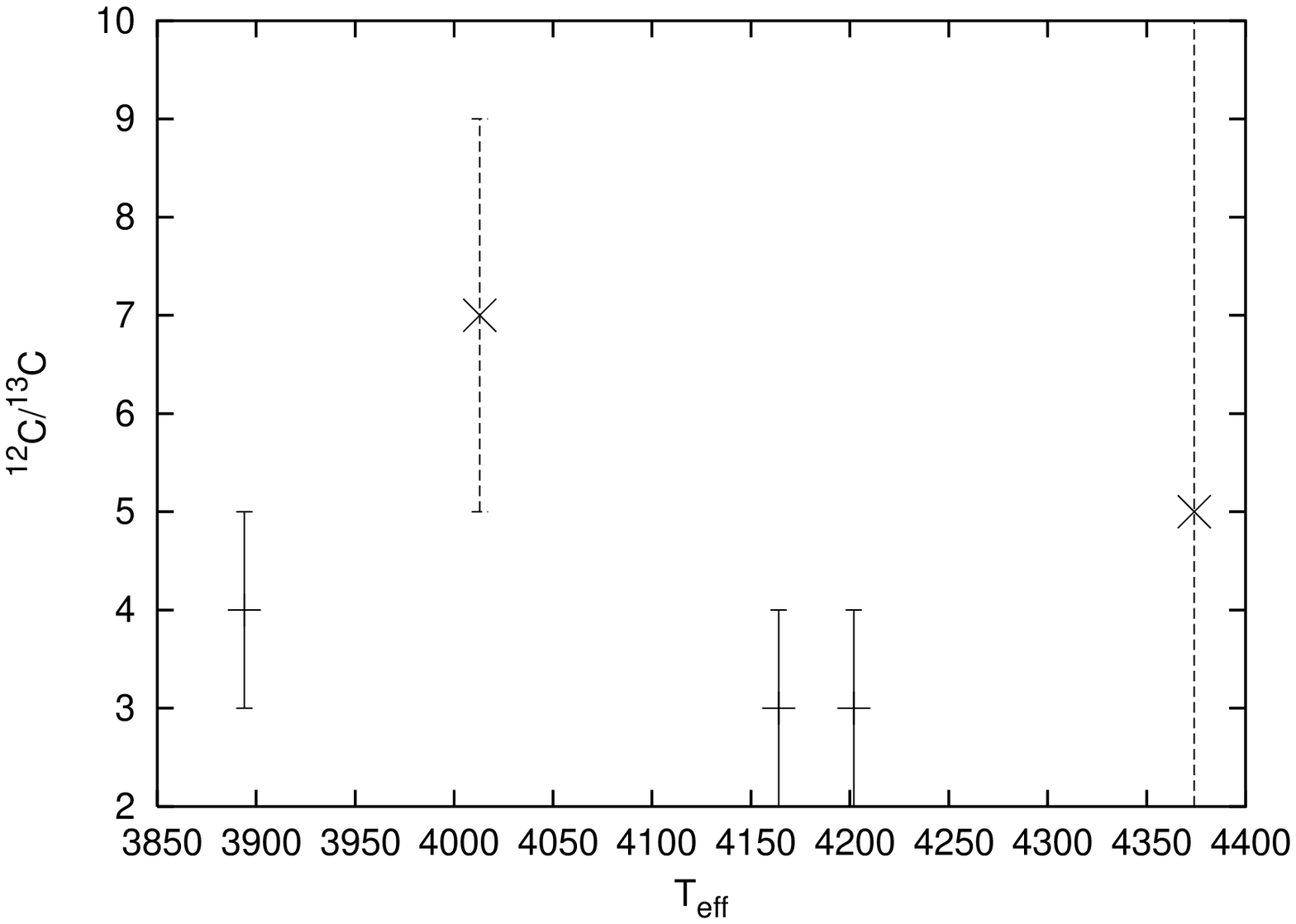}
\includegraphics [width=62mm, height=32mm]
{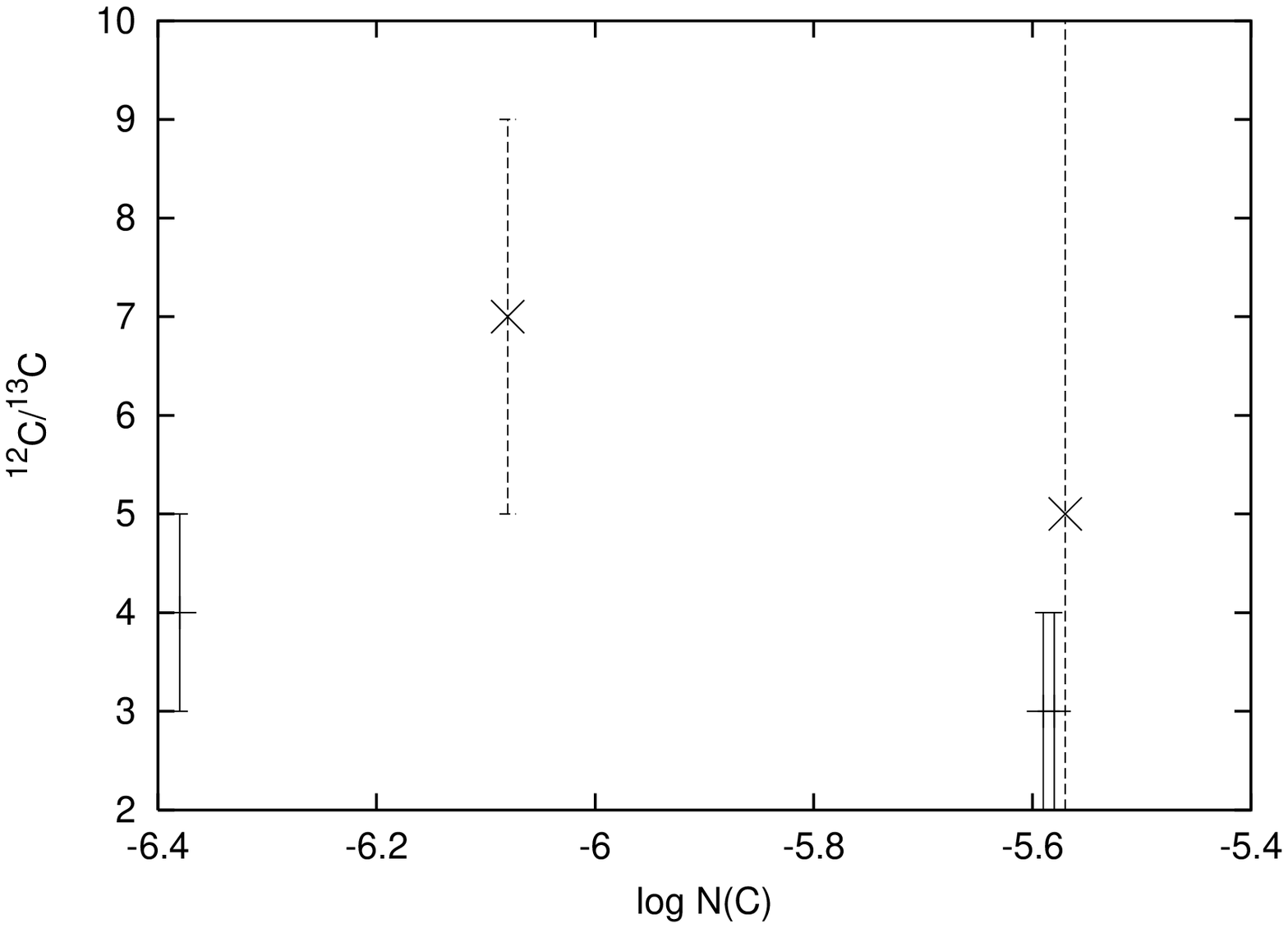} \caption[]{\label{_M13__} The interdependence of carbon
abundance (log N(C)), isotopic ratio (\CDC), effective temperature
(\Tef) and luminosities (L\L$_o$) for M13 giants. Resylts for more and less
confident spectra are showed by (+) and ($\times$), respectively. }
\end{center}
\end{figure*}

Fig. \ref{_M13__} shows a negative slope in the dependence of log N(C) on
f(L/L$_{\odot}$). This result was found by other
authors for different clusters
(e.g., Briley et al. 2002 and references therein).
In the atmospheres of the majority of M13 giants \CDC $<$ 10.

A few determinations of log N(C) for some giants of  M13
are known from the literature. In
general, our determinations agree with
them well (see table \ref {__abund}).

\begin{figure*}
\begin{center}
\includegraphics [width=62mm]
{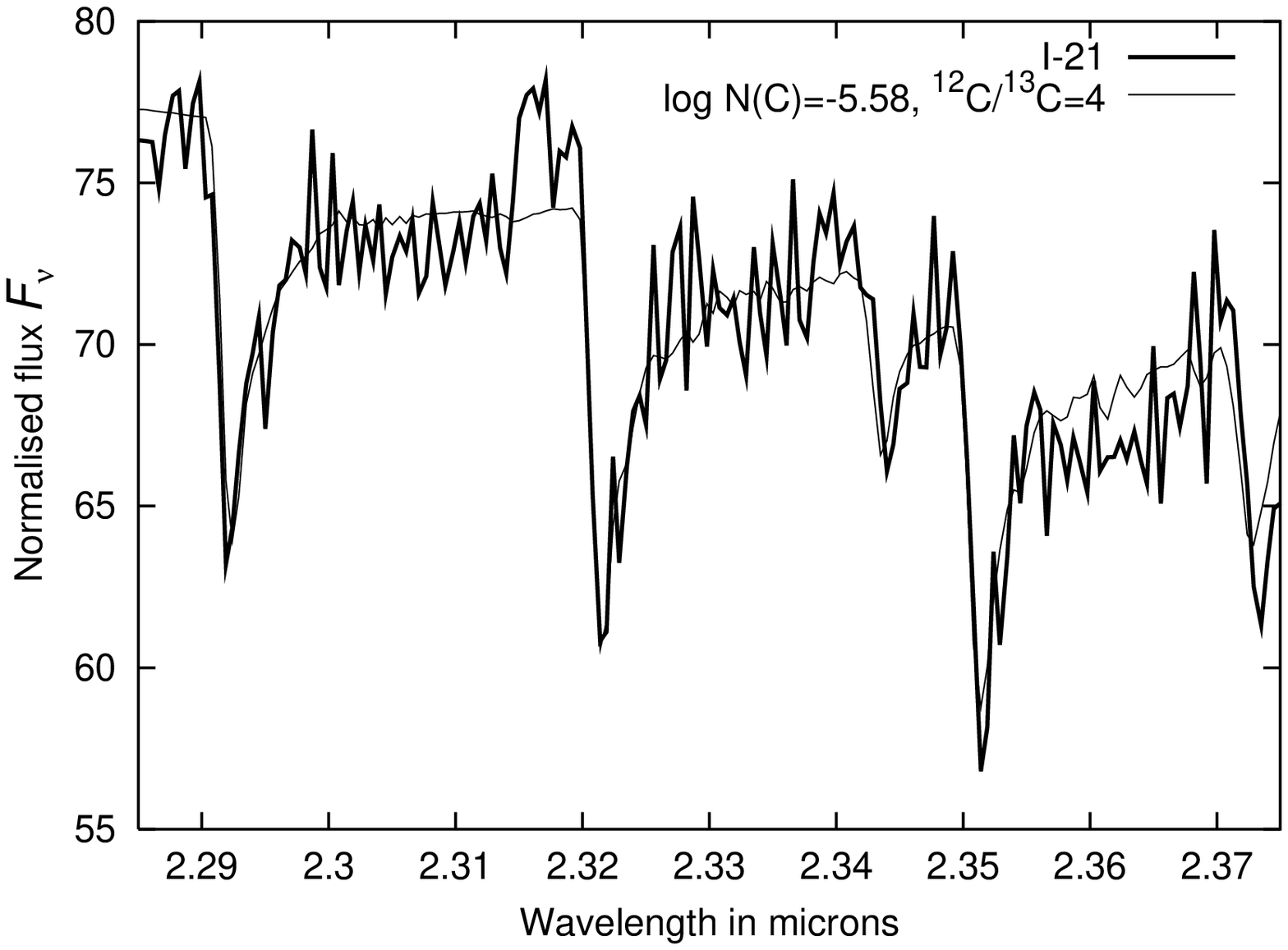}
\includegraphics [width=62mm]
{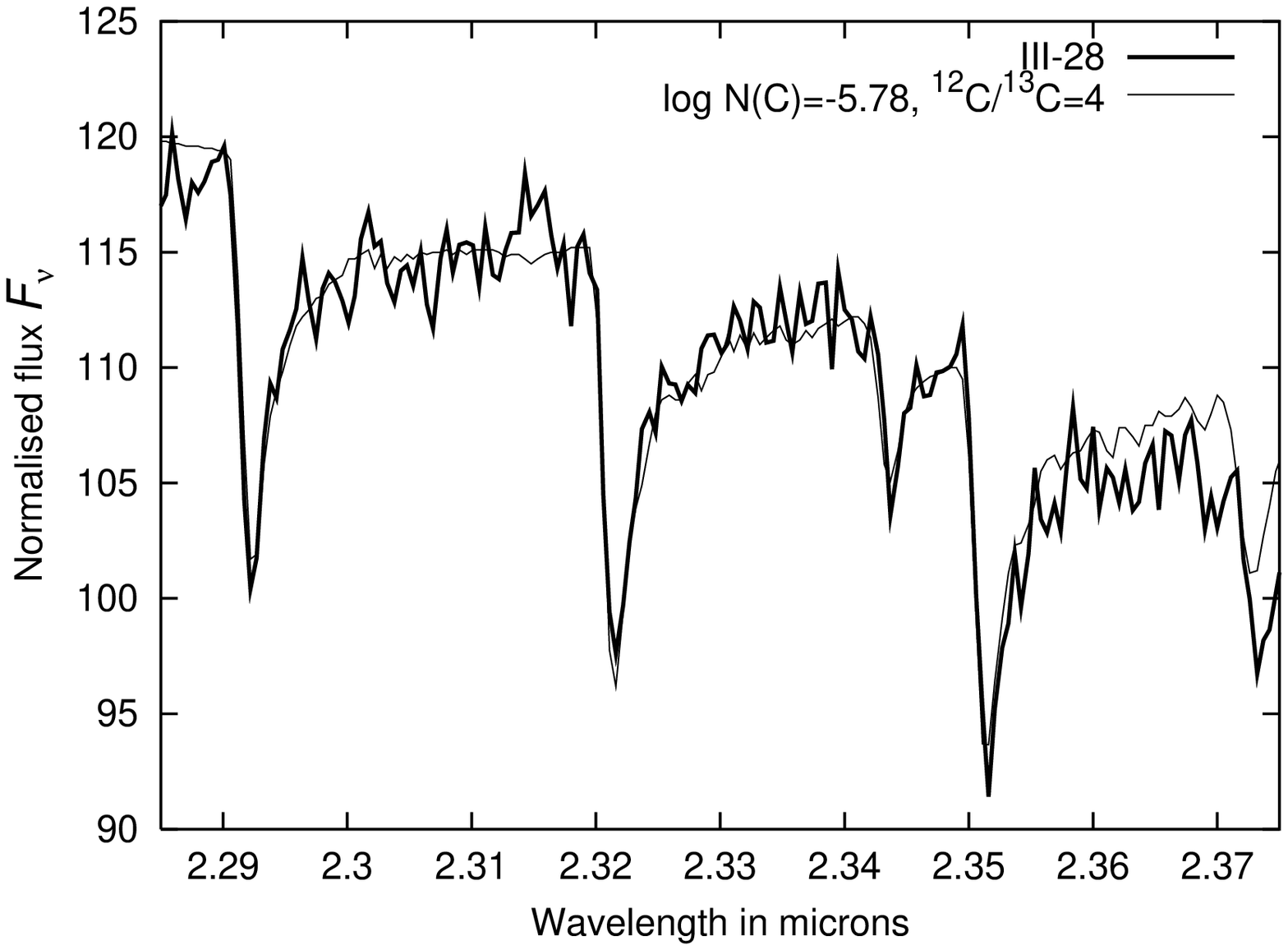}
\includegraphics [width=62mm]
{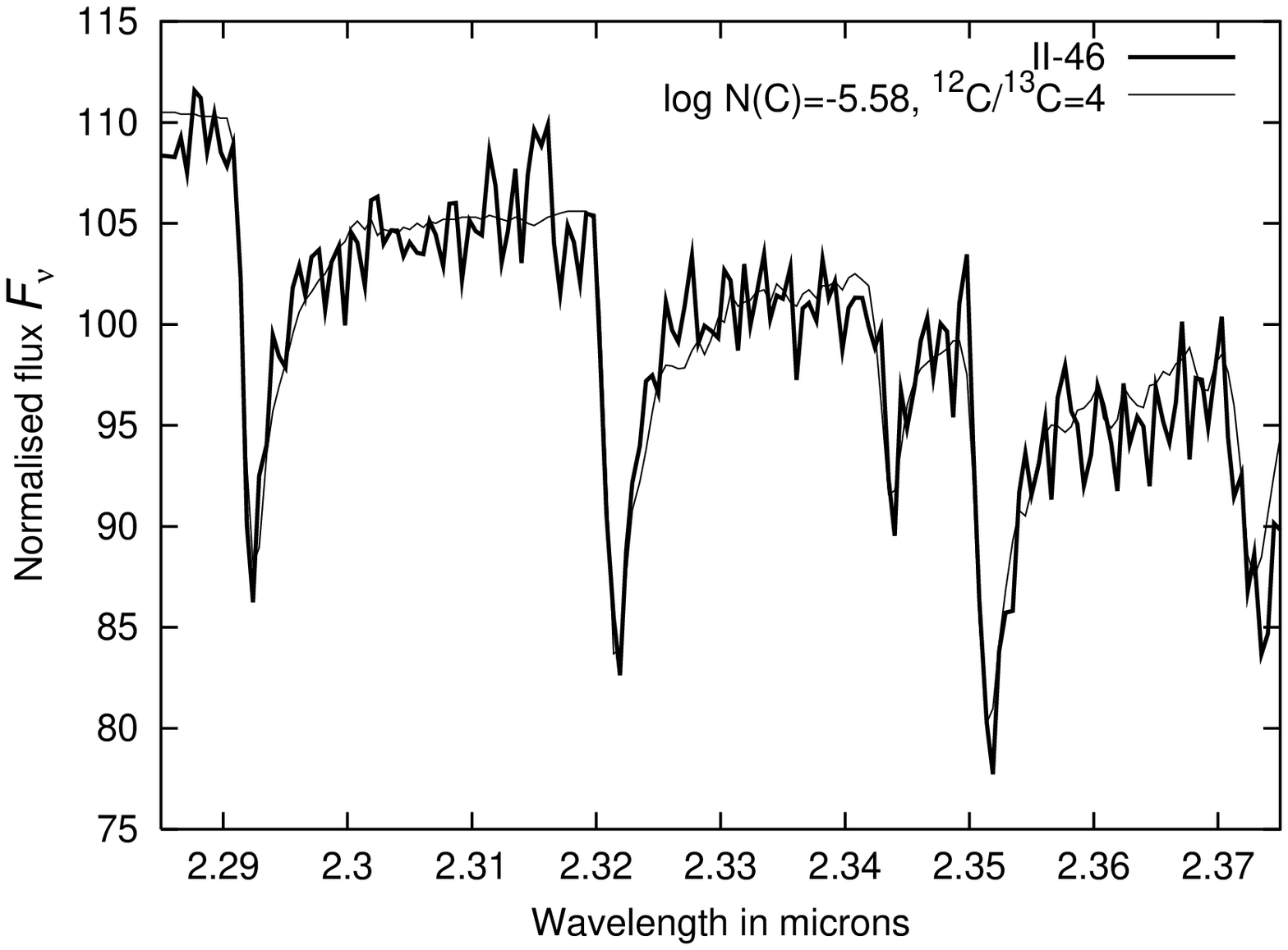}
\includegraphics [width=62mm]
{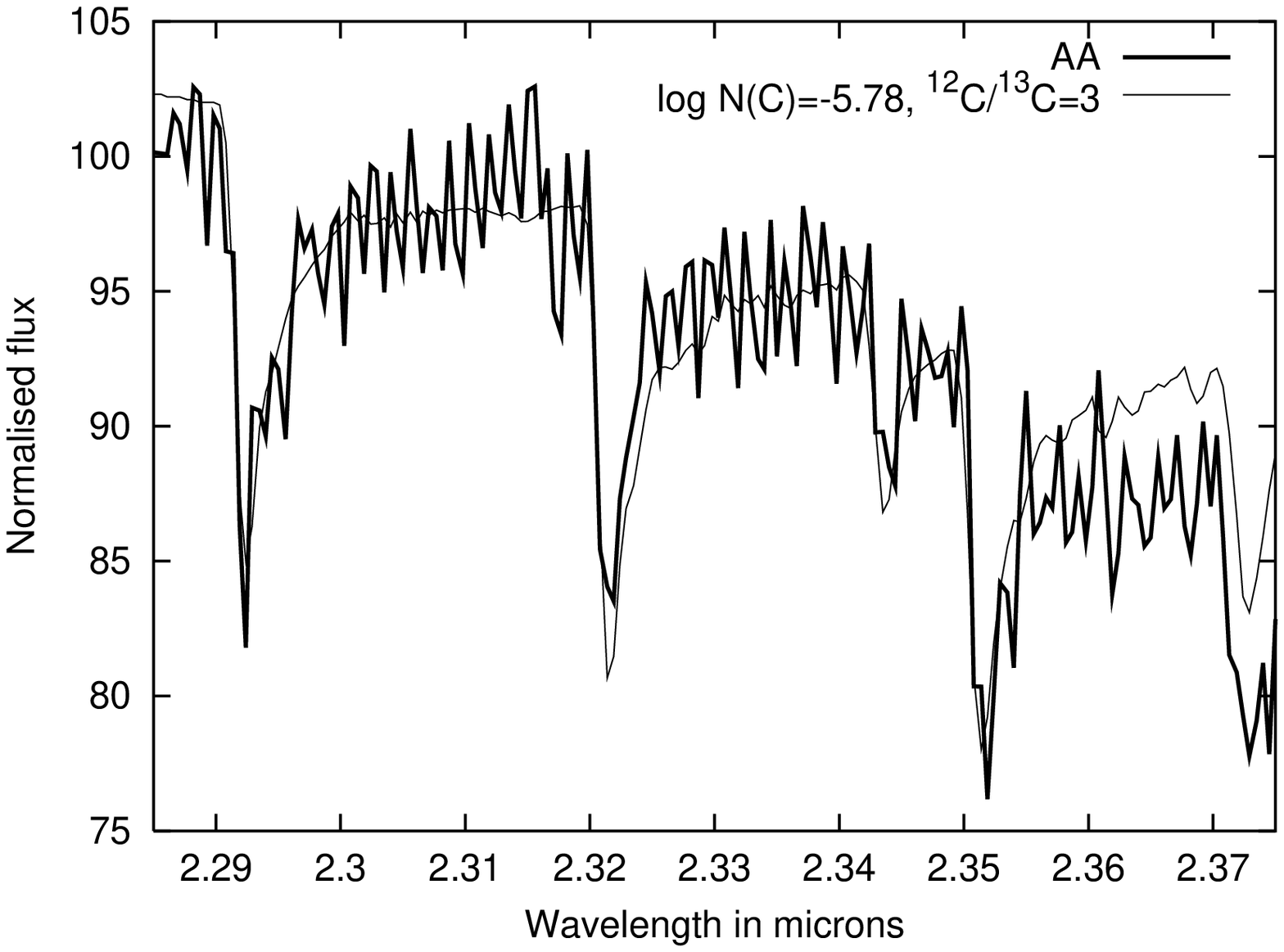}
\includegraphics [width=62mm]
{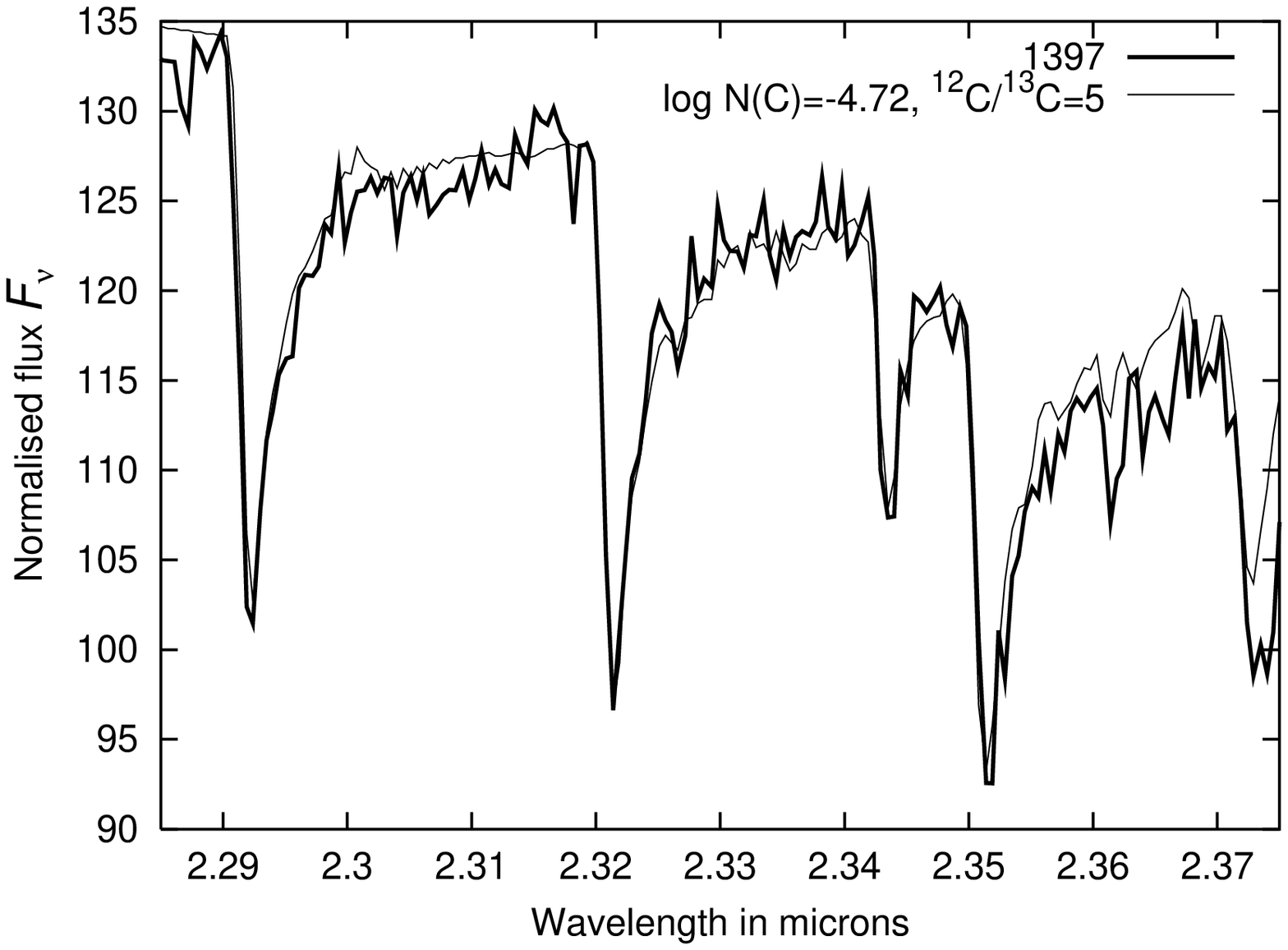}
\end{center}
\caption[]{\label{_M3_} Fits to observed spectra in M3.}
\end{figure*}

\subsection{M3 (NGC 6272)}

Globular cluster M3 is the most metal deficient from our sample (\MU=
--1.6). According to Cudworth (1979b), all our stars are members of M3
(with probability 99 \%), except for 1397. We found a carbon-rich
solution for this star, which
seems to be an artefact (like M71-N and M71-79), most probably caused
by adopting (incorrectly) a metallicity the same as that of the cluster.

Fits to observed spectra are shown in Fig. \ref{_M3_}. Our
sample of giants has a small range of temperature around 4000K
though a large range of log N(C) from  --5.58 to --6.08. Nonetheless
all have a very similar
\CDC ratio of  4 --- 5. Our carbon abundance
determinations agree well with the M3 giants in the list of
Smith et al. (1997).

\begin{figure*}
\begin{center}
\includegraphics [width=62mm]
{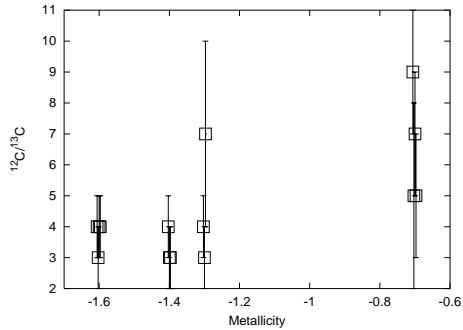}
\end{center}
\caption[]{\label{_muCC} Isotopic ratios \CDC in giants of
different clusters. Two giants with \CDC $>$ 90 are not showed to
simplify the plot.}
\end{figure*}

\section{Discussion}

We have analyzed photospheric carbon isotope ratios for a sample
of 25 giants in different galactic globular clusters
using fits to infrared spectra dominated by \HHO and CO bands in the
wavelength range 2.28--2.39 \mum.
We used \Tef and log g values from Alonso et
al. (1999) and cluster metallicities from the
literature. Effective temperatures in Alonso et al. are given with
an accuracy of $ < \pm$ 100 K. Our numerical experiments indicate that
with this level of uncertainty our results remain inside our
empirical error bars: $\pm$
0.2 and 1 for carbon abundance and \CDC  respectively in the atmospheres of the
majority of our giants.

In our analysis we used model atmospheres computed using a
classical approach with opacity sampling treatments of the
absorption by atomic and molecular lines. Comparison of our
models with NEXTGEN oxygen rich model atmospheres (Hauschildt at
al. 1999) and carbon-rich models of Erikson (1994) shows good
agreement (Pavlenko 2002, 2003). For the most luminous giants
the plane-parallel models may be not valid. Nevertheless,
due to the lower opacities in the atmospheres of metal-deficient
stars, their photospheres move toward higher pressure regions, in
which sphericity effects should be smaller. Therefore we believe
that our results would not change significantly if analysed with
model atmospheres computed taking sphericity effects into account. We
also note that for luminous stars we should in principle
compute effects caused by mass loss processes, i.e. stellar
wind, peculiar convection processes, inhomogeneities of their
photospheres. In practice, consideration of these effects lies beyond the
scope of this paper.

Although nitrogen and oxygen abundances can vary in the
atmospheres of globular cluster giants, initially we set fixed
abundances of oxygen and nitrogen. Our numerical experiments
(section 4) then indicated that the dependence of our results on
the adopted value of log N(C) is much stronger than the dependence
on log N(O). This may be understood in terms of molecular
abundances -- N(O) $>>$ N(C), and N(CO) $>>$ N(CN) in our cases.
We did not try varying the abundance of nitrogen because there
were no transitions involving nitrogen in our spectra.

Our carbon abundances are in good agreement with other
authors despite the different numerical procedures inherent to the
various models. We therefore confirm the deficiency in carbon
abundances and the general nature of low \CDC in the
atmospheres of the giants of globular clusters. We determined log N(C)
and \CDC for relatively luminous giants which are above the bump of
luminosity function. We note that an
increase of \CDC with decreasing luminosity towards the bump has been
observed for giants in M4 and NGC 6528 (Shetrone 2003). In general,
our results are also in good qualitative agreement with the theoretical
predictions of Boothroyd \& Sackman (1999) for the connection of the
surface abundances and the still hypothetical deep circulation mixing
below the base of the standard convective envelope and the
consequent ``cool bottom processing'' of CNO isotopes.
We note that in these theories of the final stage of evolution of
a single star, a number of important parameters remain poorly
known. These include mass loss, rotation rates (see Denissenkov et
al. 1997 and reference therein), the temperature difference
between the bottom of mixing and the bottom of H-burning shell
(Boothroyd \& Sackman 1999).

Some mechanisms that have been suggested to transport evolved matter into the
surface layers include: \\
a) meridional circulation driven by rotation of the
interior (Sweigart and Mengel 1979), \\
b) turbulent diffusion
processes (Bienayme et al 1984, Denissenkov \& Weiss 1996), \\
c) deep circulation mixing (Boothroyd \& Sackman 1999). \\
In general the efficiency of such mixing mechanisms
depends on the individual characteristic of each star ---
rotation, multiplicity, etc. Moreover,
the problem is
complicated by the need to take
into account a contribution from pollution processes on the early
stages of evolution of globular clusters, which have their own history.

\section{Conclusions}

We confirm the results of previous observational studies that the \CDC
ratio in metal-poor stars is a factor of 3-5 lower than predicted by
standard models of low-mass stellar evolution that do not include mixing
beyond the first dredge-up.

$\bullet$ Within both M3 and M13, with \MU $<$ - 1.3, the  giants we
observed have about the same \CDC ratio.

$\bullet$  Fig. \ref{_muCC} provides some evidence for lower \CDC
in giants of globular clusters of lower metallicity. This result was
predicted by theory. Indeed,  all theories of mixing to date predict a
metallicity dependence for mixing efficiency - the molecular weight
gradient/barrier will be steeper with increasing \MU and the H-shell
will be hotter/wider with decreasing \MU, so a plot like Fig.
\ref{_muCC} was expected. 

$\bullet$ Giants of more metal-rich clusters show larger
dispersion of \CDC (see Table \ref {__table1__}).

$\bullet$ More evolved red giants with lower log g ($<1$)
show reduced log N(C).

Our measurements confirm the conclusions of work on field giants.
However, the  carbon isotope ratios of  the most evolved population I
giants are even
lower than our typical values of \CDC = 4 -- 9.
Nevertheless, our data in
concert with carbon isotope ratios measured in other studies, for
field and cluster stars spanning a large range of metallicities,
illustrate the existence of an unidentified extra mixing
processes inside globular cluster giants after the the first
dredge-up event. To distinguish when this process begins it
is important to conduct similar measurements on less evolved
stars in these clusters.

\section*{Acknowledgments}

 We thank our anonymous referee for a very helpful review
of the paper.
This work was partially supported by a PPARC visitors grants from
PPARC and the Royal Society.
YPs studies are partially supported by a Small Research
Grant from American Astronomical Society. We thank David
Schwenke and David Goorvitch (AMES) for providing \HHO and CO line
lists in
digital form and Roger Bell and Paul Butler for useful discussions.

This research has made use of the SIMBAD database,
operated at CDS, Strasbourg, France.

\bsp

\label{lastpage}


\begin{thebibliography}{99}




\bibitem[\protect\astroncite{Alonso}{2000}]{Alonso1999}
Alonso, A., Arribas, S., Martinez-Roger, C, 1999, A\&AS., 139, 335.

\bibitem[\protect\astroncite{Alonso}{2000}]{Alonso2000}
Alonso A., Salaris M., Arribas S., Martinez-Roger C., Asensio Ramos A.
2000, A\&A, 355,1060.


\bibitem[\protect\astroncite{anders}{1989}]{Anders1989}
  Anders, E. \& Grevesse, N. 1989, Geochimica et
    Cosmochimica Acta, 53, 197.


\bibitem[\protect\astroncite{Asida}{2000}]{Asida2000}
Asida, S.M. 2000 ApJ 528, 896.




\bibitem[\protect\astroncite{Bell}{1991}]{Bell1991}
Bell, R.A., Briley, M.M. 1991, AJ, 102, 763.


\bibitem[\protect\astroncite{Bienayme}{1984}]{Bienayme1984}
Bienayme, O., Maeder, A., Schatzman, E.
1984, A\&A, 131, 316.

\bibitem[\protect\astroncite{Boothroyd}{1999}]{Boothroyd}
Boothroyd, A.I., Sackman, I-J. 1999, ApJ, 510, 232.

\bibitem[\protect\astroncite{Boyarchuk}{1991}]{Boyarchuk1991}
Boyarchuk, M. E., Pavlenko, Y. V., Shavrina, A. V. 1991.
         Sov. Astron., 35, 143.


\bibitem[\protect\astroncite{Briley}{1997}]{Briley1997}
Briley, M.M., Smith, V.V., Lambert, D.L., 1994, ApJ, 429, L119.

\bibitem[\protect\astroncite{Briley}{1997}]{Briley_1997}
Briley, M.M., Smith, V.V., King, J., Lambert, D.L. 1997,
AJ, 113, 306.


\bibitem[\protect\astroncite{Briley}{2001}]{Briley_2001}
Briley, M.M., Cohen, J.G., 2001, AJ, 122, 242

\bibitem[\protect\astroncite{Briley}{2001}]{Briley_2001a}
Briley, M.M., Smith, C.H., Claver, C.F. 2001, AJ, 122, 2561.

\bibitem[\protect\astroncite{Briley}{2002}]{Briley_2002}
Briley, M.M., Cohen, G.C., Stetson, P.B. 2002,
ApJ, 579, L17.




\bibitem[\protect\astroncite{Caretta}{1997}]{Caretta1997}
Caretta, E. \& Gratton, R.G. 1997, A\&AS, 121, 95.


\bibitem[\protect\astroncite{Charbonnel}{1994}]{Charbonnel1994}
Charbonnel, C., 1994, A\&A, 282, 811.


\bibitem[\protect\astroncite{Cotrell}{1994}]{Cotrell1978}
Cottrell, P. L., 1978. Apj, 223, 544


\bibitem[\protect\astroncite{Cudworth, K.M.}{1979}]{Cudworth1979}
Cudworth, K.M., 1979, AJ, 84, 1866.

\bibitem[\protect\astroncite{Cudworth,
K.M.}{1979a}]{Cudworth1979a}
Cudworth, K.M., 1979a, AJ, 84, 1866.

\bibitem[\protect\astroncite{Cudworth,
K.M.}{1979b}]{Cudworth1979b}
Cudworth, K.M., 1979b, AJ, 84, 1312.

\bibitem[\protect\astroncite{Cudworth, K.M.}{1985}]{Cudworth1985}
Cudworth, K.M., 1985, AJ, 90, 65.



\bibitem[\protect\astroncite{Denissenkov}{1996}]{DEnissenkov1996}
Denissenkov, P.A., Weiss, A. 1996, A.A. 1996, A\&A, 308, 773.

\bibitem[\protect\astroncite{Doyle}{1968}]{Doyle1968}
 Doyle, R.O. 1968, ApJ, 153, 987.


\bibitem[\protect\astroncite{Gilroy}{1991}]{Gilroy1991}
Gilroy, K.K., Brown, J.A., 1991, ApJ, 371, 578.



\bibitem[\protect\astroncite{Gohen}{2002}]{Cohen2002}
Cohen, J.G., Briley, M.M., Stetson, P.B. 2002, AJ, 123, 2525.


\bibitem[\protect\astroncite{Erikson}{2002}]{Erikson2002}
 Eriksson, K., Gustafsson, B.,
J\o rgensen, U. G., and Nordlung, \AA.  1984.
A\&A,  132, 37.


\bibitem[\protect\astroncite{Goorvitch}{1994}]{Goorvitch1994}
Goorvitch, D. 1994, Ap.J.Suppl.Ser., 95, 535.

\bibitem[\protect\astroncite{Gratton}{2001}]{Gratton2001}
Gratton, R.G., Bonifacio, P., Bragaglia, A., Carretta, E.,
Castellani, V.,
Centurion, M., Chieffi, A., Claudi, R., Clementini, G., D'antona, F.,
Desidera,
S., Francois, P., Grundahl, F., Lucatello, S., Molaro, P.,
Pasquini, L., Sneden,
C., Spite, F., Straniero, O. 2001, A\&A, 369, 87.

\bibitem[\protect\astroncite{Hauschildt}{1999}]{Hauschildt1999}
Hauschildt, Peter H.; Allard, F., Baron, E.
 1999, ApJ, 512, 377.


\bibitem[\protect\astroncite{Heiter}{2002}]{Heiter2002}
Heiter, U., Kupka, F., van't Veer- Menneret, C., Barban, C., Weiss, W.
W., Goupil, M.-J., Schmidt, W., Katz, D., Garrido, R.
2002, A\&A, 392, 619


\bibitem[\protect\astroncite{Iben}{1983}]{Iben1983}
Iben, Jr.I., Renzini, A. 1983, ARA\&A, 21, 271.

\bibitem[\protect\astroncite{Iben}{1964}]{Iben1964}
Iben, Jr.I. 1964, ApJ, 140, 1631.

\bibitem[\protect\astroncite{Ivans}{2001}]{Ivans2001}
Ivans, I.I., Kraft, R.P., Snenen, C., Smith, G.H.,
Rich, R.M., Shetrone, M. 2001, AJ, 122, 1438.

\bibitem[\protect\astroncite{Jones}{1994}]{Jones1994}
 Jones, H.R.A., Longmore, A. J., Jameson, R. F., Mountain, C. M.,
1994, MNRAS, 267, 413


\bibitem[\protect\astroncite{Jones}{2001}]{Jones2001}
Jones, H.R.A., Pavlenko, Y.V., Tennyson, J.,
Viti, S. 2002, MNRAS, 330, 675.



\bibitem[\protect\astroncite{Kraft}{1992}]{Kraft1992} Kraft,
R.P., Sneden, C., Langer, G.E. \& Prosser, C.F. 1992, Astron. J., 104,
645.

\bibitem[\protect\astroncite{Kraft}{1994}]{Kraft1994} Kraft, R.P.
1994, PASP, 106, 553.


\bibitem[\protect\astroncite{Kraft}{1997}]{Kraft1997} Kraft, R.P.,
Sneden, C., Smith, G.H., Shetrone, M.D., Langer, G.E., Pilachowski, C.A.
1997, AJ, 113, 279.


\bibitem[\protect\astroncite{Kupka}{1999}]{Kupka1999}
Kupka, F., Piskunov, N., Ryabchikova, T. A.,
                     Stempels, H. C., Weiss, W. W. 1999.
                  A\&A Suppl., 138, 119.




\bibitem[\protect\astroncite{Kurucz}{1993}]{Kurucz1993}
Kurucz, R.L., 1993, CD-ROM 18.


\bibitem[\protect\astroncite{Kurucz}{1999}]{Kurucz1999}
Kurucz, R.L., 1999. http://cfa5.harvard.edu


\bibitem[\protect\astroncite{Lambert}{1991}]{Lambert1981}
Lambert, D.L., Ries, L.M. 1981. ApJ, 248, 228.

\bibitem[\protect\astroncite{Malkan}{2002}]{metal2002}
Malkan M.A., Hicks E.K., Teplitz H.I., McLean I.M., Sugai H.,
Guichard J., 2002, ApJS, 142, 79


\bibitem[\protect\astroncite{Myerscough}{1968}]{Myerscough1968}
Myerscough, V. 1968, ApJ, 153, 421.

\bibitem[\protect\astroncite{Messenger}{2002}]{Messenger2002}
Messenger, B.B., Lattanzio, J.C. 2002, MNRAS, 331, 684.

\bibitem[\protect\astroncite{Pavlenko}{1991}]{Pavlenko1991}
Pavlenko, Y. V. 1991. Sov. Astron., v.35, p.212.


\bibitem[\protect\astroncite{Pavlenko}{1997}]{Pavlenko1997}
Pavlenko, Ya.V. 1997.
Astrophys. Space Sci., 253, 43.


\bibitem[\protect\astroncite{Pavlenko}{2000}]{Pavlenko2000}
Pavlenko, Ya. V. 2000. Astron. Rept. 44, 219

\bibitem[\protect\astroncite{Pavlenko}{2002}]{Pavlenko2002}
Pavlenko, Y. V. 2002. Proc. of IAU210, Uppsala, ed. N.Piskunov, in press
(astro-ph 0209022).

\bibitem[\protect\astroncite{Pavlenko}{2002}]{Pavlenko2002a}
Pavlenko, Y.V., Jones, H.R.A. 2002. A\&A., 396, 967.

\bibitem[\protect\astroncite{Pavlenko}{2003}]{Pavlenko2003}
Pavlenko, Y. V. 2003. Astron. Rept., 47, 59.

\bibitem[\protect\astroncite{Pavlenko}{2003a}]{Pavlenko2003a}
Pavlenko, Ya.V. \& Zhukovska, S.V. 2003, Kimemat. Phys. Celest.
Bodies, 19, 28.

\bibitem[\protect\astroncite{Pavlenko}{2003b}]{Pavlenko2003b}
Pavlenko, Ya.V. 2003b,
http://www.mao.kiev.ua/staff/yp/Results/CNO\_bf.tar.gz

\bibitem[\protect\astroncite{Ramirez}{2002}]{Ramirez2002}
Ramirez, S.V., Cohen, J.G. 2002, Astron. J., 123, 3277.

\bibitem[\protect\astroncite{Ramirez}{2002a}]{Ramirez2002a}
Ramirez, S.V., Cohen, J.G. 2002a, Astron. J., 125, 224.

\bibitem[\protect\astroncite{Partrige}{1997}]{Partrige1997}
Partrige, H., Schwenke, D.J. 1997,
Chem. Phys. 106, 4618.


\bibitem[\protect\astroncite{Salaris}{2002}]{Salaris2002}
Salaris, M., Cassisi, S., Weiss, A.,
 2002, PASP, 114, 375.

\bibitem[\protect\astroncite{Scwenke}{1997}]{Scwenke1997}
Schwenke, D.J. 1997.
http://george.arc.nasa.gov/\~dschwenke



\bibitem[\protect\astroncite{Pilacowski}{1997}]{Pilachowski1997}
Pilachowski, C., Sneden, C., Hinkle, K., Joyce, R.
 1997, AJ, 114, 819.

\bibitem[\protect\astroncite{Seaton}{1992}]{Seaton1992}
Seaton, M.J., Zeipen, C.J., Tully, J.A. 1992, Rev. Mexic. Astron.
Astrophys. 23, 107.



\bibitem[\protect\astroncite{Shetrone}{2003}]{Shetrone2003}
Shetrone, M.D. 2003, ApJ, 585, L45.

\bibitem[\protect\astroncite{Smith}{1989}]{Smith1989}
Smith,V.V., Suntzeff, N.B. 1989, AJ, 97, 1699.

\bibitem[\protect\astroncite{Smith}{1996}]{Smith1996}
Smith, G.H., Schetrone, M.D., Bell, R.A., Churchill, C.W.,
Briley, M.M. 1996, AJ, 112, 1511.

\bibitem[\protect\astroncite{Smith}{1997}]{Smith1997}
Smith, G.H., Shetrone, M.D., Briley, M.M., Churchill, C.W.,
Bell, R.A. 1997. PASP, 109, 236.


\bibitem[\protect\astroncite{Smith}{2000}]{Smith2000}
Smith, V.V., Suntzeff, N.B., Cunha, K.,
Allino, R., Busso M., Lambert, D.L., Straniero, O. 2000, AJ, 119, 1239.

\bibitem[\protect\astroncite{Smith}{2002}]{Smith2002}
Smith, G.H. 2002, PASP, 114, 1097.

\bibitem[\protect\astroncite{Smith}{2002}]{Smith2002a}
Smith, G.H., Terndrup, D.M., Suntzeff, N.B. 2002, AJ,
in press (astro-ph 020743).

\bibitem[\protect\astroncite{Sneden}{1976}]{Sneden1976}
Sneden, C.; Johnson, H. R.; Krupp, B. M. 1976, ApJ, 204, 218.


\bibitem[\protect\astroncite{Sneden}{1986}]{Sneden1986}
Sneden, C., Pilachovski, C.A. and VandenBerg, D.A. 1986,
ApJ, 311, 826.

\bibitem[\protect\astroncite{Sneden}{1992}]{Sneden1992}
Sneden, C., Kraft, R,P., Prosser, C., Langer, G.E. 1992,
AJ, 104, 2121.

\bibitem[\protect\astroncite{Sweigart}{1979}]{Sweigart1979}
Sweigart, A.V., \& Mengel, J.G. 1979, ApJ, 229, 624.

\bibitem[\protect\astroncite{Tsuji}{1973}]{Tsuji1973}
  Tsuji, T. 1973 A\&A, 23, 411.


\bibitem[\protect\astroncite{Unsold}{1955}]{Unsold1955}
 Unsold, A. 1955. Physik der
Sternatmospharen, springer Verlag, 2-nd ed.


\bibitem[\protect\astroncite{Vanture}{2002}]{Vanture2002}
Vanture, A.D., Wallerstein, G., Suntzeff, N.B. 2002, ApJ, 569, 984.

\bibitem[\protect\astroncite{Yakovina}{1984}]{Yakovina} Yakovina, L.A.
\& Pavlenko, Ya. V.  1998, Kimem. Phys. Celest. Bodies, 14, 195.



\bibitem[\protect\astroncite{Zinn}{1984}]{Zinn1994} Zinn, R. \&
West, M.J. 1984 ApJS, 55, 45.




\end{thebibliography}
\end{document}